\documentclass[a4paper,11pt]{article}

\usepackage{amsmath,amsfonts,amssymb,amsthm,graphicx,epsfig,multirow}
\usepackage[usenames]{color}
\usepackage{pstricks}
\usepackage{arydshln,cite}
\usepackage{hyperref}
\hypersetup{
colorlinks=true,
citecolor=red,
linkcolor=darkblue
}
\definecolor{darkblue}{rgb}{0,0,.7}
 \graphicspath{{./eps/}}
\numberwithin{equation}{section}
\usepackage{subfig}

\definecolor{purple}{rgb}{1,0,1}
\definecolor{darkpurple}{rgb}{1,.2,1}
\definecolor{pink}{rgb}{1,.7,.7}

\definecolor{apricot}{rgb}{1,0.9,0.7}

\newcommand{\nc}{\newcommand}
\nc\disp{\displaystyle}
\nc{\fh}{\hat{f}}
\nc{\muh}{\hat{\mu}}
\nc{\nuh}{\hat{\nu}}
\nc{\spos}[2]{\makebox(0,0)[#1]{$\sm{#2}$}}
\nc{\sm}[1]{{\scriptstyle #1}}
\nc{\qbar}{\overline{q}}

\nc{\bib}{\bibitem}
\nc{\al}{\alpha}
\nc{\g}{\gamma}
\nc{\G}{\Gamma}
\nc{\D}{\Delta}
\nc{\eps}{\epsilon}
\nc{\la}{\lambda}
\nc{\La}{\Lambda}
\nc{\var}{\varphi}
\nc{\pa}{\partial}
\nc{\nn}{\nonumber \\ }
\nc{\hf}{\frac{1}{2}}
\nc{\dz}{\frac{dz}{2\pi i}}
\nc{\bin}[2]{\left(\!\!\!\begin{array}{c} {#1}\\ {#2} \end{array}\!\!\!\right)}
\nc{\be}{\begin{equation}}
\nc{\ee}{\end{equation}}
\nc{\bea}{\begin{eqnarray}}
\nc{\eea}{\end{eqnarray}}
\nc{\bra}[1]{\langle {#1}|}
\nc{\ket}[1]{|{#1}\rangle}
\nc{\ketw}[1]{({#1})^{\phantom{a}}_{{\cal W}}}
\nc{\chit}{\raisebox{0.25ex}{$\chi$}}
\nc{\chih}{\raisebox{0.25ex}{$\hat\chi$}}
\nc{\Db}{\mbox{\boldmath $D$}}
\nc{\Hb}{\mbox{\boldmath $H$}}
\nc{\calH}{{\cal H}}
\nc{\calR}{{\cal R}}
\nc{\calL}{{\cal L}}
\nc{\calV}{{\cal V}}
\nc{\Hc}{{\cal H}}
\nc{\Rc}{{\cal R}}
\nc{\Lc}{{\cal L}}
\nc{\Vc}{{\cal V}}
\nc{\Ib}{\mbox{\boldmath $I$}}
\nc{\qb}{\bar{q}}
\nc{\Ac}{\mathcal{A}}
\nc{\Bc}{\mathcal{B}}
\nc{\Cc}{\mathcal{C}}
\nc{\Dc}{\mathcal{D}}
\nc{\Ec}{\mathcal{E}}
\nc{\Gc}{\mathcal{G}}
\nc{\Ic}{\mathcal{I}}
\nc{\Jc}{\mathcal{J}}
\nc{\Oc}{\mathcal{O}}
\nc{\Pc}{\mathcal{P}}
\nc{\Sc}{\mathcal{S}}
\nc{\Tc}{\mathcal{T}}
\nc{\Wc}{\mathcal{W}}
\nc{\Xc}{\mathcal{X}}
\nc{\Yc}{\mathcal{Y}}
\nc{\Zc}{\mathcal{Z}}
\nc{\fus}{\mbox{}\,\hat\otimes\,\mbox{}}
\nc{\Pch}{\hat{\Pc}}
\nc{\Rch}{\hat{\Rc}}
\nc{\Dh}{\hat{\Delta}}
\nc{\rh}{\hat{r}}
\nc{\sh}{\hat{s}}
\nc{\taub}{\bar{\tau}}
\nc{\Jcb}{\Jc_{\mathrm{b}}}
\nc{\rtt}{\mathtt{r}}
\nc{\stt}{\mathtt{s}}
\nc{\cosR}{\cos\frac{\pi p'rr'}{p}}
\nc{\cosS}{\cos\frac{\pi pss'}{p'}}
\nc{\sinR}{\sin\frac{\pi p'rr'}{p}}
\nc{\sinS}{\sin\frac{\pi pss'}{p'}}
\def\vvdots{\mathinner{\mkern1mu\raise1pt\vbox{\kern7pt\hbox{.}}\mkern2mu
  \raise4pt\hbox{.}\mkern2mu\raise7pt\hbox{.}\mkern1mu}}
\nc{\gauss}[2]{\big[\!\!\begin{array}{c} {#1}\\ {#2} \end{array}\!\!\big]}
\nc{\bgauss}[2]{\Big[\!\!\begin{array}{c} {#1}\\ {#2} \end{array}\!\!\Big]}
\nc{\sbin}[2]{\left\{\!\!\!\begin{array}{c} {#1}\\ {#2}
\end{array}\!\!\!\right\}}
\nc{\sbinlr}[2]{\Big\langle\!\!\begin{array}{c} {#1}\\ {#2}
\end{array}\!\!\Big\rangle}
\nc{\bino}[2]{\left(\!\!\begin{array}{c} {#1}\\ {#2} \end{array}\!\!\right)}
\def\half {\mbox{$\textstyle \frac{1}{2}$}}
\def\vec#1{\mbox {\boldmath $#1$}}

\definecolor{lightblue}{rgb}{.61,.61,1}
\definecolor{midblue}{rgb}{.7,.7,1}
\definecolor{lightlightblue}{rgb}{.85,.85,1}
\definecolor{lightestblue}{rgb}{.96,.96,1}
\definecolor{lightpurple}{rgb}{1,.65,1}

\def\facegrid#1#2{
\psframe[fillstyle=solid,fillcolor=lightlightblue,linewidth=0pt]#1#2
\psgrid[gridlabels=0pt,subgriddiv=1]#1#2}

\nc{\ch}{{\rm ch}}
\nc{\R}{{\cal R}}
\nc{\dkk}{\delta_{j,\{k,k'\}}^{(2)}}
\nc{\drr}{\delta_{j,\{r,r'\}}^{(2)}}
\nc{\ddkk}{\delta_{j,\{k,k'\}}^{(4)}}
\nc{\dddkk}{\delta_{j,\{k,k'\}}^{(8)}}
\nc{\dnn}{\delta_{j,\{n,n'\}}^{(2)}}
\nc{\ddnn}{\delta_{j,\{n,n'\}}^{(4)}}
\nc{\dddnn}{\delta_{j,\{n,n'\}}^{(8)}}

\definecolor{pink}{rgb}{1,.65,.65}

\thicklines

\def\faceu#1#2#3#4#5{\ \
\begin{pspicture}[shift=-.6](0,-.25)(1,1.25)
\pspolygon[linewidth=.5pt,fillstyle=solid,fillcolor=lightlightblue](0,0)(1,0)(1,1)(0,1)(0,0)
\rput[tr](0,0){\scriptsize $#1$}
\rput[tl](1,0){\scriptsize $#2$}
\rput[bl](1,1){\scriptsize $#3$}
\rput[br](0,1){\scriptsize $#4$}
\rput(.5,.5){\small $#5$}
\end{pspicture}}

\nc{\Wthree}[5]{W^{3,3}\!\!\left.\left(
\begin{matrix}
#4 & #3\\
#1 & #2
\end{matrix}
\right|#5\right)}

\nc{\Wn}[5]{W^{n,n}\Big.\Big(\,
\begin{matrix}
#4 & #3\\
#1 & #2
\end{matrix}\,
\Big|#5\Big)\,}

\nc{\Wtwo}[5]{W^{2,2}\Big.\Big(
\begin{matrix}
#4 & #3\\
#1 & #2
\end{matrix}
\Big|#5\Big)}

\nc{\Wt}[4]{W^{2,2}\!\left(
\begin{matrix}
#4 & #3\\
#1 & #2
\end{matrix}
\right)}

\nc{\Wtt}[6]{#1\Big.\Big(
\begin{matrix}
#5 & #4\\
#2 & #3
\end{matrix}
\Big|#6\Big)}

\nc{\chh}{\widehat{\mathrm{ch}}}
\nc{\ph}{\hat{p}}
\def\equiveq{=}

\nc{\thf}{s}
\nc{\lam}{\lambda}
\nc{\round}[1]{\left(#1\right)}
\nc{\therm}{t}
\def\sfloor#1{\lfloor #1\rfloor}
\def\floor#1{\left\lfloor #1\right\rfloor}
\def\bfloor#1{\big\lfloor #1\big\rfloor}
\def\Bfloor#1{\Big\lfloor #1\Big\rfloor}
\nc{\flexpr}[1]{
h_{#1}
}

\def \trinomial[#1][#2][#3][#4]{\left[{#1\atop #2,#3,#4}\right]}

\def \superTrinomial[#1][#2]{
\left({#1 \atop #2} \right)_{\!2}
}

\def \qTrinomial[#1][#2][#3][#4]{\left[{#1\atop #2,#3,#4}\right]_{\!q}}

\nc{\mydot}[0]{
\pscircle[fillstyle=solid,fillcolor=black](0,0){0.08}
}

\begin{document}

\topmargin -5mm \oddsidemargin 5mm

\setcounter{page}{1}

\mbox{}\vspace{12mm}
\begin{center}
{\huge {\bf Fused RSOS Lattice Models as Higher-Level\\[4pt] Nonunitary Minimal Cosets}}

\vspace{7mm}
{\Large Elena Tartaglia and Paul A.\;Pearce}\\[.3cm]
{\em School of Mathematics and Statistics, University of Melbourne}\\
{\em Parkville, Victoria 3010, Australia}\\[.4cm]
{\tt elena.tartaglia\,@\,unimelb.edu.au}
\qquad
{\tt p.pearce\,@\,ms.unimelb.edu.au}
\end{center}

\vspace{15mm}
\centerline{{\bf{Abstract}}}
\vskip.3cm
\noindent
We consider the Forrester-Baxter RSOS lattice models with crossing parameter $\lambda=(m'\!-\!m)\pi/m'$ in Regime~III. In the continuum scaling limit, these models are described by the minimal models ${\cal M}(m,m')$. 
We conjecture that, for $\lambda<\pi/n$, the $n\times n$ fused RSOS models with $n\ge 2$ are described by the higher-level coset 
$(A^{(1)}_1)_k\otimes (A^{(1)}_1)_n/(A^{(1)}_1)_{k+n}$ at fractional level $k=nM/(M'\!-\!M)-2$ with
$(M,M')=\big(nm-(n\!-\!1)m',m'\big)$. To support this conjecture, we investigate the one-dimensional sums arising from Baxter's off-critical corner transfer matrices. In unitary cases ($m=m'\!-\!1$) it is known  that, up to leading powers of $q$, these coincide with the branching functions $b_{r,s,\ell}^{m'\!-n,m'\!,n}(q)$. For general nonunitary cases ($m<m'\!-\!1$), we identify the ground state one-dimensional RSOS paths and relate them to the quantum numbers $(r,s,\ell)$ in the various sectors. For $n=1,2,3$, we obtain the local energy functions $H(a,b,c)$ in a suitable gauge and verify that the associated one-dimensional sums produce finitized forms that converge, as $N$ becomes large, to the fractional level branching functions $b_{r,s,\ell}^{M,M'\!,n}(q)$. Extending the work of Schilling, we also conjecture finitized bosonic branching functions $b_{r,s,\ell}^{M,M'\!,n;(N)}(q)$ for general $n$ and check that these agree with the one-dimensional sums for $n=1,2,3$ out to system sizes $N=14$. Lastly, the finitized Kac characters $\chi_{r,s,\ell}^{P,P'\!,n;(N)}(q)$ of the $n\times n$ fused logarithmic minimal models ${\cal LM}(p,p')$ are obtained by taking the {\em logarithmic limit\/} $m,m'\to\infty$ with $m'/m\to p'/p+$.

\renewcommand{\thefootnote}{\arabic{footnote}}
\setcounter{footnote}{0}

\newpage
\tableofcontents

\newpage
\section{Introduction}

Solid-On-Solid (SOS) models with unbounded heights and Restricted Solid-On-Solid (RSOS) models with bounded heights were originally used~\cite{Beijeren,Knops,ChuiWeeks,Luck} to study fluctuating surfaces and the roughening transition. The SOS condition ensures columns of solid with no overhangs. 
The $A_{m'-1}$ $\mbox{RSOS}(m,m')$ models of Andrews, Baxter and Forrester~\cite{ABF84,FB84} in Regime III, with heights $1\le a\le m'-1$, are exactly solvable RSOS models in the sense that they are Yang-Baxter integrable~\cite{BaxBook}. The RSOS$(m,m')$ models have crossing parameter $\lambda=(m'-m)\pi/m'$ with $2\le m<m'$ and $\mbox{gcd}(m,m')=1$. The $\mbox{RSOS}(3,4)$ model is the Ising model and the $\mbox{RSOS}(m'\!-\!1,m')$ models with $m'\ge 4$ realize~\cite{Huse84} the ${\Bbb Z}_2$ universality classes of multicritical Ising models. The universal behaviour of such systems is described in the continuum by Conformal Field Theories (CFTs)~\cite{FMS1987}. 
Indeed it is now well established that, in the continuum scaling limit, the $\mbox{RSOS}(m,m')$ models in Regime III realize the unitary~\cite{FQS84} $(m=m'\!-\!1)$ and nonunitary~\cite{ISZ86,Riggs89,Nakanishi90,FodaW} $(m<m'\!-\!1$) minimal models 
${\cal M}(m,m')$ of Belavin, Polyakov and Zamolodchikov~\cite{BPZ84}.  In fact, the minimal models ${\cal M}(m,m')$ represent the simplest family of rational coset CFTs~\cite{GKO1,GKO2}.

On the one hand, lattice fusion \cite{KRS81,AndrewsBaxter,DJMO,DJKMO} of $n\times n$ blocks of elementary face weights of the RSOS$(m,m')$ models yields the fused Yang-Baxter integrable lattice models $\mbox{RSOS}(m,m')_{n\times n}$.  On the other hand, higher fusion level minimal models ${\cal M}(M,M'\!,n)$ at integer fusion level  $n\ge 1$ and fractional level $k=nM/(M'\!-\!M)-2$ can be constructed~\cite{KacPeterson84,KastorMQ88,BaggerNY88,Ravanini88,ACT91,BMSW97} as the Goddard-Kent-Olive (GKO) cosets {$(A^{(1)}_1)_k\otimes (A^{(1)}_1)_n/(A^{(1)}_1)_{k+n}$} (\ref{cosetAAA}). For $n=1$, these are the minimal models~\cite{BPZ84,GKO1} ${\cal M}(m,m')={\cal M}(M,M'\!,1)$ with $(M,M')=(m,m')$. For $n=2$, these are the superconformal minimal models~\cite{GKO2} ${\cal M}(M,M'\!,2)$.
For unitary cases ($m=m'\!-\!1$) the known identification~\cite{AndrewsBaxter,DJMO,DJKMO,BR89,KP92,Schilling96}, 
in the continuum scaling limit, is given by $(M,M')=(m'-n,m')$. In particular, the one-dimensional sums arising from Baxter's Corner Transfer Matrices (CTMs)~\cite{CTM} coincide, up to leading powers of $q$, with the coset branching functions $b_{r,s,\ell}^{M'-n,M'\!,n}(q)$~\cite{KacPeterson84}. This observation is in agreement with the {\it correspondence principle} of the Kyoto school~\cite{JimboMiwa} which is generally valid in Regime~III. 

The consequences of the identification of the unitary 
$\mbox{RSOS}(m'\!-\!1,m')_{n\times n}$ model with the coset CFT ${\cal M}(m'-n,m',n)$ are twofold. First the conformal data, and therefore the universality class and all of the universal critical exponents, are exactly determined for these $\mbox{RSOS}(m,m')$ lattice models. Second, the integrable lattice regularizations can be used~\cite{PearceNienhuis} to study further properties of these coset CFTs and their integrable off-critical $\varphi_{1,3}$ thermal perturbations.

In this paper, we generalize the identification of $n\times n$ fused $\mbox{RSOS}(m,m')$ lattice models with coset CFTs to the nonunitary cases. Specifically we conjecture that, for $\lambda<\pi/n$, the general nonunitary $n\times n$ fused $\mbox{RSOS}(m,m')$ lattice models with $m<m'-1$ are described, in the continuum scaling limit, by the higher-level coset $(A^{(1)}_1)_k\otimes (A^{(1)}_1)_n/(A^{(1)}_1)_{k+n}$ at fractional level $k=nM/(M'\!-\!M)-2$ with $(M,M')=\mbox{$(nm-(n\!-\!1)m',m')$}$
\bea
\mbox{RSOS}(m,m')_{n\times n}\simeq {\cal M}(M,M'\!,n),\qquad (M,M')=\mbox{$(nm-(n\!-\!1)m',m')$}\label{conjecture}
\eea
To support this conjecture, we investigate the one-dimensional sums arising from Baxter's off-critical CTMs. We identify the ground state one-dimensional RSOS paths and relate them to the quantum numbers $(r,s,\ell)$ in the various sectors. For $n=1,2,3$, we obtain local energy functions {$H(\sigma_{j-1},\sigma_j,\sigma_{j+1})$} in a suitable gauge and, using Mathematica~\cite{Wolfram}, we verify that the one-dimensional sums produce finitized fermionic forms which, for large $N$, converge to the fractional level branching functions $b_{r,s,\ell}^{M,M'\!,n}(q)$. 
Lastly, extending the results of Schilling~\cite{Schilling96}, we conjecture finitized bosonic branching functions $b_{r,s,\ell}^{M,M'\!,n;(N)}(q)$ for general $n$ using $q$-multinomials~\cite{Schilling96,Warnaar97} and check that these agree with the one-dimensional sums for $n=1,2,3$.

There are a number mathematical and physical motivations for studying nonunitary RSOS models arising from different areas of application. Although nonunitary RSOS models in two dimensions have some negative Boltzmann face weights and lack a strict probabilistic interpretation, the associated CFTs and $1$-dimensional quantum Hamiltonians are well defined and physical. Indeed, the form of the critical Hamiltonians in Regime~III was recently generalized to off-critical Hamiltonians in \cite{BianchiniEtAl}. In statistical mechanics, the Lee-Yang ${\cal M}(2,5)$ theory~\cite{LeeYang} describes~\cite{Fisher,Cardy,CardyMussardo}
the closing, in the complex magnetic field plane, of the gap in the distribution of Lee-Yang
zeros of the two-dimensional ferromagnetic Ising model ${\cal M}(3,4)$. In Quantum Field Theory (QFT), ${\cal M}(2,5)$~\cite{Zamolodchikov,BajnokEtAl} and ${\cal M}(3,5)$~\cite{Deeb} describe simple one-particle massive scattering theories. More general theories describe scattering theories with kinks and breathers. The $\mbox{RSOS}(m,m')$ models provide lattice regularizations of these field theories. In the context of anyons and the Fractional Quantum Hall Effect (FQHE), ${\cal M}(3,5)$~\cite{FQHE} has been used to describe spinless electrons at filling fraction $\nu=2/5$. In condensed matter physics, nonunitary RSOS models have also been studied recently to shed light on the properties of quantum entanglement~\cite{BianchiniRavanini}. It is therefore generally important to understand the physical consequences of the loss of unitarity. This is particularly relevant in the study of statistical systems with nonlocal degrees of freedom which are described by logarithmic CFTs~\cite{Gurarie,specialissue} and are invariably nonunitary. Such theories, exemplified by the logarithmic minimal models ${\cal LM}(p,p')$~\cite{PRZ} and their fused counterparts~\cite{PR2013,PRT2014}, can be studied by taking the logarithmic limit~\cite{RasLogLimit} of rational nonunitary minimal models as in Section~\ref{SectLogLimit}. These exactly solvable families contain many generalized lattice models of critical polymers and critical percolation in two dimensions. Perhaps most importantly, from a mathematical perspective, the fusion hierarchies RSOS$(m,m')_{n\times n'}$ encode~\cite{KP92} the integrability of these nonunitary minimal models through their $T$- and $Y$-systems.

\goodbreak
The layout of the paper is as follows. 
In Section~2, we describe cosets with integer fusion level $n$ and fractional level $k$ and present explicit formulas for their central charges, conformal weights and branching functions. 
In Section~3, we use techniques coming from CTMs  to set up the associated one-dimensional sums of the $n\times n$ fused $\mbox{RSOS}(m,m')$ lattice models in
Regime~III. For $n=1,2,3$, we explicitly calculate the local energies. 
In Section~4, generalizing the results of Schilling~\cite{Schilling96} to nonunitary cases, we present a conjecture for the finitized bosonic branching functions for general $n$. For $n=1,2,3$ and out to order $N=14$, we verify that (up to the leading terms involving the
central charges and conformal dimensions) the one-dimensional 
sums agree with the bosonic forms and give finitized forms of the fractional level branching functions $b_{r,s,\ell}^{M,M'\!,n}(q)$. 
In Appendix~A, we collect relevant properties of elliptic functions. In Appendix~B, we establish the counting and 
properties of contiguous shaded or unshaded bands.  In Appendix~C, we list the explicit fused RSOS face weights for $n=2$ and $n=3$. 
In Appendix~D, we list the explicit conjugate modulus forms of the diagonal fused face weights for $n=2$ and $n=3$. 
We finish with some concluding remarks.

\section{Higher-Level Nonunitary Minimal Cosets ${\cal M}(M,M'\!,n)$}

\subsection{Coset construction and central charges}

The minimal models ${\cal M}(m,m')$~\cite{BPZ84}, with coprime integers $m,m'$ satisfying $2\le m<m'$, are rational Conformal Field Theories (CFTs) with central charges
\bea
 c=1-{6(m-m')^2\over mm'},\qquad 2\le m<m',\qquad\mbox{gcd}(m,m')=1
\eea
The conformal weights and associated Virasoro characters are
\begin{align}
\Delta_{r,s}^{m,m'}&={(rm'-sm)^2-(m-m')^2\over 4mm'},\quad 1\le r\le m-1,\ \ 1\le s\le m'-1\\
\mbox{ch}^{m,m'}_{r,s}(q)&={q^{-c/24+\Delta_{r,s}^{m,m'}}\over
(q)_\infty}\!\!\! \sum_{k=-\infty}^\infty \big[q^{k(k m
m'+rm'-sm)}-q^{(km+r)(km'+s)}\big] \label{vira}
\end{align}
where the $q$-factorials are
\bea
(q)_n =
\prod_{k=1}^{n}(1-q^k),\qquad (q)_\infty =
\prod_{k=1}^{\infty}(1-q^k) \label{qn}
\eea
In these expressions,
$q=e^{\pi i\tau}$ is the modular nome.

Algebraically, the higher-level minimal models are constructed~\cite{ACT91,BMSW97} as cosets
\bea
 {\cal M}(M,M'\!,n)\simeq \mbox{COSET}\Big(\frac{nM}{M'\!-\!M}-2,n\Big),\ \mathrm{gcd}\Big(\frac{M'\!-\!M}{n},M'\Big)=1,\  
    2\leq M<M',\ n,M,M'\in\mathbb{N}
\label{MinCoset}
\eea
where $n=1,2,\ldots$ is the integer fusion level, $k=\frac{nM}{M'\!-\!M}-2$ is the fractional fusion level and $\mathbb{N}$ denotes the set of positive integers. The diagonal GKO coset~\cite{GKO1,GKO2} takes the form
\be
 \mbox{COSET}(k,n):\quad \frac{(A_1^{(1)})_k\oplus(A_1^{(1)})_n}{(A_1^{(1)})_{k+n}},
 \qquad k=\frac{\ph}{\ph'}-2,\qquad \mathrm{gcd}(\ph,\ph')=1,\qquad n,\ph,\ph'\in\mathbb{N}
\label{cosetAAA}
\ee
where the subscripts on the 
affine $su(2)$ current algebra $A_1^{(1)}$ denote the respective levels $k$, $n$ and $k+n$.
The central charge of the coset Virasoro algebra is thus given by
\be
 c=c_k+c_n-c_{k+n}
  =\frac{3kn(k+n+4)}{(k+2)(n+2)(k+n+2)}
\ee
where the central charge of the affine current algebra $(A_1^{(1)})_k$ is 
\be
 c_k=\frac{3k}{k+2}
\label{ck}
\ee
The central charges of the minimal models ${\cal M}(M,M'\!,n)$ are thus
\bea
c^{M,M'\!,n}={3n\over n+2} \Big[1-{2(n+2)(M'-M)^2\over n^2 M M'}\Big],\qquad 2\le M<M',\qquad \mbox{gcd}\Big({M'-M\over n},M'\Big)=1
\label{centralcharges}
\eea
The usual minimal models~\cite{BPZ84} are given by $n=1$. The superconformal minimal models are given by the specialization $n=2$ with central charges
\bea
c^{M,M'\!,2}={3\over 2} \Big[1-{2(M'-M)^2\over M M'}\Big],\qquad 2\le M<M',\qquad \mbox{gcd}\Big({M'-M\over 2},M'\Big)=1
\label{supercentralcharges}
\eea

\subsection{Branching functions}

The characters of the higher fusion level minimal models ${\cal M}(M,M'\!,n)$ are given by the branching functions  $b_{r,s,\ell}^{M,M'\!,n}(q)$~\cite{ACT91,BMSW97} of the coset (\ref{MinCoset}).
These are expressible in terms of the
string functions~\cite{KacPeterson84,JM84,GQ87,DQ90,HNY90} of $\mathbb{Z}_n$ 
parafermions with central charge $c=\frac{2n-2}{n+2}$. For the fundamental domain
\be
 0\leq m\leq\ell\leq n,\qquad \ell-m\in2\mathbb{Z},\qquad m,\ell=0,1,\ldots,n,\qquad 
n\in {\mathbb N}
\label{stringdomain}
\ee 
the string functions are given by
\bea
 c_m^\ell(q)&=&\frac{q^{-\frac{1}{24}\frac{2n-2}{n+2}+\frac{\ell(\ell+2)}{4(n+2)}-\frac{m^2}{4n}}}{(q)_\infty^3}
  \sum_{i,j=0}^\infty(-1)^{i+j}q^{ij(n+1)+\frac{1}{2}i(i+1)+\frac{1}{2}j(j+1)}\nn
&&\qquad\quad\mbox{}\times\big[q^{\frac{i}{2}(\ell+m)+\frac{j}{2}(\ell-m)}-q^{n-\ell+1+\frac{i}{2}(2n+2-\ell-m)+\frac{j}{2}(2n+2-\ell+m)}\big]\label{string}
\eea
where the dependence on $n$ has been suppressed. The $\mathbb{Z}_n$ parafermionic index $m$ should not be confused with the minimal model label $m$ in $(m,m',n)$. 
The fundamental domain of definition (\ref{stringdomain}) of the string functions
is extended to the domain
\be
 \ell=0,1,\ldots,n,
 \qquad m\in\mathbb{Z},\qquad n\in\mathbb N
\label{elldomain}
\ee
by setting $c_m^\ell(q)=0$ for $\ell-m\notin 2\mathbb{Z}$ and using the symmetries
\bea
 c_m^\ell(q)=c_{-m}^\ell(q)=c_{n-m}^{n-\ell}(q)=c_{m+2n}^\ell(q)\label{stringsym}
\eea
so that $c_m^\ell(q)$ is even and periodic in $m$ with period $2n$. 

Explicitly, on the checkerboard $r+s=\ell$ mod 2, the branching functions are given by
\bea
\begin{array}{rcl}
b_{r,s,\ell}^{M,M'\!,n}(q)&\!\!=\!&\!\disp q^{\Delta_{r,s}^{M,M'\!,n}-\frac{c^{M,M'\!,n}}{24}+\frac{n-1}{12(n+2)}}\\
 &&\times\!\!\disp \sum_{0\le m\le n/2\atop m\equiveq\ell/2\;\text{\tiny mod 1}} \!\!\!\!\!c_{2m}^\ell(q) 
\bigg[\!\!\sum_{j\in\mathbb{Z}\atop m_{r-s}(j)= \pm m\;\text{\tiny mod}\;n}\!\!\!\!\!\!\!\!\!\!\!\!\! q^{{j\over n}(jM M'+rM'-sM)}
-\!\!\!\!\!\!\sum_{j\in\mathbb{Z}\atop m_{r+s}(j)= \pm m\;\text{\tiny mod}\;n}\!\!\!\!\!\!\!\!\!\!\!\!\! q^{{1\over n}(jM'+s)(jM+r)}\bigg],\label{branchfns}\\[22pt]
&&\ \ 1\le r\le M-1,\qquad 1\le s\le M'-1,\qquad \ell=0,1,\ldots,n
\end{array}\\[-12pt]
\nonumber
\eea
where the first sum (on $m$) runs over integers ($\ell$ even) or half odd integers ($\ell$ odd) with
\bea
 m_a(j):=a/2+jM'
\eea
The cosets are quotients of the algebra $(A_1^{(1)})_k\oplus(A_1^{(1)})_n$ by the subalgebra $(A_1^{(1)})_{k+n}$. 
It follows that products of the characters of the algebras $(A_1^{(1)})_k$ and $(A_1^{(1)})_n$ decompose as linear sums of the characters 
of $(A_1^{(1)})_{k+n}$. The coefficients are branching functions which play the role of the multiplicities in the restriction of classical groups. 
These branching functions satisfy the decomposition or {\em branching rules}~\cite{KacPeterson84,ACT91,BMSW97}
\bea
\chh_{r,s}^{\ph,\ph'}(q,z)\,\chh_{r',0}^{n+2,1}(q,z)
  =\sum_{\mbox{\tiny$\sigma=1$}\atop \mbox{\tiny $\sigma= r\!+\!\ell$ mod 2}}^{\ph+n\ph'-1}
    b_{r,\sigma,\ell}^{\,\ph,\ph+n\ph'\!,n}(q)\,\chh_{\sigma,s}^{\ph+n\ph'\!,\ph'}\!(q,z),\quad
    \ph=M,\ \ \ph'=\frac{M'\!-\!M}{n}
\eea
relating admissible characters $\chh_{r,s}^{\ph,\ph'}(q,z)$ of affine current algebras $(A_1^{(1)})_k$, $(A_1^{(1)})_n$, $(A_1^{(1)})_{k+n}$ with
\be
 r'=
\begin{cases}
 n+1-\ell,&\mbox{$s$ odd}
\\
 \ell+1,&\mbox{$s$ even}
\end{cases}\qquad\ \  \ell=0,1,\ldots,n
\ee
For $n=1$ and $n=2$, the branching functions reduce to the Virasoro minimal and superconformal characters respectively.

\subsection{Conformal weights and Kac tables}

Explicitly, for $ r+s\equiveq \ell\ \mathrm{mod}\ 2$, the ${\cal M}(M,M'\!,n)$ conformal weights are \cite{PR2013}
\bea
\D_{r,s,\ell}^{M,M'\!,n}= \D_{r,s}^{M,M'\!,n}+\D_{r-s}^{\ell,n}+\mbox{Max}[\half(\ell\!+\!2\!-\!r\!-\!s),0]+\mbox{Max}[\half\big(\ell'\!+\!2\!-\!(M\!-r)\!-\!(M'\!-\!s)\big),0]
\label{genConfWts}
\eea
The first term on the right and $\ell'$ are given by
\bea
\D_{r,s}^{M,M'\!,n}=\frac{(rM'-sM)^2-(M'-M)^2}{4nM M'},\qquad\ 
\ell'=\begin{cases}
\ell,&\mbox{$\frac{M'-M}{n}$ even}\\[2pt]
n-\ell,&\mbox{$\frac{M'-M}{n}$ odd}
\end{cases}
\eea
Setting $\overline{m}=m$ mod $2n$, the second term is the conformal weight of the string function $c_m^\ell(q)$ 
\be
\Delta_m^{\ell,n}=\mbox{Max}[\Delta(\overline{m},\ell,n),\Delta(2n\!-\!\overline{m},\ell,n),\Delta(n\!-\!\overline{m},n\!-\!\ell,n)],\qquad
\Delta(m,\ell,n)={\ell(\ell+2)\over 4(n+2)}-{m^2\over 4n}
\label{stringConfWts}
\ee
folded into the fundamental domain (\ref{stringdomain}). 
The third term only gives a nonzero contribution for $r+s\le\ell\le n$. 
The fourth term is the counterpart of the third term under the Kac table symmetry. 
It only contributes for $r+s\ge M+M'-\ell'\ge M+M'-n$. 
The conformal weights are thus conveniently organized into $n+1$ layered Kac tables each displaying the 
checkerboard pattern and satisfying the Kac table symmetry
\bea
 \D_{r,s,\ell}^{M,M'\!,n}= \begin{cases}     
 \D_{M-r,M'-s,\ell}^{M,M'\!,n}&\ \ \mbox{$\frac{M'-M}{n}$ even}\\[8pt]
 \D_{M-r,M'-s,n-\ell}^{M,M'\!,n}&\ \ \mbox{$\frac{M'-M}{n}$ odd}
 \end{cases}
\eea
The Kac tables of ${\cal M}(3,7,2)$ and ${\cal M}(5,7,2)$ are shown in Table~\ref{KacTables}.

\begin{table}[htb]
{\vspace{.5cm}\psset{unit=1cm}
{\small
\begin{center}
\small
\qquad
\begin{pspicture}(0,0)(2,6)
\psframe[linewidth=0pt,fillstyle=solid,fillcolor=lightlightblue](0,0)(2,6)
\multirput(0,0)(0,2){3}{\multirput(0,0)(2,0){1}{\psframe[linewidth=0pt,fillstyle=solid,fillcolor=lightestblue](0,0)(1,1)}}
\multirput(0,0)(0,2){3}{\multirput(1,1)(2,0){1}{\psframe[linewidth=0pt,fillstyle=solid,fillcolor=lightestblue](0,0)(1,1)}}
\psgrid[gridlabels=0pt,subgriddiv=1]
\rput(.5,5.5){$\frac{11}{16}$}\rput(1.5,5.5){$0,\!\frac 32$}
\rput(.5,4.5){$\frac 27,\frac{11}{14}$}\rput(1.5,4.5){${-}\frac 3{112}$}
\rput(.5,3.5){$\frac{13}{112}$}\rput(1.5,3.5){$\frac 37,\!\!{-}\frac{1}{14}$}
\rput(.5,2.5){$\frac 37,\!{-}\frac 1{14}$}\rput(1.5,2.5){$\frac{13}{112}$}
\rput(.5,1.5){${-}\frac{3}{112}$}\rput(1.5,1.5){$\frac 27,\frac{11}{14}$}
\rput(.5,.5){$0,\frac 32$}\rput(1.5,.5){$\frac{11}{16}$}
{\color{blue}
\rput(.5,-.5){$1$}
\rput(1.5,-.5){$2$}
\rput(2.25,-.5){$r$}
\rput(-.5,.5){$1$}
\rput(-.5,1.5){$2$}
\rput(-.5,2.5){$3$}
\rput(-.5,3.5){$4$}
\rput(-.5,4.5){$5$}
\rput(-.5,5.5){$6$}
\rput(-.5,6.25){$s$}}
\end{pspicture}
\qquad
\qquad\qquad\ \ 
\begin{pspicture}(0,0)(4,6)
\psframe[linewidth=0pt,fillstyle=solid,fillcolor=lightlightblue](0,0)(4,6)
\multirput(0,0)(0,2){3}{\multirput(0,0)(2,0){2}{\psframe[linewidth=0pt,fillstyle=solid,fillcolor=lightestblue](0,0)(1,1)}}
\multirput(0,0)(0,2){3}{\multirput(1,1)(2,0){2}{\psframe[linewidth=0pt,fillstyle=solid,fillcolor=lightestblue](0,0)(1,1)}}
\psgrid[gridlabels=0pt,subgriddiv=1]
\rput(.5,5.5){$\frac{31}{16}$}\rput(1.5,5.5){$\frac 9{10},\!\frac 75$}\rput(2.5,5.5){$\frac {27}{80}$}\rput(3.5,5.5){$\frac 32,0$}
\rput(.5,4.5){$\frac 87,\frac {23}{14}$}\rput(1.5,4.5){$\frac {269}{560}$}\rput(2.5,4.5){$\frac{19}{35},\!\frac 3{70}$}\rput(3.5,4.5){$\frac{9}{112}$}
\rput(.5,3.5){$\frac{73}{112}$}\rput(1.5,3.5){$\frac {43}{70},\!\frac{4}{35}$}\rput(2.5,3.5){$\frac{29}{560}$}\rput(3.5,3.5){$\frac3{14},\frac 57$}
\rput(.5,2.5){$\frac 57,\!\frac 3{14}$}\rput(1.5,2.5){$\frac{29}{560}$}\rput(2.5,2.5){$\frac 4{35},\!\frac{43}{70}$}\rput(3.5,2.5){$\frac{73}{112}$}
\rput(.5,1.5){$\frac{9}{112}$}\rput(1.5,1.5){$\frac 3{70},\!\frac{19}{35}$}\rput(2.5,1.5){$\frac{269}{560}$}\rput(3.5,1.5){$\frac{23}{14},\!\frac 87$}
\rput(.5,.5){$0,\frac 32$}\rput(1.5,.5){$\frac{27}{80}$}\rput(2.5,.5){$\frac 75,\!\frac 9{10}$}\rput(3.5,.5){$\frac{31}{16}$}
{\color{blue}
\rput(.5,-.5){$1$}
\rput(1.5,-.5){$2$}
\rput(2.5,-.5){$3$}
\rput(3.5,-.5){$4$}
\rput(4.25,-.5){$r$}
\rput(-.5,.5){$1$}
\rput(-.5,1.5){$2$}
\rput(-.5,2.5){$3$}
\rput(-.5,3.5){$4$}
\rput(-.5,4.5){$5$}
\rput(-.5,5.5){$6$}
\rput(-.5,6.25){$s$}}
\end{pspicture}\qquad\qquad
\raisebox{3cm}{\begin{pspicture}[shift=-.4](0,0)(1,1)
\psframe[linewidth=.7pt,fillstyle=solid,fillcolor=white](0,0)(1,1)
\end{pspicture}\;=\;NS,}
\ \ 
\raisebox{3cm}{\begin{pspicture}[shift=-.4](0,0)(1,1)
\psframe[linewidth=.7pt,fillstyle=solid,fillcolor=lightlightblue](0,0)(1,1)
\end{pspicture}\;=\;R}
\vspace{.2cm}
\end{center}}}
\caption{\label{KacTables}Kac tables of conformal weights (\ref{genConfWts}) for the superconformal minimal models ${\cal M}(3,7,2)$ and ${\cal M}(5,7,2)$ with central charges $c=-\frac{11}{14},\frac{81}{70}$ respectively. The Neveu-Schwarz (NS) sectors (not shaded) with $r+s$ even correspond to $\ell=0,2$ and are shown as the pair $\Delta_{r,s,0}^{M,M'\!,2},\Delta_{r,s,2}^{M,M'\!,2}$. The conformal weights in these two sectors differ by half-odd integers. The Ramond (R) sectors (shaded) with $r+s$ odd correspond to $\ell=1$.}
\end{table}

\section{One-Dimensional Sums of Fused $\mbox{RSOS}(m,m')$ Lattice Models}
\subsection{Forrester-Baxter $\mbox{RSOS}(m,m')$ lattice models}

\setlength{\unitlength}{28pt}
\psset{unit=28pt}
The Forrester-Baxter $\mbox{RSOS}(m,m')$ lattice models~\cite{FB84}, with $2\le m<m'$ and $m,m'$ coprime, are defined on a square lattice with heights
{$a=1,2,\ldots,\mbox{$m'\!-\!1$}$} restricted so that nearest neighbour heights differ by $\pm 1$. The heights thus live on the $A_{m'-1}$ Dynkin diagram.
The nonzero Boltzmann face weights in Regime~III are\vspace{-6pt}
\begin{align}
&\Wtt{W}{a}{a\mp1}{a}{a\pm1}u=\ \  \faceu {a}{a\!\mp\!1}{a}{a\!\pm\!1}u\ \ \ =\;\thf(\lam-u)
\label{Boltzmann1}\\
&\Wtt{W}{a\mp1}{a}{a\pm1}{a}u=
\ \ \faceu {a\!\mp\!1}{a}{a\!\pm\!1}{a}u\ \ \ =\;-{g_{a\pm 1}\over g_{a\mp 1}}\,
\frac{\thf((a\pm 1)\lam)}{\thf(a\lam)}\,
\thf(u)
\label{Boltzmann2}\\
&\Wtt{W}{a\pm1}{a}{a\pm1}{a}u=
\ \ \faceu {a\!\pm\!1}{a}{a\!\pm\!1}{a}u\ \ \ =\;\frac{\thf(a\lam\pm u)}{s(a\lam)}
\label{Boltzmann3}
\end{align}
where $\thf(u)=\vartheta_1(u,\therm)/\vartheta_1(\lambda,\therm)$ is a quotient of the standard elliptic theta
functions~\cite{GR}
\bea
\vartheta_{1}(u,t)=2t^{1/4}\sin u\prod_{n=1}^{\infty}(1-2t^{2n}\cos2u+t^{4n})(1-t^{2n}),\qquad 0<u<\lambda,\quad 0<t<1
\eea
$u$ is the spectral parameter and $g_a$ are arbitrary gauge factors. Unless stated otherwise, we work in the gauge $g_a=1$. The elliptic nome $\therm=e^{-\epsilon}$
is a temperature-like variable, with $\therm^2$ measuring the
departure from criticality corresponding to the $\varphi_{1,3}$
integrable perturbation~\cite{ZamPert}. The crossing parameter is
\begin{equation}
\lambda={(m'-m)\pi\over m'},\qquad 1\le m<m',\qquad\mbox{$m,m'$ coprime}
\label{crossing}
\end{equation}
where $m\ge 2$ relates to minimal models and $m=1$ relates to the $\mathbb{Z}_{m'-2}$ parafermions.

\subsection{Fused $\mbox{RSOS}(m,m')$ lattice models}
\subsubsection{Construction of $n\times n$ fused face weights}

The RSOS$(m,m')_{n\times n}$ face weights are constructed by fusing $n\times n$ blocks of elementary faces
\bea
\begin{array}{rcl}
W^{n,n}\Big(\!\!\Big.\begin{array}{cc}
d & c\\
a & b
\end{array}
\end{array}\!\!\Big|u\!\Big)=\frac{1}{\eta^{n,n}(u)}\;
\psset{unit=.8cm}
\begin{pspicture}[shift=-2.75](-.4,-.4)(5.4,5.4)
\facegrid{(0,0)}{(5,5)}
\rput(4.5,.5){\small $u_0$}
\rput(4.5,1.5){\small $u_1$}
\rput(4.5,2.5){\small $\vdots$}
\rput(4.5,3.5){\small $\vdots$}
\rput[l](4.1,4.5){\small $u_{n\!-\!1}$}
\rput(3.5,.5){\small $u_{-1}$}
\rput(3.5,1.5){\small $u_0$}
\rput(3.5,2.5){\small $\vdots$}
\rput(3.5,3.5){\small $\vdots$}
\rput[r](3.9,4.5){\small $u_{n\!-\!2}$}
\rput(.5,.5){\small $u_{1\!-\!n}$}
\rput(.5,1.5){\small $u_{2\!-\!n}$}
\rput(.5,2.5){\small $\vdots$}
\rput(.5,3.5){\small $\vdots$}
\rput(.5,4.5){\small $u_{0}$}
\multiput(0,0)(1,0){2}{\rput(1.5,.5){\small $\cdots$}}
\multiput(0,1)(1,0){2}{\rput(1.5,.5){\small $\cdots$}}
\multiput(0,4)(1,0){2}{\rput(1.5,.5){\small $\cdots$}}
\rput(1.5,3.5){\small $\ddots$}
\rput(2.5,2.5){\small $\ddots$}
\psarc[linewidth=1pt,linecolor=red](0,0){.15}{0}{90}
\multiput(0,0)(0,1){5}{\multirput(1,0)(1,0){4}{\pscircle*{.07}}}
\multirput(0,1)(0,1){4}{\pscircle*{.07}}
\multirput(5,1)(0,1){4}{$\times$}
\multirput(1,5)(1,0){4}{$\times$}
\rput[tr](0,0){$a$}
\rput[tl](5,0){$b$}
\rput[bl](5,5){$c$}
\rput[br](0,5){$d$}
\end{pspicture}\!\!\!\!, 
\qquad u_j = u + j\lambda
\label{nxnface}
\eea
where the solid dots indicate free sums over the allowed values of the heights. In the gauge $g_a=1$, the fused weights are independent of the heights at the sites marked with a cross.
The fused weights are set to zero unless the adjacent pairs of heights $a,b=1,\ldots,m'\!-\!1$ on each edge satisfy the restrictions
	\be
	|a-b| = \begin{cases}
	0,2,4,\ldots, n, & n \text{ even}\\
	1, 3,5,\ldots, n, & n \text{ odd}
	\end{cases}
	\qquad\quad
	n+2\le a+b\le 2m'-n-2
	\label{adjacency}
	\ee
We therefore see that $\half(a-b)=-\frac{n}{2},-\frac{n-1}{2},\ldots,\frac{n-1}{2},\frac{n}{2}$ is a spin-$\frac{n}{2}$ variable.

\goodbreak
The $n\times n$ fused $\mbox{RSOS}(m,m')$ models exhibit a duality under the involution
\bea
\lambda\leftrightarrow \pi\!-\!\lambda\qquad\mbox{or}\qquad m\leftrightarrow m'\!-\!m\label{duality}
\eea
in the sense that both the spectrum (set of eigenvalues) of the row and corner transfer matrices are invariant. For finite-size systems, the effect of this involution is to turn the eigenvalue spectrum upside down 
and to interchange the ground states. In particular, the fused RSOS models with $m=m'\!-\!1$, related to the unitary minimal models, are dual to those with $m=1$ which is not allowed as a minimal model. 
In fact, unlike the cases $m\ge 2$, the ground states of the fused RSOS models with $m=1$ possess a $\mathbb{Z}_{m'-2}$ symmetry and relate, in the continuum scaling limit, to $\mathbb{Z}_{m'-2}$ parafermions. 
In nonunitary cases with $m\ge 2$, this duality maps between RSOS models related to pairs of nonunitary minimal models.

\subsubsection{Local properties of $n\times n$ fused face weights}

The $n\times n$ fused weights satisfy local relations in the form of the initial condition, the inversion relation and the Yang-Baxter equation 
\begin{align}
\Wn{a}{b}{c}{d}{0}&=\delta(a,c)\nonumber\\[0pt]
\disp\sum_g \Wn abgdu \Wn gbcd{-\!u}&=\delta(a,c) \disp\prod_{k=1}^n \frac{s(k\lambda-u)s(k\lambda+u)}{s(k\lambda)^2}\\[2pt]
\disp\sum_g\Wn{\!a\!}{\!b}{\!g}{\!f\!}{u\!}\!\Wn{\!f\!}{\!g}{\!d}{\!e\!}{u\!+\!v\!}\!\Wn{\!g\!}{\!b}{\!c}{\!d\!}{v\!}
&=\disp\sum_g\Wn{\!f\!}{\!a}{\!g}{\!e\!}{v\!}\!\Wn{\!a\!}{\!b}{\!c}{\!g\!}{u\!+\!v\!}\!\Wn{\!g\!}{\!c}{\!d}{\!e\!}{u\!}\nonumber
\end{align}
A number of identities in elliptic theta functions are needed to verify these relations. These all follow from the fundamental identity listed in Appendix \ref{AppA}. 
Together, these local relations imply commuting row and corner transfer matrices and exact integrability. The weights are also symmetric under reflection about the leading diagonal and under height reversal
\be
\Wn{a}{b}{c}{d}{u} = \Wn{a}{d}{c}{b}{u},\qquad \Wn{a}{b}{c}{d}{u} = \Wn{m'\!-\!a}{m'\!-\!b}{m'\!-\!c}{m'\!-\!d}{u}
\ee

\subsection{Fused RSOS paths, shaded bands and ground states}

In this section, we recall the shaded band diagrams of \cite{FodaW} and posit the ground states of the $n\times n$ fused RSOS models. 
The considerations in this section and the next section on one-dimensional sums are purely combinatorial.

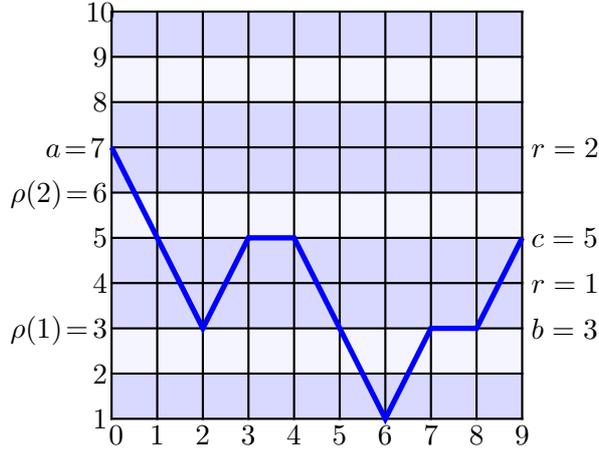
\begin{figure}[htbp]
\centering
\psset{unit=0.6cm}
\begin{pspicture}(-0.4,-0.4)(9,9)
\pspolygon[linewidth=0pt,fillstyle=solid,fillcolor=lightestblue](0,0)(9,0)(9,9)(0,9)
\rput(0,0){\pspolygon[linewidth=0pt,fillstyle=solid,fillcolor=lightlightblue](0,0)(9,0)(9,1)(0,1)}
\rput(0,2){\pspolygon[linewidth=0pt,fillstyle=solid,fillcolor=lightlightblue](0,0)(9,0)(9,1)(0,1)}
\rput(0,3){\pspolygon[linewidth=0pt,fillstyle=solid,fillcolor=lightlightblue](0,0)(9,0)(9,1)(0,1)}
\rput(0,5){\pspolygon[linewidth=0pt,fillstyle=solid,fillcolor=lightlightblue](0,0)(9,0)(9,1)(0,1)}
\rput(0,6){\pspolygon[linewidth=0pt,fillstyle=solid,fillcolor=lightlightblue](0,0)(9,0)(9,1)(0,1)}
\rput(0,8){\pspolygon[linewidth=0pt,fillstyle=solid,fillcolor=lightlightblue](0,0)(9,0)(9,1)(0,1)}
\psgrid[gridlabels=0pt,subgriddiv=1]
\rput(-.25,0){1}\rput(-.25,1){2}\rput(-.25,2){3}\rput(-.25,3){4}\rput(-.25,4){5}
\rput(-.25,5){6}
\rput[r](-.15,6){$a\!=\!7$}
\rput(-.25,7){8}\rput(-.25,8){9}\rput(-.25,9){10}
\rput(0.1,-0.35){0}\rput(1,-0.35){1}\rput(2,-0.35){2}\rput(3,-0.35){3}\rput(4,-0.35){4}\rput(5,-0.35){5}\rput(6,-0.35){6}\rput(7,-0.35){7}\rput(8,-0.35){8}\rput(9,-0.35){9}
\psline[linewidth=2pt,linecolor=blue](0,6)(1,4)(2,2)(3,4)(4,4)(5,2)(6,0)(7,2)(8,2)(9,4)
\rput[l](9.2,6){$r=2$}
\rput[l](9.2,3){$r=1$}
\rput[l](9.2,2){$b=3$}
\rput[l](9.2,4){$c=5$}
\rput[r](-.5,1.95){$\rho(1)\!=$}
\rput[r](-.5,4.95){$\rho(2)\!=$}
\end{pspicture}
\caption{Band diagram showing a typical path $\sigma = \{7,5,3,5,5,3,1,3,3,5\}$ in the sector $(\sigma_0,\sigma_N,\sigma_{N+1})=(7,3,5)$ for the superconformal minimal model ${\cal M}(3,11,2)$ with $(m,m',n)=(7,11,2)$ and $N=8$. Shaded 1-bands occur at heights $a=1,3,4,6,7,9$ and shaded 2-bands occur at $\rho=\rho(r)=3,6$. Since $M-1=2$, the two shaded 2-bands are labelled by $r=1,2$. We always take $N$ to be even.}
\label{fig:patheg}
\end{figure}

\subsubsection{Paths, shaded band diagrams and sectors}

A path $\sigma=\{\sigma_0,\sigma_1,\ldots,\sigma_N,\sigma_{N+1}\}$ of the $n\times n$ fused RSOS lattice models is an $(N+1)$-step walk, with $\sigma_j\in A_{m'-1}$, on the level $n$ fused adjacency diagram given by the adjacency rules (\ref{adjacency}). In this paper, we always take $N$ to be even. 
If $n$ is even, all of the heights $\sigma_j$ have the same parity (all even or all odd). If $n$ is odd, the heights $\sigma_j$ alternate in parity along the path. 
The $(N\!+\!1)$-step RSOS paths are separated into various sectors labelled by the boundary conditions
\be
(\sigma_0,\sigma_N,\sigma_{N+1})=(a,b,c)
\ee
In the continuum scaling limit, the heights $(a,b,c)$ are related to the quantum numbers $(r,s,\ell)$.
A typical path is shown in Figure~\ref{fig:patheg}. 
Combinatorially, it is convenient to describe these paths as walks on the $A_{m'-1}$ shaded band diagram~\cite{FodaW}. The band $(a,a\!+\!1)$ between heights $a$ and $a\!+\!1$ is shaded if
\be
a = \Bfloor{\frac{r m'}{m}}, \qquad r = 1,\ldots, m-1\label{shadedbands}
\ee
and is otherwise unshaded. Shaded and unshaded bands are interchanged under duality \mbox{$m\leftrightarrow m'\!-\!m$}.  
An \emph{$n$-band} consists of $n$ contiguous bands, where each band is shaded or unshaded. If all the 1-bands in an $n$-band are shaded, we call it a \emph{shaded $n$-band}. If all the 1-bands in an $n$-band are unshaded, we call it an \emph{unshaded $n$-band}. Otherwise, it a \emph{mixed $n$-band}.

For fixed $(m,m'\!,n)$, the heights of the shaded $n$-bands $(\rho,\rho\!+\!n)$ are labelled by the sequences
\bea
\rho=\rho(r)=\rho^{m,m'\!,n}(r),\qquad r=1,2,\ldots, M\!-\!1\label{rhodef}
\eea
Since these are monotonically increasing sequences the inverse exists
\bea
r=r(\rho)=r^{m,m'\!,n}(\rho),\qquad \rho=\rho(1),\rho(2),\ldots,\rho(M\!-\!1)
\eea
Here $M=M(m,m',n)$ counts the length of the finite sequence. 
For $n=1$, $\rho=\rho(r)$ is given by the sequence (\ref{shadedbands}). Although these finite sequences are easily enumerated diagrammatically, as in Figure~\ref{fig:patheg}, we have been unable to find explicit expressions for these sequences for $n>1$. Nevertheless in Appendix~\ref{sec:shadedNbands} we show that, for $0<\lambda<\pi/n$, the number of shaded $n$-bands is
\be
\mbox{\# shaded $n$-bands}=M\!-\!1=nm-(n-1)m'-1
\ee
which coincides with the maximum value of the Kac label $r$. For $\lambda>\pi/n$, there are no shaded $n$-bands. For $\lambda<\pi/n$, since $n'<n$ implies $\pi/n<\pi/n'$, it follows that for each $n'<n$ the number of shaded $n'$-bands is also given by $M(m,m',n')-1=n'm-(n'-1)m'-1$. 

Following \cite{FB84}, we also use the sequences
\bea
h_a=\bfloor{\tfrac{a(m'-m)}{m'}}=\bfloor{\tfrac{a\lambda}{\pi}}=\mbox{\# unshaded 1-bands below the height $a$}
\label{floorha}
\eea
The value of $h_a$ remains unchanged within any shaded $n$-band. The value $h_a=0,1,2,\ldots$ thus labels, from the bottom, the contiguous shaded bands (independent of the width of the individual shaded bands) separated by unshaded 1-bands.

\subsubsection{Fused RSOS ground state boundary conditions}

The initial height $a$ in the fused RSOS paths is to be identified with the Kac label $s$. We posit further that, in the sector $(a,b,c)$, $(b,c)$ is a ground state boundary condition if the heights $b$ and $c$ lie within the same shaded $n$-band $(\rho,\rho+n)$ (labelled by $r$), that is $b,c\in(\rho,\rho+n)$, and they are symmetrically placed about its center $\half(b+c)=\rho+\half n$. Defining 
\be
\ell=\half[n+(-1)^{h_b}(b-c)]=0,1,2,\ldots,n,\qquad \tilde{\ell}=\half(n+b-c)=\begin{cases}\ell,&\mbox{$h_b$ even}\\ n-\ell,&\mbox{$h_b$ odd}\end{cases}
\label{elldef}
\ee
we see that $h_b=h_c$ and
\be
s=a,\quad \rho=\half(b+c-n),\quad b=\rho+\tilde{\ell},\quad c=\rho+n-\tilde{\ell}
\label{elltilde}
\ee
Given $(m,m'\!,n)$, these relations allow to uniquely map back and forth between the boundary conditions $(a,b,c)$ and the Kac label quantum numbers $(r,s,\ell)$ with $r=r^{m,m'\!,n}(\rho)$. Boundary conditions $(b,c)$ not satisfying these conditions are non-ground state boundary conditions.

\subsection{Local energies and one-dimensional sums}

In this section, we consider the local energy functions $H(d,a,b)$ and their associated one-dimensional sums. In particular, restricting to the interval $0<\lambda<\pi/n$, we give exhaustive lists of the values of the local energies according to the shading of the internal bands for $n=1,2,3$. As explained in Appendix~\ref{sec:shadedNbands}, not all patterns of shaded bands actually occur for  $0<\lambda<\pi/n$. Specifically, in this smaller interval considered in this paper, any $n$ contiguous bands must have at most one unshaded 1-band. In the unitary cases ($m=m'\!-\!1$), all of the 1-bands are shaded. Accordingly, in agreement with \cite{DJKMO}, we find that the local energy functions with all internal 1-bands shaded (that is 1-bands between $\mbox{Min}[d,a,b]$ and $\mbox{Max}[d,a,b]$) are given by
\be
H(d,a,b)=\tfrac{1}{4}|b-d|
\ee
Moreover, the local energies possess reflection and height reversal symmetries
\be
H(d,a,b)=H(b,a,d)=H(m'\!-\!d,m'\!-\!a,m'\!-\!b)
\ee
which are inherited from the face weights. 
Noting that the physical quantities of interest are unchanged if the local energy functions are shifted by an additive constant, we use this and the gauge freedom to ensure that
\be
0\le H(d,a,b)\le \tfrac{n}{2}
\ee
Lastly, we impose the ground state requirement
\bea
H(b,c,b)=H(c,b,c)=0,\qquad \mbox{whenever $(b,c)$ is a ground state boundary condition}\label{bcGS}
\eea

\begin{figure}[p]
\centering
\subfloat[$n=1$]{
\psset{unit=.9cm}
\begin{pspicture}(-1,-0.25)(6.5,1.5)
\pspolygon[fillstyle=solid,fillcolor=lightlightblue](0,0)(5,0)(5,1)(0,1)
\psgrid[gridlabels=0pt,subgriddiv=1](5,1)
\psline[linecolor=blue,linewidth=2pt](0,0)(1,1)(2,0)(3,1)(4,0)(5,1)
\rput(-0.5,0){\small$\tilde{\ell}=0$}
\rput(-0.5,1){\small$\tilde{\ell}=1$}
\rput[l](5.1,0){$\rho$}
\rput[l](5.1,1){$\rho\!+\!1$}
\rput(0,-.3){\scriptsize $N$}
\rput(1,-.3){\scriptsize $N\!+\!1$}
\rput(2,-.3){$\cdots$}
\end{pspicture}
}
\subfloat[$n=2$]{
\psset{unit=.9cm}
\begin{pspicture}(-1,-0.25)(6.5,2.5)
\pspolygon[fillstyle=solid,fillcolor=lightlightblue](0,0)(5,0)(5,2)(0,2)
\psgrid[gridlabels=0pt,subgriddiv=1](5,2)
\psline[linecolor=blue,linewidth=2pt](0,1)(1,1)(2,1)(3,1)(4,1)(5,1)
\psline[linecolor=blue,linewidth=2pt](0,0)(1,2)(2,0)(3,2)(4,0)(5,2)
\rput(-0.5,0){\small$\tilde{\ell}=0$}
\rput(-0.5,1){\small$\tilde{\ell}=1$}
\rput(-0.5,2){\small$\tilde{\ell}=2$}
\rput[l](5.1,0){$\rho$}
\rput[l](5.1,1){$\rho\!+\!1$}
\rput[l](5.1,2){$\rho\!+\!2$}
\rput(0,-.3){\scriptsize $N$}
\rput(1,-.3){\scriptsize $N\!+\!1$}
\rput(2,-.3){$\cdots$}
\end{pspicture}
}\\
\subfloat[$n=3$]{
\psset{unit=.9cm}
\begin{pspicture}(-1,-0.25)(6.5,3.5)
\pspolygon[fillstyle=solid,fillcolor=lightlightblue](0,0)(5,0)(5,3)(0,3)
\psline[linecolor=blue,linewidth=2pt](0,1)(1,2)(2,1)(3,2)(4,1)(5,2)
\psline[linecolor=blue,linewidth=2pt](0,0)(1,3)(2,0)(3,3)(4,0)(5,3)
\psgrid[gridlabels=0pt,subgriddiv=1](5,3)
\rput(-0.5,0){\small$\tilde{\ell}=0$}
\rput(-0.5,1){\small$\tilde{\ell}=1$}
\rput(-0.5,2){\small$\tilde{\ell}=2$}
\rput(-0.5,3){\small$\tilde{\ell}=3$}
\rput[l](5.1,0){$\rho$}
\rput[l](5.1,1){$\rho\!+\!1$}
\rput[l](5.1,2){$\rho\!+\!2$}
\rput[l](5.1,3){$\rho\!+\!3$}
\rput(0,-.3){\scriptsize $N$}
\rput(1,-.3){\scriptsize $N\!+\!1$}
\rput(2,-.3){$\cdots$}
\end{pspicture}
}
\subfloat[$n=4$]{
\psset{unit=.9cm}
\begin{pspicture}(-1,-0.25)(6.5,4.5)
\pspolygon[fillstyle=solid,fillcolor=lightlightblue](0,0)(5,0)(5,4)(0,4)
\psgrid[gridlabels=0pt,subgriddiv=1](5,4)
\psline[linecolor=blue,linewidth=2pt](0,2)(1,2)(2,2)(3,2)(4,2)(5,2)
\psline[linecolor=blue,linewidth=2pt](0,1)(1,3)(2,1)(3,3)(4,1)(5,3)
\psline[linecolor=blue,linewidth=2pt](0,0)(1,4)(2,0)(3,4)(4,0)(5,4)
\rput(-0.5,0){\small$\tilde{\ell}=0$}
\rput(-0.5,1){\small$\tilde{\ell}=1$}
\rput(-0.5,2){\small$\tilde{\ell}=2$}
\rput(-0.5,3){\small$\tilde{\ell}=3$}
\rput(-0.5,4){\small$\tilde{\ell}=4$}
\rput[l](5.1,0){$\rho$}
\rput[l](5.1,1){$\rho\!+\!1$}
\rput[l](5.1,2){$\rho\!+\!2$}
\rput[l](5.1,3){$\rho\!+\!3$}
\rput[l](5.1,4){$\rho\!+\!4$}
\rput(0,-.3){\scriptsize $N$}
\rput(1,-.3){\scriptsize $N\!+\!1$}
\rput(2,-.3){$\cdots$}
\end{pspicture}
}
\caption{Extended ground state boundary conditions $(\sigma_N,\sigma_{N+1})=(b,c)=(\rho\!+\!\tilde{\ell},\rho\!+\!n\!-\!\tilde{\ell})$ for $\tilde{\ell}=0,1,\ldots,\sfloor{\frac{n}{2}}$ and $n=1,2,3,4$. 
The values of $\tilde{\ell}$ are shown on the left and the heights are shown on the right. The ground states for $\sfloor{\frac{n}{2}}<\tilde{\ell}\le n$ are obtained by applying the height reversal $\sigma_j\mapsto 2\rho+n-\sigma_j$ within the 
shaded $n$-band. In all cases, the local energies satisfy $H(b,c,b)=H(c,b,c)=0$. Here $\tilde{\ell}=\ell$ or $n-\ell$ according to (\ref{elldef}).}
\label{fig:fusedGS}
\end{figure}
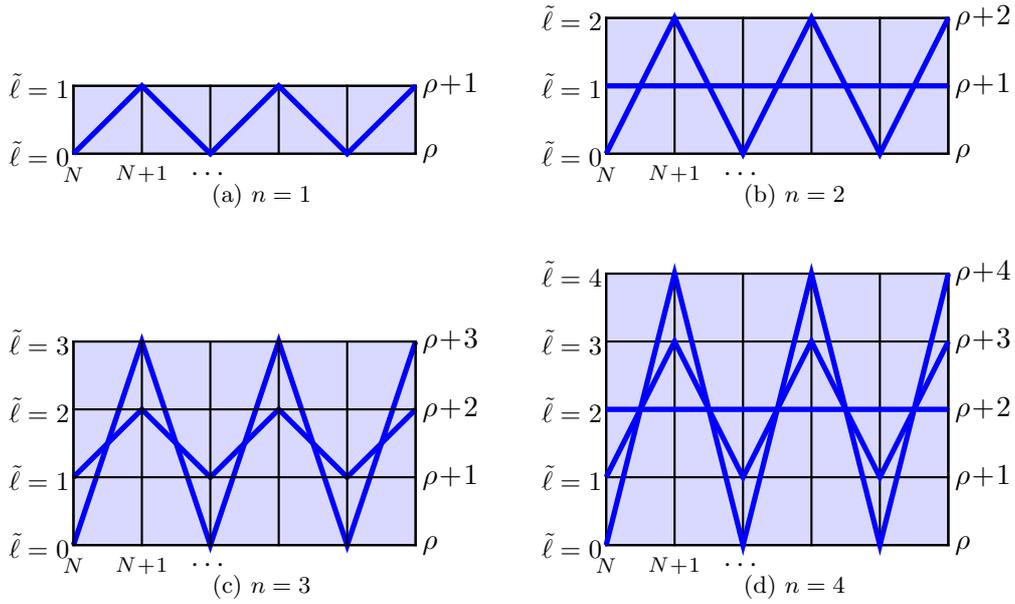

\begin{figure}[p]
\centering
\subfloat[Alternating ground states in Neveu-Schwarz sectors $\ell=0,2$.]{
\psset{unit=0.75cm}
\begin{pspicture}(-0.4,-0.5)(6,9)
\pspolygon[linewidth=0pt,fillstyle=solid,fillcolor=lightestblue](0,0)(6,0)(6,9)(0,9)
\pspolygon[linewidth=0pt,fillstyle=solid,fillcolor=lightlightblue](0,0)(6,0)(6,1)(0,1)
\pspolygon[linewidth=0pt,fillstyle=solid,fillcolor=lightlightblue](0,2)(6,2)(6,3)(0,3)
\pspolygon[linewidth=0pt,fillstyle=solid,fillcolor=lightlightblue](0,3)(6,3)(6,4)(0,4)
\pspolygon[linewidth=0pt,fillstyle=solid,fillcolor=lightlightblue](0,5)(6,5)(6,6)(0,6)
\pspolygon[linewidth=0pt,fillstyle=solid,fillcolor=lightlightblue](0,6)(6,6)(6,7)(0,7)
\pspolygon[linewidth=0pt,fillstyle=solid,fillcolor=lightlightblue](0,8)(6,8)(6,9)(0,9)
\psgrid[gridlabels=0pt,subgriddiv=1]
\rput(-.25,0){1}\rput(-.25,1){2}\rput(-.25,2){3}\rput(-.25,3){4}\rput(-.25,4){5}
\rput(-.25,5){6}\rput(-.25,6){7}\rput(-.25,7){8}\rput(-.25,8){9}\rput(-.35,9){10}
\rput(0.1,-0.35){0}\rput(1,-0.35){1}\rput(2,-0.35){2}\rput(3,-0.35){3}\rput(4,-0.35){4}\rput(5,-0.35){5}\rput(6,-0.35){6}
\psline[linewidth=1.5pt,linecolor=purple](0,4)(1,2)(2,4)(3,2)(4,4)(5,2)(6,4)
\psline[linewidth=1.5pt,linecolor=blue](0,2)(1,4)(2,2)(3,4)(4,2)(5,4)(6,2)
\psline[linewidth=1.5pt,linecolor=purple](0,5)(1,7)(2,5)(3,7)(4,5)(5,7)(6,5)
\psline[linewidth=1.5pt,linecolor=blue](0,7)(1,5)(2,7)(3,5)(4,7)(5,5)(6,7)
\rput(6.75,3){$r=1$}
\rput(6.75,6){$r=2$}
\end{pspicture}
}\hspace{2cm}
\subfloat[Flat ground states in the Ramond sector $\ell=1$.]{
\psset{unit=0.75cm}
\begin{pspicture}(-0.4,-0.5)(6,9)
\pspolygon[linewidth=0pt,fillstyle=solid,fillcolor=lightestblue](0,0)(6,0)(6,9)(0,9)
\pspolygon[linewidth=0pt,fillstyle=solid,fillcolor=lightlightblue](0,0)(6,0)(6,1)(0,1)
\pspolygon[linewidth=0pt,fillstyle=solid,fillcolor=lightlightblue](0,2)(6,2)(6,3)(0,3)
\pspolygon[linewidth=0pt,fillstyle=solid,fillcolor=lightlightblue](0,3)(6,3)(6,4)(0,4)
\pspolygon[linewidth=0pt,fillstyle=solid,fillcolor=lightlightblue](0,5)(6,5)(6,6)(0,6)
\pspolygon[linewidth=0pt,fillstyle=solid,fillcolor=lightlightblue](0,6)(6,6)(6,7)(0,7)
\pspolygon[linewidth=0pt,fillstyle=solid,fillcolor=lightlightblue](0,8)(6,8)(6,9)(0,9)
\psgrid[gridlabels=0pt,subgriddiv=1]
\rput(-.25,0){1}\rput(-.25,1){2}\rput(-.25,2){3}\rput(-.25,3){4}\rput(-.25,4){5}
\rput(-.25,5){6}\rput(-.25,6){7}\rput(-.25,7){8}\rput(-.25,8){9}\rput(-.35,9){10}
\rput(0.1,-0.35){0}\rput(1,-0.35){1}\rput(2,-0.35){2}\rput(3,-0.35){3}\rput(4,-0.35){4}\rput(5,-0.35){5}\rput(6,-0.35){6}
\psline[linewidth=1.5pt,linecolor=blue](0,3)(6,3)
\psline[linewidth=1.5pt,linecolor=blue](0,6)(6,6)
\rput(6.75,3){$r=1$}
\rput(6.75,6){$r=2$}
\end{pspicture}
}
\caption{Extended ground state boundary conditions $(\sigma_N,\sigma_{N+1})=(b,c)=(\rho\!+\!\tilde{\ell},\rho\!+\!n\!-\!\tilde{\ell})$ for $\tilde{\ell}=0,1,\ldots,n$ for the superconfomal minimal model ${\cal M}(3,11,2)$ with $(m,m'\!,n)=(7,11,2)$ and 2-bands at $\rho=3,6$ with $r=1,2$. Here $\tilde{\ell}=\ell$ or $n-\ell$ according to (\ref{elldef}). The ground states are separated into Neveu-Schwarz ($\ell=0,2;\,\mbox{$r+s$ even}$) and Ramond ($\ell=1;\,\mbox{$r+s$ odd}$) sectors.}
\label{fig:GSpaths}
\end{figure}
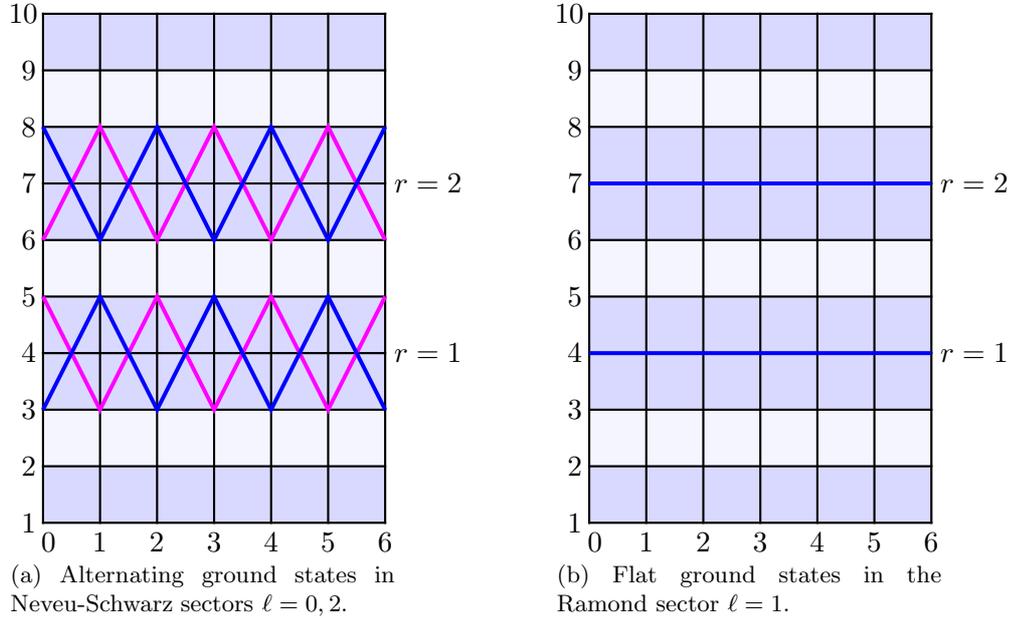

\subsubsection{Local energy functions}

Following Baxter, after fixing a suitable normalization and gauge $g_a$, the local energy functions $H(d,a,b)$ are given by the low-temperature limit 
\be
t=e^{-\varepsilon}\to 1,\quad u,\varepsilon\to 0,\quad \mbox{$u/\varepsilon$ fixed}
\label{tLowT}
\ee 
of the face weights  (\ref{nxnface}) 
\bea
\Wtt{W^{n,n}}{a}{b\,}{c\,}{d}{\,u\!}\;\sim\; \frac{g_a g_c}{g_b g_d}\,w^{H(d,a,b)}\,\delta(a,c),\qquad w=e^{-2\pi u/\varepsilon}
\label{lowT}
\eea
This limit is evaluated by performing a \emph{conjugate modulus transformation} on the elliptic theta functions
\bea
\theta_1(u,t)= \Big(\frac{\pi}{\varepsilon} \Big)^{1/2}e^{-(u-\pi/2)^2/\varepsilon} E(w,p)
\eea
where
\be
E(w,p) = E(w) = \sum_{n=-\infty}^{\infty} (-1)^n p^{n(n-1)/2} w^n = \prod_{n=1}^\infty (1-p^{n-1}w)(1-p^n w^{-1})(1-p^n)
\ee
with variables
\be
w= e^{-2 \pi u/\varepsilon},\qquad p=e^{-2\pi^2/\varepsilon}, \qquad x = p^{\lambda/\pi} = e^{-2\pi \lambda/\varepsilon}
\ee

To evaluate the low-temperature limit $x\rightarrow0$, or equivalently $p\rightarrow0$, we use the following elementary properties of the $E$-functions
\begin{align}
E(w,p) &= E(p\,w^{-1},p) = -w\,E(w^{-1},p)\\
E(p^n w,p) &= (-w)^{-n} p^{-n(n-1)/2} E(w,p)\\
\lim_{p\rightarrow0} E(p^a w,p^b) &= 
\begin{cases}
1, & 0<a<b\\
1-w, & a=0
\end{cases}
\end{align}
where $n$ is an arbitrary integer. Using these properties, we deduce the further useful limit
\be
\lim_{x\rightarrow0} \frac{E(x^a w^{-1})}{E(x^a)} = w^{\lfloor a \lambda/\pi \rfloor}=w^{h_a}
\ee

\subsubsection{Energy statistic and one-dimensional sums}

The energy statistic and the associated one-dimensional sums were introduced by Baxter in the context of Corner Transfer Matrices (CTMs)~\cite{BaxBook,CTM}. 
The energy statistic associated with a one-dimensional RSOS path $\sigma\!=\!\{\sigma_0,\sigma_1,\ldots,\sigma_N,\sigma_{N+1}\}$ is
\bea
E(\sigma)=\sum_{j=1}^N j\,H(\sigma_{j-1},\sigma_j,\sigma_{j+1})\label{Estatistic}
\eea
The associated one-dimensional sums are
\bea
X_{abc}^{(N)}(q)=\sum_\sigma q^{E(\sigma)}
\label{oned}
\eea
where the sum is over all allowed RSOS paths satisfying the boundary conditions
\bea
(\sigma_0,\sigma_N,\sigma_{N+1})\!=\!(a,b,c)\label{bc}
\eea
Due to the requirement (\ref{bcGS}), the ground state boundary condition $(b,c)$ can be extended beyond $j=N\!+\!1$ to infinity, without changing the energy statistic (\ref{Estatistic}) 
associated to paths, simply by alternating $\sigma_j$ between heights $b$ and $c$ for $j\ge N$. It is in this sense that the energy-weighted finite RSOS paths give a truncated set 
of conformal energies of the infinite system. Examples of extended ground states are shown in Figures~\ref{fig:fusedGS} and \ref{fig:GSpaths}.

The local energy function $H(\sigma_{j-1},\sigma_j,\sigma_{j+1})$ is not unique. Its form is changed by incorporating the gauge factors $g_a=w^{\tilde g_a}$ in (\ref{lowT})
\begin{align}
H'(\sigma_{j-1},\sigma_j,\sigma_{j+1})&=H(\sigma_{j-1},\sigma_j,\sigma_{j+1})+2\tilde g_{\sigma_j}-\tilde g_{\sigma_{j-1}}-\tilde g_{\sigma_{j+1}}\nonumber\\
&=H(\sigma_{j-1},\sigma_j,\sigma_{j+1})+(\tilde g_{\sigma_j}-\tilde g_{\sigma_{j-1}})-(\tilde g_{\sigma_{j+1}}-\tilde g_{\sigma_j})
\end{align}
In a given sector with $(\sigma_0,\sigma_N,\sigma_{N+1})\!=\!(a,b,c)$, it follows that 
\begin{align}
E'(\sigma)&=\sum_{j=1}^N j H'(\sigma_{j-1},\sigma_j,\sigma_{j+1})=E(\sigma)+\sum_{j=1}^N j (2\tilde g_{\sigma_j}-\tilde g_{\sigma_{j-1}}-\tilde g_{\sigma_{j+1}})\nonumber\\
&=E(\sigma)+[(N+1)\tilde g_b-N\tilde g_c-\tilde g_a]=E(\sigma)+ N(\tilde g_b-\tilde g_c)+(\tilde g_b-\tilde g_a)
\end{align}
independent of the path $\sigma$. Since this amounts to a shift in the ground state energy by a constant amount, the two energy statistics are equivalent. 

Summing over allowed neighbours of $b$, the one-dimensional sums satisfy the linear recursion relations
\bea
X_{abc}^{(N)}(q)=\sum_{d\sim b} q^{N H(d,b,c)} \,X_{adb}^{(N-1)}(q)\label{1drecursion}
\eea
subject to the initial and boundary conditions
\bea
X_{abc}^{(0)}(q)=\delta(a,b),\qquad X_{a0c}^{(N)}(q)=X_{a\{m'\}c}^{(N)}(q)=0
\eea
where $b$ and $c$ are neighbours. This data uniquely determines $X_{abc}^{(N)}(q)$. Our conjecture is that the one-dimensional sums coincide with finitized branching functions up to the leading powers of $q$
\bea
X_{abc}^{(N)}(q)\cong b_{r,s,\ell}^{M,M'\!,n;(N)}(q),\qquad \lim_{N\to\infty}X_{abc}^{(N)}(q)\cong b_{r,s,\ell}^{M,M'\!,n}(q)
\eea
Duality (\ref{duality}) is implemented on finite one-dimensional sums by $q\leftrightarrow q^{-1}$ or $H(a,b,c)\leftrightarrow -H(a,b,c)$ and interchanges the intervals $\lambda\in (0,\pi/2)$ and $\lambda\in (\pi/2,\pi)$.

\subsubsection{$n=1$ local energies}

\begin{figure}[p]
\psset{unit=.6cm}
\begin{align}
\qquad\qquad
&\begin{pspicture}(0,.3)(2,1.5)
\psframe[linewidth=0pt,fillstyle=solid,fillcolor=lightlightblue](0,0)(2,1)
\multirput(1,0)(1,0){2}{\psline[linewidth=.5pt](0,0)(0,1)}
\psline[linewidth=2pt](0,1)(1,0)(2,1)
\end{pspicture}\ =\ 
\begin{pspicture}(0,.3)(2,1.5)
\psframe[linewidth=0pt,fillstyle=solid,fillcolor=lightlightblue](0,0)(2,1)
\multirput(1,0)(1,0){2}{\psline[linewidth=.5pt](0,0)(0,1)}
\psline[linewidth=2pt](0,0)(1,1)(2,0)
\end{pspicture}\ =\ 0,
\qquad\qquad
\begin{pspicture}(0,.3)(2,1.5)
\psframe[linewidth=0pt,fillstyle=solid,fillcolor=lightestblue](0,0)(2,1)
\multirput(1,0)(1,0){2}{\psline[linewidth=.5pt](0,0)(0,1)}
\psline[linewidth=2pt](0,1)(1,0)(2,1)
\end{pspicture}\ =\ 
\begin{pspicture}(0,.3)(2,1.5)
\psframe[linewidth=0pt,fillstyle=solid,fillcolor=lightestblue](0,0)(2,1)
\multirput(1,0)(1,0){2}{\psline[linewidth=.5pt](0,0)(0,1)}
\psline[linewidth=2pt](0,0)(1,1)(2,0)
\end{pspicture}\ =\ \tfrac{1}{2}\nonumber\\[0pt]
&\begin{pspicture}[shift=-.6](0,.3)(2,2.5)
\psframe[linewidth=0pt,fillstyle=solid,fillcolor=lightlightblue](0,1)(2,2)
\psframe[linewidth=0pt,fillstyle=solid,fillcolor=lightlightblue](0,0)(2,1)
\multirput(1,0)(1,0){2}{\psline[linewidth=.5pt](0,0)(0,2)}
\psline[linewidth=2pt](0,0)(1,1)(2,2)
\end{pspicture}\ =\ 
\begin{pspicture}[shift=-.6](0,.3)(2,2.5)
\psframe[linewidth=0pt,fillstyle=solid,fillcolor=lightlightblue](0,1)(2,2)
\psframe[linewidth=0pt,fillstyle=solid,fillcolor=lightlightblue](0,0)(2,1)
\multirput(1,0)(1,0){2}{\psline[linewidth=.5pt](0,0)(0,2)}
\psline[linewidth=2pt](0,2)(1,1)(2,0)
\end{pspicture}\ =\ \tfrac{1}{2},\qquad\qquad
\begin{pspicture}[shift=-.6](0,.3)(2,2.5)
\psframe[linewidth=0pt,fillstyle=solid,fillcolor=lightestblue](0,1)(2,2)
\psframe[linewidth=0pt,fillstyle=solid,fillcolor=lightestblue](0,0)(2,1)
\multirput(1,0)(1,0){2}{\psline[linewidth=.5pt](0,0)(0,2)}
\psline[linewidth=2pt](0,0)(1,1)(2,2)
\end{pspicture}\ =\ 
\begin{pspicture}[shift=-.6](0,.3)(2,2.5)
\psframe[linewidth=0pt,fillstyle=solid,fillcolor=lightestblue](0,1)(2,2)
\psframe[linewidth=0pt,fillstyle=solid,fillcolor=lightestblue](0,0)(2,1)
\multirput(1,0)(1,0){2}{\psline[linewidth=.5pt](0,0)(0,2)}
\psline[linewidth=2pt](0,2)(1,1)(2,0)
\end{pspicture}\ =\ 0\nonumber\\[4pt]
&\begin{pspicture}[shift=-.6](0,.3)(2,2.5)
\psframe[linewidth=0pt,fillstyle=solid,fillcolor=lightestblue](0,1)(2,2)
\psframe[linewidth=0pt,fillstyle=solid,fillcolor=lightlightblue](0,0)(2,1)
\multirput(1,0)(1,0){2}{\psline[linewidth=.5pt](0,0)(0,2)}
\psline[linewidth=2pt](0,0)(1,1)(2,2)
\end{pspicture}\ =\ 
\begin{pspicture}[shift=-.6](0,.3)(2,2.5)
\psframe[linewidth=0pt,fillstyle=solid,fillcolor=lightestblue](0,1)(2,2)
\psframe[linewidth=0pt,fillstyle=solid,fillcolor=lightlightblue](0,0)(2,1)
\multirput(1,0)(1,0){2}{\psline[linewidth=.5pt](0,0)(0,2)}
\psline[linewidth=2pt](0,2)(1,1)(2,0)
\end{pspicture}\ =\ \tfrac{1}{4},\qquad\qquad
\begin{pspicture}[shift=-.6](0,.3)(2,2.5)
\psframe[linewidth=0pt,fillstyle=solid,fillcolor=lightlightblue](0,1)(2,2)
\psframe[linewidth=0pt,fillstyle=solid,fillcolor=lightestblue](0,0)(2,1)
\multirput(1,0)(1,0){2}{\psline[linewidth=.5pt](0,0)(0,2)}
\psline[linewidth=2pt](0,0)(1,1)(2,2)
\end{pspicture}\ =\ 
\begin{pspicture}[shift=-.6](0,.3)(2,2.5)
\psframe[linewidth=0pt,fillstyle=solid,fillcolor=lightlightblue](0,1)(2,2)
\psframe[linewidth=0pt,fillstyle=solid,fillcolor=lightestblue](0,0)(2,1)
\multirput(1,0)(1,0){2}{\psline[linewidth=.5pt](0,0)(0,2)}
\psline[linewidth=2pt](0,2)(1,1)(2,0)
\end{pspicture}\ =\ \tfrac{1}{4}\qquad\qquad\quad\nonumber\\[-16pt]\nonumber
\end{align}
\caption{\label{n=1H}The gauged local energies of the $n=1$ RSOS models in the interval $0<\lambda<\pi$. These local energies satisfy duality $H^{m,m'}(a,b,c)= \half-H^{m'-m,m'}(a,b,c)$ under interchange of shaded and unshaded bands. In this gauge, the local energies take the values $0,\tfrac{1}{4},\tfrac{1}{2}$.}
\end{figure}
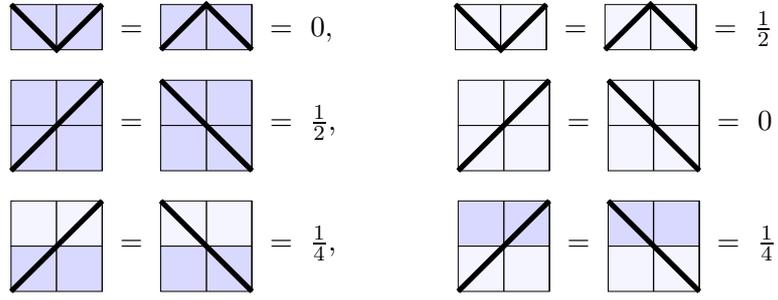

Working in the gauge $g_a=1$ for $n=1$, the local energy function obtained by Forrester-Baxter~\cite{FB84} is 
\begin{subequations}
\begin{align}
H^\text{FB}(a,a\mp 1,a)&=\pm h_a\\
H^\text{FB}(a\!\pm\!1,a,a\!\mp\!1)&=\half
\end{align}
\end{subequations}
where $h_a$ is given by (\ref{floorha}).

Starting with $H^\text{FB}(a,b,c)$, we apply a specific additive gauge transformation given by
\bea
G_a-G_b=\tfrac{1}{4} (a-b)(h_a+h_b),\quad b=a\pm 1;\qquad 
G_a=\tfrac{1}{4}\sum_{c=1}^a (h_c+h_{c-1})+G_0\label{gauge}
\eea
where we choose $G_0=0$ and solve by iterating with $b=a\!-\!1$. This gives the equivalent gauged local energy function
\begin{subequations}
\begin{align}
H(a\!+\!1,a,a\!+\!1)&=\half(h_{a+1}-h_a)\\
H(a\!-\!1,a,a\!-\!1)&=\half(h_{a}-h_{a-1})\\
H(a\!\pm\!1,a,a\!\mp\!1)&=\half-\tfrac{1}{4}(h_{a+1}-h_{a-1})
\end{align}\label{minH}
\end{subequations}
where 
\bea
h_{a+1}-h_a=\begin{cases}
0,&\mbox{$a$ labels a ground state: $(a,a\!+\!1)$ is a shaded band}\\
1,&\mbox{$a$ is not a ground state: $(a,a\!+\!1)$ is not a shaded band}
\end{cases}
\eea
It follows that $H(a,b,c)$ is nonnegative with values shown in Figure~\ref{n=1H}.
This particular choice of gauge respects the duality
\bea
m\mapsto m'\!-\!m,\quad \mbox{shaded bands}\leftrightarrow \mbox{unshaded bands},\quad H^{m,m'}(a,b,c)\mapsto \half-H^{m'-m,m'}(a,b,c)
\eea
Anticipating the cases $n=2,3$, we further note that
\bea
h_{a+n}-h_a=\sum_{k=0}^{n-1} (h_{a+k+1}-h_{a+k})=\begin{cases}
0,&\mbox{$(a,a\!+\!n)$ is a shaded $n$-band}\\
1,&\mbox{$(a,a\!+\!n)$ is not a shaded $n$-band}
\end{cases}
\eea
since, from Appendix~\ref{sec:shadedNbands}, any $n$-band contains at most one unshaded 1-band.

\subsubsection{$n=2$ local energies}

\begin{figure}[p]
\begin{align}
\qquad\ \ 
\psset{unit=0.6cm}
\begin{pspicture}[shift=-1.85](0,0)(2,4)
\pspolygon[fillstyle=solid,fillcolor=apricot](0,1)(2,1)(2,4)(0,4)
\pspolygon[fillstyle=solid,fillcolor=lightlightblue](0,2)(2,2)(2,3)(0,3)
\pspolygon[fillstyle=solid,fillcolor=lightestblue](0,1)(2,1)(2,2)(0,2)
\pspolygon[fillstyle=solid,fillcolor=lightlightblue](0,0)(2,0)(2,1)(0,1)
\psgrid[gridlabels=0pt,subgriddiv=1]
\psline[linewidth=2pt,linecolor=black](0,0)(2,4)
\end{pspicture} = 
\psset{unit=0.6cm}
\begin{pspicture}[shift=-1.85](0,0)(2,4)
\pspolygon[fillstyle=solid,fillcolor=apricot](0,0)(2,0)(2,4)(0,4)
\pspolygon[fillstyle=solid,fillcolor=lightlightblue](0,2)(2,2)(2,3)(0,3)
\pspolygon[fillstyle=solid,fillcolor=lightestblue](0,1)(2,1)(2,2)(0,2)
\pspolygon[fillstyle=solid,fillcolor=lightlightblue](0,0)(2,0)(2,1)(0,1)
\psgrid[gridlabels=0pt,subgriddiv=1]
\psline[linewidth=2pt,linecolor=black](0,4)(2,0)
\end{pspicture} = 
\psset{unit=0.6cm}
\begin{pspicture}[shift=-1.85](0,0)(2,4)
\pspolygon[fillstyle=solid,fillcolor=apricot](0,0)(2,0)(2,4)(0,4)
\pspolygon[fillstyle=solid,fillcolor=lightlightblue](0,3)(0,4)(2,4)(2,3)
\pspolygon[fillstyle=solid,fillcolor=lightestblue](0,2)(2,2)(2,3)(0,3)
\pspolygon[fillstyle=solid,fillcolor=lightlightblue](0,1)(2,1)(2,2)(0,2)
\psgrid[gridlabels=0pt,subgriddiv=1]
\psline[linewidth=2pt,linecolor=black](0,0)(2,4)
\end{pspicture} = 
\psset{unit=0.6cm}
\begin{pspicture}[shift=-1.85](0,0)(2,4)
\pspolygon[fillstyle=solid,fillcolor=apricot](0,0)(2,0)(2,4)(0,4)
\pspolygon[fillstyle=solid,fillcolor=lightlightblue](0,3)(0,4)(2,4)(2,3)
\pspolygon[fillstyle=solid,fillcolor=lightestblue](0,2)(2,2)(2,3)(0,3)
\pspolygon[fillstyle=solid,fillcolor=lightlightblue](0,1)(2,1)(2,2)(0,2)
\psgrid[gridlabels=0pt,subgriddiv=1]
\psline[linewidth=2pt,linecolor=black](0,4)(2,0)
\end{pspicture} =
0
\qquad\qquad\ 
&
\psset{unit=0.6cm}
\begin{pspicture}[shift=-1.85](0,0)(2,4)
\pspolygon[fillstyle=solid,fillcolor=apricot](0,0)(2,0)(2,4)(0,4)
\pspolygon[fillstyle=solid,fillcolor=lightlightblue](0,1)(2,1)(2,3)(0,3)
\psgrid[gridlabels=0pt,subgriddiv=1]
\psline[linewidth=2pt,linecolor=black](0,0)(2,4)
\end{pspicture} =
\psset{unit=0.6cm}
\begin{pspicture}[shift=-1.85](0,0)(2,4)
\pspolygon[fillstyle=solid,fillcolor=apricot](0,0)(2,0)(2,4)(0,4)
\pspolygon[fillstyle=solid,fillcolor=lightlightblue](0,1)(2,1)(2,3)(0,3)
\psgrid[gridlabels=0pt,subgriddiv=1]
\psline[linewidth=2pt,linecolor=black](0,4)(2,0)
\end{pspicture} =  1\nonumber
\end{align}
\be
\psset{unit=0.6cm}
\begin{pspicture}[shift=-0.85](0,0)(2,2)
\pspolygon[fillstyle=solid,fillcolor=apricot](0,0)(2,0)(2,2)(0,2)
\psgrid[gridlabels=0pt,subgriddiv=1]
\psline[linewidth=2pt,linecolor=black](0,0)(1,0)
\psline[linewidth=2pt,linecolor=black](1,0)(2,2)
\end{pspicture}  =
\psset{unit=0.6cm}
\begin{pspicture}[shift=-0.85](0,0)(2,2)
\pspolygon[fillstyle=solid,fillcolor=apricot](0,0)(2,0)(2,2)(0,2)
\psgrid[gridlabels=0pt,subgriddiv=1]
\psline[linewidth=2pt,linecolor=black](0,2)(1,2)
\psline[linewidth=2pt,linecolor=black](1,2)(2,0)
\end{pspicture} = 
\psset{unit=0.6cm}
\begin{pspicture}[shift=-0.85](0,0)(2,2)
\pspolygon[fillstyle=solid,fillcolor=apricot](0,0)(2,0)(2,2)(0,2)
\psgrid[gridlabels=0pt,subgriddiv=1]
\psline[linewidth=2pt,linecolor=black](1,2)(2,2)
\psline[linewidth=2pt,linecolor=black](0,0)(1,2)
\end{pspicture} =
\psset{unit=0.6cm}
\begin{pspicture}[shift=-0.85](0,0)(2,2)
\pspolygon[fillstyle=solid,fillcolor=apricot](0,0)(2,0)(2,2)(0,2)
\psgrid[gridlabels=0pt,subgriddiv=1]
\psline[linewidth=2pt,linecolor=black](1,0)(2,0)
\psline[linewidth=2pt,linecolor=black](0,2)(1,0)
\end{pspicture} = \half \nonumber \\
\ee
\begin{align}
\psset{unit=0.6cm}
\qquad\quad
\begin{pspicture}[shift=-0.85](0,0)(2,2)
\pspolygon[fillstyle=solid,fillcolor=lightlightblue](0,0)(2,0)(2,2)(0,2)
\psgrid[gridlabels=0pt,subgriddiv=1]
\psline[linewidth=2pt,linecolor=black](0,0)(1,2)
\psline[linewidth=2pt,linecolor=black](1,2)(2,0)
\end{pspicture} &= 
\psset{unit=0.6cm}
\begin{pspicture}[shift=-0.85](0,0)(2,2)
\pspolygon[fillstyle=solid,fillcolor=lightlightblue](0,0)(2,0)(2,2)(0,2)
\psgrid[gridlabels=0pt,subgriddiv=1]
\psline[linewidth=2pt,linecolor=black](0,2)(1,0)
\psline[linewidth=2pt,linecolor=black](1,0)(2,2)
\end{pspicture} = 
\psset{unit=0.6cm}
\begin{pspicture}[shift=-0.85](0,0)(2,2)
\pspolygon[fillstyle=solid,fillcolor=lightestblue](0,1)(2,1)(2,2)(0,2)
\pspolygon[fillstyle=solid,fillcolor=lightlightblue](0,0)(2,0)(2,1)(0,1)
\psgrid[gridlabels=0pt,subgriddiv=1]
\psline[linewidth=2pt,linecolor=black](0,0)(1,2)
\psline[linewidth=2pt,linecolor=black](1,2)(2,0)
\end{pspicture} =
\psset{unit=0.6cm}
\begin{pspicture}[shift=-0.85](0,0)(2,2)
\pspolygon[fillstyle=solid,fillcolor=lightlightblue](0,1)(2,1)(2,2)(0,2)
\pspolygon[fillstyle=solid,fillcolor=lightestblue](0,0)(2,0)(2,1)(0,1)
\psgrid[gridlabels=0pt,subgriddiv=1]
\psline[linewidth=2pt,linecolor=black](0,2)(1,0)
\psline[linewidth=2pt,linecolor=black](1,0)(2,2)
\end{pspicture} =0
& 
\psset{unit=0.6cm}
\begin{pspicture}[shift=-0.85](0,0)(2,2)
\pspolygon[fillstyle=solid,fillcolor=lightlightblue](0,1)(2,1)(2,2)(0,2)
\pspolygon[fillstyle=solid,fillcolor=lightestblue](0,0)(2,0)(2,1)(0,1)
\psgrid[gridlabels=0pt,subgriddiv=1]
\psline[linewidth=2pt,linecolor=black](0,0)(1,2)
\psline[linewidth=2pt,linecolor=black](1,2)(2,0)
\end{pspicture} = 
\psset{unit=0.6cm}
\begin{pspicture}[shift=-0.85](0,0)(2,2)
\pspolygon[fillstyle=solid,fillcolor=lightestblue](0,1)(2,1)(2,2)(0,2)
\pspolygon[fillstyle=solid,fillcolor=lightlightblue](0,0)(2,0)(2,1)(0,1)
\psgrid[gridlabels=0pt,subgriddiv=1]
\psline[linewidth=2pt,linecolor=black](0,2)(1,0)
\psline[linewidth=2pt,linecolor=black](1,0)(2,2)
\end{pspicture} = 
 1\nonumber
\end{align}
\bea
\psset{unit=0.6cm}
\begin{pspicture}[shift=-0.85](0,0)(2,2)
\pspolygon[fillstyle=solid,fillcolor=lightlightblue](0,0)(2,0)(2,2)(0,2)
\psgrid[gridlabels=0pt,subgriddiv=1]
\psline[linewidth=2pt,linecolor=black](0,1)(2,1)
\end{pspicture} = 0  \qquad\ \ \ \ 
\psset{unit=0.6cm}
\begin{pspicture}[shift=-0.85](0,0)(2,2)
\pspolygon[fillstyle=solid,fillcolor=lightlightblue](0,1)(2,1)(2,2)(0,2)
\pspolygon[fillstyle=solid,fillcolor=lightestblue](0,0)(2,0)(2,1)(0,1)
\psgrid[gridlabels=0pt,subgriddiv=1]
\psline[linewidth=2pt,linecolor=black](0,1)(2,1)
\end{pspicture} = 
\psset{unit=0.6cm}
\begin{pspicture}[shift=-0.85](0,0)(2,2)
\pspolygon[fillstyle=solid,fillcolor=lightestblue](0,1)(2,1)(2,2)(0,2)
\pspolygon[fillstyle=solid,fillcolor=lightlightblue](0,0)(2,0)(2,1)(0,1)
\psgrid[gridlabels=0pt,subgriddiv=1]
\psline[linewidth=2pt,linecolor=black](0,1)(2,1)
\end{pspicture} =
1 \nonumber
\eea
\bea
\psset{unit=0.6cm}
\begin{pspicture}(0,.3)(2,1.5)
\psframe[linewidth=0pt,fillstyle=solid,fillcolor=apricot](0,0)(2,1)
\multirput(1,0)(1,0){2}{\psline[linewidth=.5pt](0,0)(0,1)}
\end{pspicture}\ =\ 
\begin{pspicture}(0,.3)(2,1.5)
\psframe[linewidth=0pt,fillstyle=solid,fillcolor=white](0,0)(2,1)
\multirput(1,0)(1,0){2}{\psline[linewidth=.5pt](0,0)(0,1)}
\end{pspicture}\quad\mbox{or}\quad \begin{pspicture}(0,.3)(2,1.5)
\psframe[linewidth=0pt,fillstyle=solid,fillcolor=lightlightblue](0,0)(2,1)
\multirput(1,0)(1,0){2}{\psline[linewidth=.5pt](0,0)(0,1)}
\end{pspicture}
\nonumber
\eea
\caption{\label{n=2H}The gauged local energies of the $n=2$ RSOS models in the interval $0<\lambda<\pi/2$. The light orange bands indicate that the result holds whether the band is shaded or unshaded. As shown in Appendix~\ref{sec:shadedNbands}, contiguous unshaded bands do not occur in the interval $0<\lambda<\pi/2$. In this gauge, the local energies take the values $0,\half,1$.}
\end{figure}
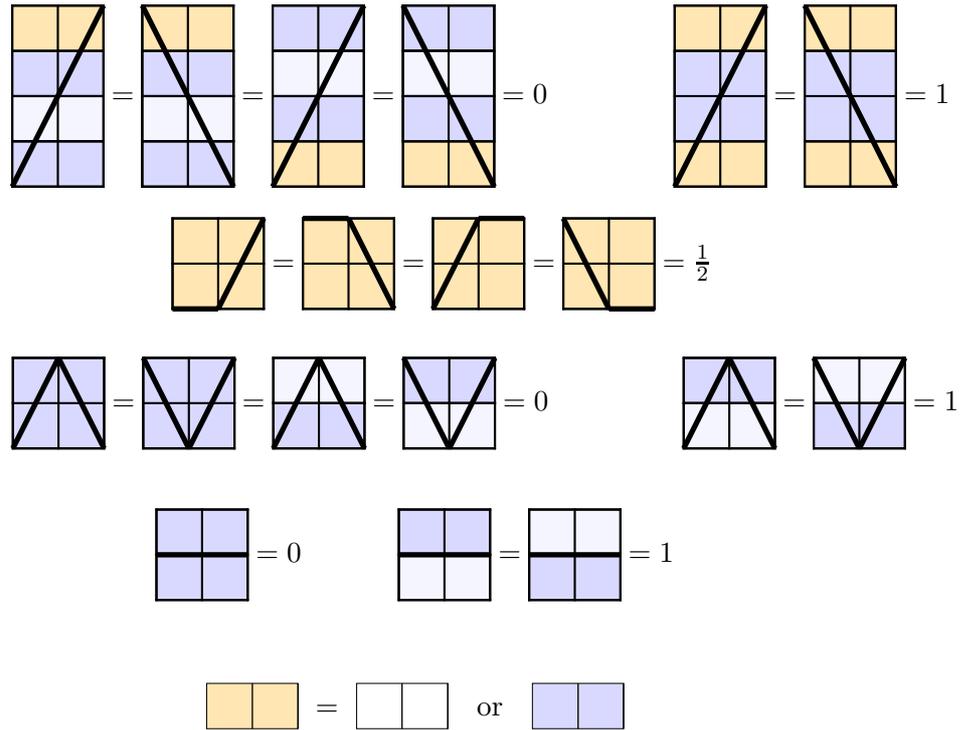

For $n=2$, the low temperature limit is similarly obtained by taking $x\rightarrow0$. 
After a renormalization by $\exp(2u(\lambda-u)/\varepsilon)$ and a conjugate modulus transformation, the diagonal $n=2$ face weights can be rewritten as in (\ref{conjmod2}) with the multiplicative gauge 
\be
g_a = w^{a(a \lambda - \pi)/4\pi}
\ee
The local energies are found to be
\begin{subequations}
\begin{align}
H(a\pm2,a,a\mp2)&= 1\\
H(a\pm2,a,a) &= H(a,a,a\pm2) = \half \pm  \flexpr{a\pm1}\\
H(a,a\pm2,a) &= \mp \big( \flexpr{a} + \flexpr{a\pm1}\big)\\
H(a,a,a) &= 
\begin{cases}
0, &\flexpr{a-1}= \flexpr{a} =  \flexpr{a+1}\\
1, &\text{otherwise}
\end{cases}
\end{align}
\end{subequations}

We apply a further additive gauge transformation $G_a$ that satisfies
\be
G_a = 
\begin{cases}
\flexpr{1} + \flexpr{3} + \ldots \flexpr{a-1}, &a \text{ even}\\
\flexpr{2} + \flexpr{4} + \ldots \flexpr{a-1}, &a \text{ odd}
\end{cases}
\qquad
G_{a+1} - G_{a-1} = \flexpr{a}
\ee
This transformation is implemented more neatly by defining $G(a,b) = G_b - G_a$ so that
\be
H'(a,b,c) = H(a,b,c) + G(a,b) - G(b,c) \geq0, \qquad G(a,b) =\half(b-a) \flexpr{\frac{a+b}{2}}
\ee 
Rewriting the local energies using this new gauge, and omitting the prime, gives the local energies 
\begin{subequations}
\begin{align}
H(a\pm2,a,a\mp2) &= \flexpr{a-1} - \flexpr{a+1} + 1 \label{eq:localE1}\\
H(a\pm2,a,a) &= H(a,a,a\pm2) = \half \label{eq:localE2}\\
H(a,a\pm2,a) &= \pm \big(\flexpr{a\pm1} -  \flexpr{a}\big) \label{eq:localE3}\\
H(a,a,a) &= 
\begin{cases}
0, &\flexpr{a-1}= \flexpr{a} =  \flexpr{a+1}\\
1, &\text{otherwise}
\end{cases} \label{eq:localE4}
\end{align}
\label{eq:localE}
\end{subequations}
These local energies take the values 0, $1/2$ or 1 as shown in Figure~\ref{n=2H}.

\subsubsection{$n=3$ local energies}
\allowdisplaybreaks

\begin{figure}[p]
\begin{align*}
\psset{unit=0.35cm}
\begin{pspicture}[shift=-2.7](0,0)(2,6)
\pspolygon[fillstyle=solid,fillcolor=lightlightblue](0,0)(2,0)(2,2)(0,2)
\pspolygon[fillstyle=solid,fillcolor=lightestblue](0,2)(2,2)(2,3)(0,3)
\pspolygon[fillstyle=solid,fillcolor=lightlightblue](0,3)(2,3)(2,5)(0,5)
\pspolygon[fillstyle=solid,fillcolor=lightestblue](0,5)(2,5)(2,6)(0,6)
\psgrid[gridlabels=0pt,subgriddiv=1]
\psline[linewidth=2pt,linecolor=black](0,0)(2,6)
\end{pspicture} &= 
\psset{unit=0.35cm}
\begin{pspicture}[shift=-2.7](0,0)(2,6)
\rput(0,6){\psscalebox{1 -1}{
\pspolygon[fillstyle=solid,fillcolor=lightestblue](0,0)(2,0)(2,1)(0,1)
\pspolygon[fillstyle=solid,fillcolor=lightlightblue](0,1)(2,1)(2,3)(0,3)
\pspolygon[fillstyle=solid,fillcolor=lightestblue,linecolor=red](0,3)(2,3)(2,4)(0,4)
\pspolygon[fillstyle=solid,fillcolor=lightlightblue](0,4)(2,4)(2,6)(0,6)
\psgrid[gridlabels=0pt,subgriddiv=1]
\psline[linewidth=2pt,linecolor=black](0,0)(2,6)
}}
\end{pspicture} =
\psset{unit=0.35cm}
\begin{pspicture}[shift=-2.7](0,0)(2,6)
\pspolygon[fillstyle=solid,fillcolor=lightestblue](0,0)(2,0)(2,1)(0,1)
\pspolygon[fillstyle=solid,fillcolor=lightlightblue](0,1)(2,1)(2,3)(0,3)
\pspolygon[fillstyle=solid,fillcolor=lightestblue,linecolor=red](0,3)(2,3)(2,4)(0,4)
\pspolygon[fillstyle=solid,fillcolor=lightlightblue](0,4)(2,4)(2,6)(0,6)
\psgrid[gridlabels=0pt,subgriddiv=1]
\psline[linewidth=2pt,linecolor=black](0,0)(2,6)
\end{pspicture} =
\psset{unit=0.35cm}
\begin{pspicture}[shift=-2.7](0,0)(2,6)
\rput(0,6){\psscalebox{1 -1}{
\pspolygon[fillstyle=solid,fillcolor=lightlightblue](0,0)(2,0)(2,2)(0,2)
\pspolygon[fillstyle=solid,fillcolor=lightestblue](0,2)(2,2)(2,3)(0,3)
\pspolygon[fillstyle=solid,fillcolor=lightlightblue](0,3)(2,3)(2,5)(0,5)
\pspolygon[fillstyle=solid,fillcolor=lightestblue](0,5)(2,5)(2,6)(0,6)
\psgrid[gridlabels=0pt,subgriddiv=1]
\psline[linewidth=2pt,linecolor=black](0,0)(2,6)
}}
\end{pspicture} = 
\begin{pspicture}[shift=-2.7](0,0)(2,6)
\pspolygon[fillstyle=solid,fillcolor=lightlightblue](0,0)(2,0)(2,1)(0,1)
\pspolygon[fillstyle=solid,fillcolor=lightestblue](0,1)(2,1)(2,2)(0,2)
\pspolygon[fillstyle=solid,fillcolor=lightlightblue](0,2)(2,2)(2,4)(0,4)
\pspolygon[fillstyle=solid,fillcolor=lightestblue](0,4)(2,4)(2,5)(0,5)
\pspolygon[fillstyle=solid,fillcolor=lightlightblue](0,5)(2,5)(2,6)(0,6)
\psgrid[gridlabels=0pt,subgriddiv=1]
\psline[linewidth=2pt,linecolor=black](0,0)(2,6)
\end{pspicture} =
\begin{pspicture}[shift=-2.7](0,0)(2,6)
\rput(0,6){\psscalebox{1 -1}{
\pspolygon[fillstyle=solid,fillcolor=lightlightblue](0,0)(2,0)(2,1)(0,1)
\pspolygon[fillstyle=solid,fillcolor=lightestblue](0,1)(2,1)(2,2)(0,2)
\pspolygon[fillstyle=solid,fillcolor=lightlightblue](0,2)(2,2)(2,4)(0,4)
\pspolygon[fillstyle=solid,fillcolor=lightestblue](0,4)(2,4)(2,5)(0,5)
\pspolygon[fillstyle=solid,fillcolor=lightlightblue](0,5)(2,5)(2,6)(0,6)
\psgrid[gridlabels=0pt,subgriddiv=1]
\psline[linewidth=2pt,linecolor=black](0,0)(2,6)
}}
\end{pspicture} = 0
&
\psset{unit=0.35cm}
\begin{pspicture}[shift=-2.7](0,0)(2,6)
\pspolygon[fillstyle=solid,fillcolor=lightlightblue](0,0)(2,0)(2,3)(0,3)
\pspolygon[fillstyle=solid,fillcolor=lightestblue](0,3)(2,3)(2,4)(0,4)
\pspolygon[fillstyle=solid,fillcolor=lightlightblue](0,4)(2,4)(2,6)(0,6)
\psgrid[gridlabels=0pt,subgriddiv=1]
\psline[linewidth=2pt,linecolor=black](0,0)(2,6)
\end{pspicture} &=
\psset{unit=0.35cm}
\begin{pspicture}[shift=-2.7](0,0)(2,6)
\rput(2,0){\psscalebox{-1 1}{
\pspolygon[fillstyle=solid,fillcolor=lightlightblue](0,0)(2,0)(2,3)(0,3)
\pspolygon[fillstyle=solid,fillcolor=lightestblue](0,3)(2,3)(2,4)(0,4)
\pspolygon[fillstyle=solid,fillcolor=lightlightblue](0,4)(2,4)(2,6)(0,6)
\psgrid[gridlabels=0pt,subgriddiv=1]
\psline[linewidth=2pt,linecolor=black](0,0)(2,6)
}}
\end{pspicture} =
\psset{unit=0.35cm}
\begin{pspicture}[shift=-2.7](0,0)(2,6)
\pspolygon[fillstyle=solid,fillcolor=lightlightblue](0,0)(2,0)(2,2)(0,2)
\pspolygon[fillstyle=solid,fillcolor=lightestblue](0,2)(2,2)(2,3)(0,3)
\pspolygon[fillstyle=solid,fillcolor=lightlightblue](0,3)(2,3)(2,6)(0,6)
\psgrid[gridlabels=0pt,subgriddiv=1]
\psline[linewidth=2pt,linecolor=black](0,0)(2,6)
\end{pspicture} = 
\psset{unit=0.35cm}
\begin{pspicture}[shift=-2.7](0,0)(2,6)
\rput(2,0){\psscalebox{-1 1}{
\pspolygon[fillstyle=solid,fillcolor=lightlightblue](0,0)(2,0)(2,2)(0,2)
\pspolygon[fillstyle=solid,fillcolor=lightestblue](0,2)(2,2)(2,3)(0,3)
\pspolygon[fillstyle=solid,fillcolor=lightlightblue](0,3)(2,3)(2,6)(0,6)
\psgrid[gridlabels=0pt,subgriddiv=1]
\psline[linewidth=2pt,linecolor=black](0,0)(2,6)
}}
\end{pspicture} =\frac{1}{4} \\[6pt]
\psset{unit=0.35cm}
\begin{pspicture}[shift=-2.7](0,0)(2,6)
\pspolygon[fillstyle=solid,fillcolor=lightlightblue](0,0)(2,0)(2,1)(0,1)
\pspolygon[fillstyle=solid,fillcolor=lightestblue](0,1)(2,1)(2,2)(0,2)
\pspolygon[fillstyle=solid,fillcolor=lightlightblue](0,2)(2,2)(2,5)(0,5)
\pspolygon[fillstyle=solid,fillcolor=lightestblue](0,5)(2,5)(2,6)(0,6)
\psgrid[gridlabels=0pt,subgriddiv=1]
\psline[linewidth=2pt,linecolor=black](0,0)(2,6)
\end{pspicture} &=
\psset{unit=0.35cm}
\begin{pspicture}[shift=-2.7](0,0)(2,6)
\rput(2,0){\psscalebox{-1 1}{
\pspolygon[fillstyle=solid,fillcolor=lightlightblue](0,0)(2,0)(2,1)(0,1)
\pspolygon[fillstyle=solid,fillcolor=lightestblue](0,1)(2,1)(2,2)(0,2)
\pspolygon[fillstyle=solid,fillcolor=lightlightblue](0,2)(2,2)(2,5)(0,5)
\pspolygon[fillstyle=solid,fillcolor=lightestblue](0,5)(2,5)(2,6)(0,6)
\psgrid[gridlabels=0pt,subgriddiv=1]
\psline[linewidth=2pt,linecolor=black](0,0)(2,6)
}}
\end{pspicture} =
\psset{unit=0.35cm}
\begin{pspicture}[shift=-2.7](0,0)(2,6)
\pspolygon[fillstyle=solid,fillcolor=lightestblue](0,0)(2,0)(2,1)(0,1)
\pspolygon[fillstyle=solid,fillcolor=lightlightblue](0,1)(2,1)(2,4)(0,4)
\pspolygon[fillstyle=solid,fillcolor=lightestblue](0,4)(2,4)(2,5)(0,5)
\pspolygon[fillstyle=solid,fillcolor=lightlightblue](0,5)(2,5)(2,6)(0,6)
\psgrid[gridlabels=0pt,subgriddiv=1]
\psline[linewidth=2pt,linecolor=black](0,0)(2,6)
\end{pspicture} = 
\psset{unit=0.35cm}
\begin{pspicture}[shift=-2.7](0,0)(2,6)
\rput(2,0){\psscalebox{-1 1}{
\pspolygon[fillstyle=solid,fillcolor=lightestblue](0,0)(2,0)(2,1)(0,1)
\pspolygon[fillstyle=solid,fillcolor=lightlightblue](0,1)(2,1)(2,4)(0,4)
\pspolygon[fillstyle=solid,fillcolor=lightestblue](0,4)(2,4)(2,5)(0,5)
\pspolygon[fillstyle=solid,fillcolor=lightlightblue](0,5)(2,5)(2,6)(0,6)
\psgrid[gridlabels=0pt,subgriddiv=1]
\psline[linewidth=2pt,linecolor=black](0,0)(2,6)
}}
\end{pspicture} = \frac{1}{2} 
& 
\psset{unit=0.35cm}
\begin{pspicture}[shift=-2.7](0,0)(2,6)
\pspolygon[fillstyle=solid,fillcolor=lightlightblue](0,0)(2,0)(2,4)(0,4)
\pspolygon[fillstyle=solid,fillcolor=lightestblue](0,4)(2,4)(2,5)(0,5)
\pspolygon[fillstyle=solid,fillcolor=lightlightblue](0,5)(2,5)(2,6)(0,6)
\psgrid[gridlabels=0pt,subgriddiv=1]
\psline[linewidth=2pt,linecolor=black](0,0)(2,6)
\end{pspicture} &= 
\psset{unit=0.35cm}
\begin{pspicture}[shift=-2.7](0,0)(2,6)
\rput(2,0){\psscalebox{-1 1}{
\pspolygon[fillstyle=solid,fillcolor=lightlightblue](0,0)(2,0)(2,4)(0,4)
\pspolygon[fillstyle=solid,fillcolor=lightestblue](0,4)(2,4)(2,5)(0,5)
\pspolygon[fillstyle=solid,fillcolor=lightlightblue](0,5)(2,5)(2,6)(0,6)
\psgrid[gridlabels=0pt,subgriddiv=1]
\psline[linewidth=2pt,linecolor=black](0,0)(2,6)
}}
\end{pspicture} = 
\psset{unit=0.35cm}
\begin{pspicture}[shift=-2.7](0,0)(2,6)
\pspolygon[fillstyle=solid,fillcolor=lightlightblue](0,0)(2,0)(2,1)(0,1)
\pspolygon[fillstyle=solid,fillcolor=lightestblue](0,1)(2,1)(2,2)(0,2)
\pspolygon[fillstyle=solid,fillcolor=lightlightblue](0,2)(2,2)(2,6)(0,6)
\psgrid[gridlabels=0pt,subgriddiv=1]
\psline[linewidth=2pt,linecolor=black](0,0)(2,6)
\end{pspicture} = 
\psset{unit=0.35cm}
\begin{pspicture}[shift=-2.7](0,0)(2,6)
\rput(2,0){\psscalebox{-1 1}{
\pspolygon[fillstyle=solid,fillcolor=lightlightblue](0,0)(2,0)(2,1)(0,1)
\pspolygon[fillstyle=solid,fillcolor=lightestblue](0,1)(2,1)(2,2)(0,2)
\pspolygon[fillstyle=solid,fillcolor=lightlightblue](0,2)(2,2)(2,6)(0,6)
\psgrid[gridlabels=0pt,subgriddiv=1]
\psline[linewidth=2pt,linecolor=black](0,0)(2,6)
}}
\end{pspicture} = \frac{3}{4}
\end{align*}\\[-26pt]
\begin{align*}
\psset{unit=0.35cm}
\begin{pspicture}[shift=-2.7](0,0)(2,6)
\pspolygon[fillstyle=solid,fillcolor=lightestblue](0,0)(2,0)(2,1)(0,1)
\pspolygon[fillstyle=solid,fillcolor=lightlightblue](0,1)(2,1)(2,5)(0,5)
\pspolygon[fillstyle=solid,fillcolor=lightestblue](0,5)(2,5)(2,6)(0,6)
\psgrid[gridlabels=0pt,subgriddiv=1]
\psline[linewidth=2pt,linecolor=black](0,0)(2,6)
\end{pspicture} &= 
\psset{unit=0.35cm}
\begin{pspicture}[shift=-2.7](0,0)(2,6)
\rput(2,0){\psscalebox{-1 1}{
\pspolygon[fillstyle=solid,fillcolor=lightestblue](0,0)(2,0)(2,1)(0,1)
\pspolygon[fillstyle=solid,fillcolor=lightlightblue](0,1)(2,1)(2,5)(0,5)
\pspolygon[fillstyle=solid,fillcolor=lightestblue](0,5)(2,5)(2,6)(0,6)
\psgrid[gridlabels=0pt,subgriddiv=1]
\psline[linewidth=2pt,linecolor=black](0,0)(2,6)
}}
\end{pspicture} = 1
&
\psset{unit=0.35cm}
\begin{pspicture}[shift=-2.7](0,0)(2,6)
\pspolygon[fillstyle=solid,fillcolor=lightlightblue](0,0)(2,0)(2,5)(0,5)
\pspolygon[fillstyle=solid,fillcolor=lightestblue](0,5)(2,5)(2,6)(0,6)
\psgrid[gridlabels=0pt,subgriddiv=1]
\psline[linewidth=2pt,linecolor=black](0,0)(2,6)
\end{pspicture} &= 
\psset{unit=0.35cm}
\begin{pspicture}[shift=-2.7](0,0)(2,6)
\rput(2,0){\psscalebox{-1 1}{
\pspolygon[fillstyle=solid,fillcolor=lightlightblue](0,0)(2,0)(2,5)(0,5)
\pspolygon[fillstyle=solid,fillcolor=lightestblue](0,5)(2,5)(2,6)(0,6)
\psgrid[gridlabels=0pt,subgriddiv=1]
\psline[linewidth=2pt,linecolor=black](0,0)(2,6)
}}
\end{pspicture} = 
\psset{unit=0.35cm}
\begin{pspicture}[shift=-2.7](0,0)(2,6)
\rput(2,6){\psscalebox{-1 -1}{
\pspolygon[fillstyle=solid,fillcolor=lightlightblue](0,0)(2,0)(2,5)(0,5)
\pspolygon[fillstyle=solid,fillcolor=lightestblue](0,5)(2,5)(2,6)(0,6)
\psgrid[gridlabels=0pt,subgriddiv=1]
\psline[linewidth=2pt,linecolor=black](0,0)(2,6)
}}
\end{pspicture} = 
\psset{unit=0.35cm}
\begin{pspicture}[shift=-2.7](0,0)(2,6)
\rput(0,6){\psscalebox{1 -1}{
\pspolygon[fillstyle=solid,fillcolor=lightlightblue](0,0)(2,0)(2,5)(0,5)
\pspolygon[fillstyle=solid,fillcolor=lightestblue](0,5)(2,5)(2,6)(0,6)
\psgrid[gridlabels=0pt,subgriddiv=1]
\psline[linewidth=2pt,linecolor=black](0,0)(2,6)
}}
\end{pspicture} = \frac{5}{4}
&
\psset{unit=0.35cm}
\begin{pspicture}[shift=-2.7](0,0)(2,6)
\rput(0,0){\psscalebox{1 1}{
\pspolygon[fillstyle=solid,fillcolor=lightlightblue](0,0)(2,0)(2,6)(0,6)
\psgrid[gridlabels=0pt,subgriddiv=1]
\psline[linewidth=2pt,linecolor=black](0,0)(2,6)
}}
\end{pspicture} &=
\psset{unit=0.35cm}
\begin{pspicture}[shift=-2.7](0,0)(2,6)
\rput(2,0){\psscalebox{-1 1}{
\pspolygon[fillstyle=solid,fillcolor=lightlightblue](0,0)(2,0)(2,6)(0,6)
\psgrid[gridlabels=0pt,subgriddiv=1]
\psline[linewidth=2pt,linecolor=black](0,0)(2,6)
}}
\end{pspicture} = \frac{3}{2}
\end{align*}\\[-26pt]
\begin{align*}
\psset{unit=0.35cm}
\begin{pspicture}[shift=-1.7](0,0)(2,4)
\rput(0,0){\psscalebox{1 1}{
\pspolygon[fillstyle=solid,fillcolor=lightlightblue](0,0)(2,0)(2,2)(0,2)
\pspolygon[fillstyle=solid,fillcolor=lightestblue](0,2)(2,2)(2,3)(0,3)
\pspolygon[fillstyle=solid,fillcolor=lightlightblue](0,3)(2,3)(2,4)(0,4)
\psgrid[gridlabels=0pt,subgriddiv=1]
\psline[linewidth=2pt,linecolor=black](0,0)(1,1)
\psline[linewidth=2pt,linecolor=black](1,1)(2,4)
}}
\end{pspicture} &=
\psset{unit=0.35cm}
\begin{pspicture}[shift=-1.7](0,0)(2,4)
\rput(2,0){\psscalebox{-1 1}{
\pspolygon[fillstyle=solid,fillcolor=lightlightblue](0,0)(2,0)(2,2)(0,2)
\pspolygon[fillstyle=solid,fillcolor=lightestblue](0,2)(2,2)(2,3)(0,3)
\pspolygon[fillstyle=solid,fillcolor=lightlightblue](0,3)(2,3)(2,4)(0,4)
\psgrid[gridlabels=0pt,subgriddiv=1]
\psline[linewidth=2pt,linecolor=black](0,0)(1,1)
\psline[linewidth=2pt,linecolor=black](1,1)(2,4)
}}
\end{pspicture} =
\psset{unit=0.35cm}
\begin{pspicture}[shift=-1.7](0,0)(2,4)
\rput(2,4){\psscalebox{-1 -1}{
\pspolygon[fillstyle=solid,fillcolor=lightlightblue](0,0)(2,0)(2,2)(0,2)
\pspolygon[fillstyle=solid,fillcolor=lightestblue](0,2)(2,2)(2,3)(0,3)
\pspolygon[fillstyle=solid,fillcolor=lightlightblue](0,3)(2,3)(2,4)(0,4)
\psgrid[gridlabels=0pt,subgriddiv=1]
\psline[linewidth=2pt,linecolor=black](0,0)(1,1)
\psline[linewidth=2pt,linecolor=black](1,1)(2,4)
}}
\end{pspicture} =
\psset{unit=0.35cm}
\begin{pspicture}[shift=-1.7](0,0)(2,4)
\rput(0,4){\psscalebox{1 -1}{
\pspolygon[fillstyle=solid,fillcolor=lightlightblue](0,0)(2,0)(2,2)(0,2)
\pspolygon[fillstyle=solid,fillcolor=lightestblue](0,2)(2,2)(2,3)(0,3)
\pspolygon[fillstyle=solid,fillcolor=lightlightblue](0,3)(2,3)(2,4)(0,4)
\psgrid[gridlabels=0pt,subgriddiv=1]
\psline[linewidth=2pt,linecolor=black](0,0)(1,1)
\psline[linewidth=2pt,linecolor=black](1,1)(2,4)
}}
\end{pspicture} = \frac{1}{4}
&
\psset{unit=0.35cm}
\begin{pspicture}[shift=-1.7](0,0)(2,4)
\rput(0,0){\psscalebox{1 1}{
\pspolygon[fillstyle=solid,fillcolor=lightestblue](0,0)(2,0)(2,1)(0,1)
\pspolygon[fillstyle=solid,fillcolor=lightlightblue](0,1)(2,1)(2,3)(0,3)
\pspolygon[fillstyle=solid,fillcolor=lightestblue](0,3)(2,3)(2,4)(0,4)
\psgrid[gridlabels=0pt,subgriddiv=1]
\psline[linewidth=2pt,linecolor=black](0,0)(1,1)
\psline[linewidth=2pt,linecolor=black](1,1)(2,4)
}}
\end{pspicture} &=
\psset{unit=0.35cm}
\begin{pspicture}[shift=-1.7](0,0)(2,4)
\rput(2,0){\psscalebox{-1 1}{
\pspolygon[fillstyle=solid,fillcolor=lightestblue](0,0)(2,0)(2,1)(0,1)
\pspolygon[fillstyle=solid,fillcolor=lightlightblue](0,1)(2,1)(2,3)(0,3)
\pspolygon[fillstyle=solid,fillcolor=lightestblue](0,3)(2,3)(2,4)(0,4)
\psgrid[gridlabels=0pt,subgriddiv=1]
\psline[linewidth=2pt,linecolor=black](0,0)(1,1)
\psline[linewidth=2pt,linecolor=black](1,1)(2,4)
}}
\end{pspicture} =
\psset{unit=0.35cm}
\begin{pspicture}[shift=-1.7](0,0)(2,4)
\rput(2,4){\psscalebox{-1 -1}{
\pspolygon[fillstyle=solid,fillcolor=lightestblue](0,0)(2,0)(2,1)(0,1)
\pspolygon[fillstyle=solid,fillcolor=lightlightblue](0,1)(2,1)(2,3)(0,3)
\pspolygon[fillstyle=solid,fillcolor=lightestblue](0,3)(2,3)(2,4)(0,4)
\psgrid[gridlabels=0pt,subgriddiv=1]
\psline[linewidth=2pt,linecolor=black](0,0)(1,1)
\psline[linewidth=2pt,linecolor=black](1,1)(2,4)
}}
\end{pspicture} =
\psset{unit=0.35cm}
\begin{pspicture}[shift=-1.7](0,0)(2,4)
\rput(0,4){\psscalebox{1 -1}{
\pspolygon[fillstyle=solid,fillcolor=lightestblue](0,0)(2,0)(2,1)(0,1)
\pspolygon[fillstyle=solid,fillcolor=lightlightblue](0,1)(2,1)(2,3)(0,3)
\pspolygon[fillstyle=solid,fillcolor=lightestblue](0,3)(2,3)(2,4)(0,4)
\psgrid[gridlabels=0pt,subgriddiv=1]
\psline[linewidth=2pt,linecolor=black](0,0)(1,1)
\psline[linewidth=2pt,linecolor=black](1,1)(2,4)
}}
\end{pspicture} = \frac{1}{2}
\end{align*}\\[-26pt]
\begin{align*}
\psset{unit=0.35cm}
\begin{pspicture}[shift=-1.7](0,0)(2,4)
\rput(0,0){\psscalebox{1 1}{
\pspolygon[fillstyle=solid,fillcolor=lightlightblue](0,0)(2,0)(2,3)(0,3)
\pspolygon[fillstyle=solid,fillcolor=lightestblue](0,3)(2,3)(2,4)(0,4)
\psgrid[gridlabels=0pt,subgriddiv=1]
\psline[linewidth=2pt,linecolor=black](0,0)(1,1)
\psline[linewidth=2pt,linecolor=black](1,1)(2,4)
}}
\end{pspicture} &=
\psset{unit=0.35cm}
\begin{pspicture}[shift=-1.7](0,0)(2,4)
\rput(2,0){\psscalebox{-1 1}{
\pspolygon[fillstyle=solid,fillcolor=lightlightblue](0,0)(2,0)(2,3)(0,3)
\pspolygon[fillstyle=solid,fillcolor=lightestblue](0,3)(2,3)(2,4)(0,4)
\psgrid[gridlabels=0pt,subgriddiv=1]
\psline[linewidth=2pt,linecolor=black](0,0)(1,1)
\psline[linewidth=2pt,linecolor=black](1,1)(2,4)
}}
\end{pspicture} =
\psset{unit=0.35cm}
\begin{pspicture}[shift=-1.7](0,0)(2,4)
\rput(2,4){\psscalebox{-1 -1}{
\pspolygon[fillstyle=solid,fillcolor=lightlightblue](0,0)(2,0)(2,3)(0,3)
\pspolygon[fillstyle=solid,fillcolor=lightestblue](0,3)(2,3)(2,4)(0,4)
\psgrid[gridlabels=0pt,subgriddiv=1]
\psline[linewidth=2pt,linecolor=black](0,0)(1,1)
\psline[linewidth=2pt,linecolor=black](1,1)(2,4)
}}
\end{pspicture} =
\psset{unit=0.35cm}
\begin{pspicture}[shift=-1.7](0,0)(2,4)
\rput(0,4){\psscalebox{1 -1}{
\pspolygon[fillstyle=solid,fillcolor=lightlightblue](0,0)(2,0)(2,3)(0,3)
\pspolygon[fillstyle=solid,fillcolor=lightestblue](0,3)(2,3)(2,4)(0,4)
\psgrid[gridlabels=0pt,subgriddiv=1]
\psline[linewidth=2pt,linecolor=black](0,0)(1,1)
\psline[linewidth=2pt,linecolor=black](1,1)(2,4)
}}
\end{pspicture} =
\psset{unit=0.35cm}
\begin{pspicture}[shift=-1.7](0,0)(2,4)
\rput(0,0){\psscalebox{1 1}{
\pspolygon[fillstyle=solid,fillcolor=lightlightblue](0,0)(2,0)(2,1)(0,1)
\pspolygon[fillstyle=solid,fillcolor=lightestblue](0,1)(2,1)(2,2)(0,2)
\pspolygon[fillstyle=solid,fillcolor=lightlightblue](0,2)(2,2)(2,4)(0,4)
\psgrid[gridlabels=0pt,subgriddiv=1]
\psline[linewidth=2pt,linecolor=black](0,0)(1,1)
\psline[linewidth=2pt,linecolor=black](1,1)(2,4)
}}
\end{pspicture} = 
\psset{unit=0.35cm}
\begin{pspicture}[shift=-1.7](0,0)(2,4)
\rput(2,0){\psscalebox{-1 1}{
\pspolygon[fillstyle=solid,fillcolor=lightlightblue](0,0)(2,0)(2,1)(0,1)
\pspolygon[fillstyle=solid,fillcolor=lightestblue](0,1)(2,1)(2,2)(0,2)
\pspolygon[fillstyle=solid,fillcolor=lightlightblue](0,2)(2,2)(2,4)(0,4)
\psgrid[gridlabels=0pt,subgriddiv=1]
\psline[linewidth=2pt,linecolor=black](0,0)(1,1)
\psline[linewidth=2pt,linecolor=black](1,1)(2,4)
}}
\end{pspicture} = 
\psset{unit=0.35cm}
\begin{pspicture}[shift=-1.7](0,0)(2,4)
\rput(2,4){\psscalebox{-1 -1}{
\pspolygon[fillstyle=solid,fillcolor=lightlightblue](0,0)(2,0)(2,1)(0,1)
\pspolygon[fillstyle=solid,fillcolor=lightestblue](0,1)(2,1)(2,2)(0,2)
\pspolygon[fillstyle=solid,fillcolor=lightlightblue](0,2)(2,2)(2,4)(0,4)
\psgrid[gridlabels=0pt,subgriddiv=1]
\psline[linewidth=2pt,linecolor=black](0,0)(1,1)
\psline[linewidth=2pt,linecolor=black](1,1)(2,4)
}}
\end{pspicture} = 
\psset{unit=0.35cm}
\begin{pspicture}[shift=-1.7](0,0)(2,4)
\rput(0,4){\psscalebox{1 -1}{
\pspolygon[fillstyle=solid,fillcolor=lightlightblue](0,0)(2,0)(2,1)(0,1)
\pspolygon[fillstyle=solid,fillcolor=lightestblue](0,1)(2,1)(2,2)(0,2)
\pspolygon[fillstyle=solid,fillcolor=lightlightblue](0,2)(2,2)(2,4)(0,4)
\psgrid[gridlabels=0pt,subgriddiv=1]
\psline[linewidth=2pt,linecolor=black](0,0)(1,1)
\psline[linewidth=2pt,linecolor=black](1,1)(2,4)
}}
\end{pspicture} = 
\psset{unit=0.35cm}
\begin{pspicture}[shift=-1.7](0,0)(2,4)
\rput(0,0){\psscalebox{1 1}{
\pspolygon[fillstyle=solid,fillcolor=lightestblue](0,0)(2,0)(2,1)(0,1)
\pspolygon[fillstyle=solid,fillcolor=lightlightblue](0,1)(2,1)(2,4)(0,4)
\psgrid[gridlabels=0pt,subgriddiv=1]
\psline[linewidth=2pt,linecolor=black](0,0)(1,1)
\psline[linewidth=2pt,linecolor=black](1,1)(2,4)
}}
\end{pspicture} = 
\psset{unit=0.35cm}
\begin{pspicture}[shift=-1.7](0,0)(2,4)
\rput(2,0){\psscalebox{-1 1}{
\pspolygon[fillstyle=solid,fillcolor=lightestblue](0,0)(2,0)(2,1)(0,1)
\pspolygon[fillstyle=solid,fillcolor=lightlightblue](0,1)(2,1)(2,4)(0,4)
\psgrid[gridlabels=0pt,subgriddiv=1]
\psline[linewidth=2pt,linecolor=black](0,0)(1,1)
\psline[linewidth=2pt,linecolor=black](1,1)(2,4)
}}
\end{pspicture} = 
\psset{unit=0.35cm}
\begin{pspicture}[shift=-1.7](0,0)(2,4)
\rput(2,4){\psscalebox{-1 -1}{
\pspolygon[fillstyle=solid,fillcolor=lightestblue](0,0)(2,0)(2,1)(0,1)
\pspolygon[fillstyle=solid,fillcolor=lightlightblue](0,1)(2,1)(2,4)(0,4)
\psgrid[gridlabels=0pt,subgriddiv=1]
\psline[linewidth=2pt,linecolor=black](0,0)(1,1)
\psline[linewidth=2pt,linecolor=black](1,1)(2,4)
}}
\end{pspicture} = 
\psset{unit=0.35cm}
\begin{pspicture}[shift=-1.7](0,0)(2,4)
\rput(0,4){\psscalebox{1 -1}{
\pspolygon[fillstyle=solid,fillcolor=lightestblue](0,0)(2,0)(2,1)(0,1)
\pspolygon[fillstyle=solid,fillcolor=lightlightblue](0,1)(2,1)(2,4)(0,4)
\psgrid[gridlabels=0pt,subgriddiv=1]
\psline[linewidth=2pt,linecolor=black](0,0)(1,1)
\psline[linewidth=2pt,linecolor=black](1,1)(2,4)
}}
\end{pspicture} = 
\frac{3}{4}
\end{align*}\\[-26pt]
\begin{align*}
\psset{unit=0.35cm}
\begin{pspicture}[shift=-1.7](0,0)(2,4)
\rput(0,0){\psscalebox{1 1}{
\pspolygon[fillstyle=solid,fillcolor=lightlightblue](0,0)(2,0)(2,4)(0,4)
\psgrid[gridlabels=0pt,subgriddiv=1]
\psline[linewidth=2pt,linecolor=black](0,0)(1,1)
\psline[linewidth=2pt,linecolor=black](1,1)(2,4)
}}
\end{pspicture} &= 
\psset{unit=0.35cm}
\begin{pspicture}[shift=-1.7](0,0)(2,4)
\rput(2,0){\psscalebox{-1 1}{
\pspolygon[fillstyle=solid,fillcolor=lightlightblue](0,0)(2,0)(2,4)(0,4)
\psgrid[gridlabels=0pt,subgriddiv=1]
\psline[linewidth=2pt,linecolor=black](0,0)(1,1)
\psline[linewidth=2pt,linecolor=black](1,1)(2,4)
}}
\end{pspicture} = 
\psset{unit=0.35cm}
\begin{pspicture}[shift=-1.7](0,0)(2,4)
\rput(2,4){\psscalebox{-1 -1}{
\pspolygon[fillstyle=solid,fillcolor=lightlightblue](0,0)(2,0)(2,4)(0,4)
\psgrid[gridlabels=0pt,subgriddiv=1]
\psline[linewidth=2pt,linecolor=black](0,0)(1,1)
\psline[linewidth=2pt,linecolor=black](1,1)(2,4)
}}
\end{pspicture} = 
\psset{unit=0.35cm}
\begin{pspicture}[shift=-1.7](0,0)(2,4)
\rput(0,4){\psscalebox{1 -1}{
\pspolygon[fillstyle=solid,fillcolor=lightlightblue](0,0)(2,0)(2,4)(0,4)
\psgrid[gridlabels=0pt,subgriddiv=1]
\psline[linewidth=2pt,linecolor=black](0,0)(1,1)
\psline[linewidth=2pt,linecolor=black](1,1)(2,4)
}}
\end{pspicture} = 1
\end{align*}\\[-26pt]
\begin{align*}
\psset{unit=0.35cm}
\begin{pspicture}[shift=-1.2](0,0)(2,3)
\rput(0,0){\psscalebox{1 1}{
\pspolygon[fillstyle=solid,fillcolor=lightlightblue](0,0)(2,0)(2,2)(0,2)
\pspolygon[fillstyle=solid,fillcolor=lightestblue](0,2)(2,2)(2,3)(0,3)
\psgrid[gridlabels=0pt,subgriddiv=1]
\psline[linewidth=2pt,linecolor=black](0,1)(1,0)
\psline[linewidth=2pt,linecolor=black](1,0)(2,3)
}} 
\end{pspicture} &=
\psset{unit=0.35cm}
\begin{pspicture}[shift=-1.2](0,0)(2,3)
\rput(2,0){\psscalebox{-1 1}{
\pspolygon[fillstyle=solid,fillcolor=lightlightblue](0,0)(2,0)(2,2)(0,2)
\pspolygon[fillstyle=solid,fillcolor=lightestblue](0,2)(2,2)(2,3)(0,3)
\psgrid[gridlabels=0pt,subgriddiv=1]
\psline[linewidth=2pt,linecolor=black](0,1)(1,0)
\psline[linewidth=2pt,linecolor=black](1,0)(2,3)
}} 
\end{pspicture} =
\psset{unit=0.35cm}
\begin{pspicture}[shift=-1.2](0,0)(2,3)
\rput(2,3){\psscalebox{-1 -1}{
\pspolygon[fillstyle=solid,fillcolor=lightlightblue](0,0)(2,0)(2,2)(0,2)
\pspolygon[fillstyle=solid,fillcolor=lightestblue](0,2)(2,2)(2,3)(0,3)
\psgrid[gridlabels=0pt,subgriddiv=1]
\psline[linewidth=2pt,linecolor=black](0,1)(1,0)
\psline[linewidth=2pt,linecolor=black](1,0)(2,3)
}} 
\end{pspicture} =
\psset{unit=0.35cm}
\begin{pspicture}[shift=-1.2](0,0)(2,3)
\rput(0,3){\psscalebox{1 -1}{
\pspolygon[fillstyle=solid,fillcolor=lightlightblue](0,0)(2,0)(2,2)(0,2)
\pspolygon[fillstyle=solid,fillcolor=lightestblue](0,2)(2,2)(2,3)(0,3)
\psgrid[gridlabels=0pt,subgriddiv=1]
\psline[linewidth=2pt,linecolor=black](0,1)(1,0)
\psline[linewidth=2pt,linecolor=black](1,0)(2,3)
}} 
\end{pspicture} = \frac{1}{4}
&
\psset{unit=0.35cm}
\begin{pspicture}[shift=-1.2](0,0)(2,3)
\rput(0,0){\psscalebox{1 1}{
\pspolygon[fillstyle=solid,fillcolor=lightlightblue](0,0)(2,0)(2,3)(0,3)
\psgrid[gridlabels=0pt,subgriddiv=1]
\psline[linewidth=2pt,linecolor=black](0,1)(1,0)
\psline[linewidth=2pt,linecolor=black](1,0)(2,3)
}}
\end{pspicture} &=  
\psset{unit=0.35cm}
\begin{pspicture}[shift=-1.2](0,0)(2,3)
\rput(2,0){\psscalebox{-1 1}{
\pspolygon[fillstyle=solid,fillcolor=lightlightblue](0,0)(2,0)(2,3)(0,3)
\psgrid[gridlabels=0pt,subgriddiv=1]
\psline[linewidth=2pt,linecolor=black](0,1)(1,0)
\psline[linewidth=2pt,linecolor=black](1,0)(2,3)
}}
\end{pspicture} =  
\psset{unit=0.35cm}
\begin{pspicture}[shift=-1.2](0,0)(2,3)
\rput(2,3){\psscalebox{-1 -1}{
\pspolygon[fillstyle=solid,fillcolor=lightlightblue](0,0)(2,0)(2,3)(0,3)
\psgrid[gridlabels=0pt,subgriddiv=1]
\psline[linewidth=2pt,linecolor=black](0,1)(1,0)
\psline[linewidth=2pt,linecolor=black](1,0)(2,3)
}}
\end{pspicture} =  
\psset{unit=0.35cm}
\begin{pspicture}[shift=-1.2](0,0)(2,3)
\rput(0,3){\psscalebox{1 -1}{
\pspolygon[fillstyle=solid,fillcolor=lightlightblue](0,0)(2,0)(2,3)(0,3)
\psgrid[gridlabels=0pt,subgriddiv=1]
\psline[linewidth=2pt,linecolor=black](0,1)(1,0)
\psline[linewidth=2pt,linecolor=black](1,0)(2,3)
}}
\end{pspicture} =  \frac{1}{2}\\[6pt]
\psset{unit=0.35cm}
\begin{pspicture}[shift=-1.2](0,0)(2,3)
\rput(0,0){\psscalebox{1 1}{
\pspolygon[fillstyle=solid,fillcolor=lightlightblue](0,0)(2,0)(2,1)(0,1)
\pspolygon[fillstyle=solid,fillcolor=lightestblue](0,1)(2,1)(2,2)(0,2)
\pspolygon[fillstyle=solid,fillcolor=lightlightblue](0,2)(2,2)(2,3)(0,3)
\psgrid[gridlabels=0pt,subgriddiv=1]
\psline[linewidth=2pt,linecolor=black](0,1)(1,0)
\psline[linewidth=2pt,linecolor=black](1,0)(2,3)
}}
\end{pspicture} &=  
\psset{unit=0.35cm}
\begin{pspicture}[shift=-1.2](0,0)(2,3)
\rput(2,0){\psscalebox{-1 1}{
\pspolygon[fillstyle=solid,fillcolor=lightlightblue](0,0)(2,0)(2,1)(0,1)
\pspolygon[fillstyle=solid,fillcolor=lightestblue](0,1)(2,1)(2,2)(0,2)
\pspolygon[fillstyle=solid,fillcolor=lightlightblue](0,2)(2,2)(2,3)(0,3)
\psgrid[gridlabels=0pt,subgriddiv=1]
\psline[linewidth=2pt,linecolor=black](0,1)(1,0)
\psline[linewidth=2pt,linecolor=black](1,0)(2,3)
}}
\end{pspicture} =  
\psset{unit=0.35cm}
\begin{pspicture}[shift=-1.2](0,0)(2,3)
\rput(2,3){\psscalebox{-1 -1}{
\pspolygon[fillstyle=solid,fillcolor=lightlightblue](0,0)(2,0)(2,1)(0,1)
\pspolygon[fillstyle=solid,fillcolor=lightestblue](0,1)(2,1)(2,2)(0,2)
\pspolygon[fillstyle=solid,fillcolor=lightlightblue](0,2)(2,2)(2,3)(0,3)
\psgrid[gridlabels=0pt,subgriddiv=1]
\psline[linewidth=2pt,linecolor=black](0,1)(1,0)
\psline[linewidth=2pt,linecolor=black](1,0)(2,3)
}}
\end{pspicture} =  
\psset{unit=0.35cm}
\begin{pspicture}[shift=-1.2](0,0)(2,3)
\rput(0,3){\psscalebox{1 -1}{
\pspolygon[fillstyle=solid,fillcolor=lightlightblue](0,0)(2,0)(2,1)(0,1)
\pspolygon[fillstyle=solid,fillcolor=lightestblue](0,1)(2,1)(2,2)(0,2)
\pspolygon[fillstyle=solid,fillcolor=lightlightblue](0,2)(2,2)(2,3)(0,3)
\psgrid[gridlabels=0pt,subgriddiv=1]
\psline[linewidth=2pt,linecolor=black](0,1)(1,0)
\psline[linewidth=2pt,linecolor=black](1,0)(2,3)
}}
\end{pspicture} =  \frac{3}{4} 
&
\psset{unit=0.35cm}
\begin{pspicture}[shift=-1.2](0,0)(2,3)
\rput(0,0){\psscalebox{1 1}{
\pspolygon[fillstyle=solid,fillcolor=lightestblue](0,0)(2,0)(2,1)(0,1)
\pspolygon[fillstyle=solid,fillcolor=lightlightblue](0,1)(2,1)(2,3)(0,3)
\psgrid[gridlabels=0pt,subgriddiv=1]
\psline[linewidth=2pt,linecolor=black](0,1)(1,0)
\psline[linewidth=2pt,linecolor=black](1,0)(2,3)
}}
\end{pspicture} &= 
\psset{unit=0.35cm}
\begin{pspicture}[shift=-1.2](0,0)(2,3)
\rput(2,0){\psscalebox{-1 1}{
\pspolygon[fillstyle=solid,fillcolor=lightestblue](0,0)(2,0)(2,1)(0,1)
\pspolygon[fillstyle=solid,fillcolor=lightlightblue](0,1)(2,1)(2,3)(0,3)
\psgrid[gridlabels=0pt,subgriddiv=1]
\psline[linewidth=2pt,linecolor=black](0,1)(1,0)
\psline[linewidth=2pt,linecolor=black](1,0)(2,3)
}}
\end{pspicture} = 
\psset{unit=0.35cm}
\begin{pspicture}[shift=-1.2](0,0)(2,3)
\rput(2,3){\psscalebox{-1 -1}{
\pspolygon[fillstyle=solid,fillcolor=lightestblue](0,0)(2,0)(2,1)(0,1)
\pspolygon[fillstyle=solid,fillcolor=lightlightblue](0,1)(2,1)(2,3)(0,3)
\psgrid[gridlabels=0pt,subgriddiv=1]
\psline[linewidth=2pt,linecolor=black](0,1)(1,0)
\psline[linewidth=2pt,linecolor=black](1,0)(2,3)
}}
\end{pspicture} = 
\psset{unit=0.35cm}
\begin{pspicture}[shift=-1.2](0,0)(2,3)
\rput(0,3){\psscalebox{1 -1}{
\pspolygon[fillstyle=solid,fillcolor=lightestblue](0,0)(2,0)(2,1)(0,1)
\pspolygon[fillstyle=solid,fillcolor=lightlightblue](0,1)(2,1)(2,3)(0,3)
\psgrid[gridlabels=0pt,subgriddiv=1]
\psline[linewidth=2pt,linecolor=black](0,1)(1,0)
\psline[linewidth=2pt,linecolor=black](1,0)(2,3)
}}
\end{pspicture} = 1
\end{align*}\\[-26pt]
\begin{align*}
\psset{unit=0.35cm}
\begin{pspicture}[shift=-1.2](0,0)(2,3)
\pspolygon[fillstyle=solid,fillcolor=lightlightblue](0,0)(2,0)(2,3)(0,3)
\psgrid[gridlabels=0pt,subgriddiv=1]
\psline[linewidth=2pt,linecolor=black](0,3)(1,0)
\psline[linewidth=2pt,linecolor=black](1,0)(2,3)
\end{pspicture} &=
\psset{unit=0.35cm}
\begin{pspicture}[shift=-1.2](0,0)(2,3)
\pspolygon[fillstyle=solid,fillcolor=lightlightblue](0,0)(2,0)(2,3)(0,3)
\psgrid[gridlabels=0pt,subgriddiv=1]
\psline[linewidth=2pt,linecolor=black](0,0)(1,3)
\psline[linewidth=2pt,linecolor=black](1,3)(2,0)
\end{pspicture} =  0 &
\psset{unit=0.35cm}
\begin{pspicture}[shift=-1.2](0,0)(2,3)
\rput(0,0){\psscalebox{1 1}{
\pspolygon[fillstyle=solid,fillcolor=lightlightblue](0,0)(2,0)(2,2)(0,2)
\pspolygon[fillstyle=solid,fillcolor=lightestblue](0,2)(2,2)(2,3)(0,3)
\psgrid[gridlabels=0pt,subgriddiv=1]
\psline[linewidth=2pt,linecolor=black](0,0)(1,3)
\psline[linewidth=2pt,linecolor=black](1,3)(2,0)
}}
\end{pspicture} &=
\psset{unit=0.35cm}
\begin{pspicture}[shift=-1.2](0,0)(2,3)
\rput(2,3){\psscalebox{-1 -1}{
\pspolygon[fillstyle=solid,fillcolor=lightlightblue](0,0)(2,0)(2,2)(0,2)
\pspolygon[fillstyle=solid,fillcolor=lightestblue](0,2)(2,2)(2,3)(0,3)
\psgrid[gridlabels=0pt,subgriddiv=1]
\psline[linewidth=2pt,linecolor=black](0,0)(1,3)
\psline[linewidth=2pt,linecolor=black](1,3)(2,0)
}}
\end{pspicture} =
\psset{unit=0.35cm}
\begin{pspicture}[shift=-1.2](0,0)(2,3)
\rput(0,0){\psscalebox{1 1}{
\pspolygon[fillstyle=solid,fillcolor=lightlightblue](0,0)(2,0)(2,1)(0,1)
\pspolygon[fillstyle=solid,fillcolor=lightestblue](0,1)(2,1)(2,2)(0,2)
\pspolygon[fillstyle=solid,fillcolor=lightlightblue](0,2)(2,2)(2,3)(0,3)
\psgrid[gridlabels=0pt,subgriddiv=1]
\psline[linewidth=2pt,linecolor=black](0,0)(1,3)
\psline[linewidth=2pt,linecolor=black](1,3)(2,0)
}}
\end{pspicture} =
\psset{unit=0.35cm}
\begin{pspicture}[shift=-1.2](0,0)(2,3)
\rput(0,3){\psscalebox{1 -1}{
\pspolygon[fillstyle=solid,fillcolor=lightlightblue](0,0)(2,0)(2,1)(0,1)
\pspolygon[fillstyle=solid,fillcolor=lightestblue](0,1)(2,1)(2,2)(0,2)
\pspolygon[fillstyle=solid,fillcolor=lightlightblue](0,2)(2,2)(2,3)(0,3)
\psgrid[gridlabels=0pt,subgriddiv=1]
\psline[linewidth=2pt,linecolor=black](0,0)(1,3)
\psline[linewidth=2pt,linecolor=black](1,3)(2,0)
}}
\end{pspicture} =
\psset{unit=0.35cm}
\begin{pspicture}[shift=-1.2](0,0)(2,3)
\rput(0,0){\psscalebox{1 1}{
\pspolygon[fillstyle=solid,fillcolor=lightestblue](0,0)(2,0)(2,1)(0,1)
\pspolygon[fillstyle=solid,fillcolor=lightlightblue](0,1)(2,1)(2,3)(0,3)
\psgrid[gridlabels=0pt,subgriddiv=1]
\psline[linewidth=2pt,linecolor=black](0,0)(1,3)
\psline[linewidth=2pt,linecolor=black](1,3)(2,0)
}}
\end{pspicture} = 
\psset{unit=0.35cm}
\begin{pspicture}[shift=-1.2](0,0)(2,3)
\rput(0,3){\psscalebox{1 -1}{
\pspolygon[fillstyle=solid,fillcolor=lightestblue](0,0)(2,0)(2,1)(0,1)
\pspolygon[fillstyle=solid,fillcolor=lightlightblue](0,1)(2,1)(2,3)(0,3)
\psgrid[gridlabels=0pt,subgriddiv=1]
\psline[linewidth=2pt,linecolor=black](0,0)(1,3)
\psline[linewidth=2pt,linecolor=black](1,3)(2,0)
}}
\end{pspicture} = \frac{1}{2}
\end{align*}\\[-26pt]
\begin{align*}
\psset{unit=0.35cm}
\begin{pspicture}[shift=-1.2](0,0)(2,3)
\rput(0,0){\psscalebox{1 1}{
\pspolygon[fillstyle=solid,fillcolor=lightlightblue](0,0)(2,0)(2,3)(0,3)
\psgrid[gridlabels=0pt,subgriddiv=1]
\psline[linewidth=2pt,linecolor=black](0,2)(1,1)
\psline[linewidth=2pt,linecolor=black](1,1)(2,2)
}}
\end{pspicture} &=  
\psset{unit=0.35cm}
\begin{pspicture}[shift=-1.2](0,0)(2,3)
\rput(0,3){\psscalebox{1 -1}{
\pspolygon[fillstyle=solid,fillcolor=lightlightblue](0,0)(2,0)(2,3)(0,3)
\psgrid[gridlabels=0pt,subgriddiv=1]
\psline[linewidth=2pt,linecolor=black](0,2)(1,1)
\psline[linewidth=2pt,linecolor=black](1,1)(2,2)
}}
\end{pspicture} = 0
&
\psset{unit=0.35cm}
\begin{pspicture}[shift=-1.2](0,0)(2,3)
\rput(0,0){\psscalebox{1 1}{
\pspolygon[fillstyle=solid,fillcolor=lightlightblue](0,0)(2,0)(2,2)(0,2)
\pspolygon[fillstyle=solid,fillcolor=lightestblue](0,2)(2,2)(2,3)(0,3)
\psgrid[gridlabels=0pt,subgriddiv=1]
\psline[linewidth=2pt,linecolor=black](0,2)(1,1)
\psline[linewidth=2pt,linecolor=black](1,1)(2,2)
}}
\end{pspicture} &= 
\psset{unit=0.35cm}
\begin{pspicture}[shift=-1.2](0,0)(2,3)
\rput(0,3){\psscalebox{1 -1}{
\pspolygon[fillstyle=solid,fillcolor=lightlightblue](0,0)(2,0)(2,2)(0,2)
\pspolygon[fillstyle=solid,fillcolor=lightestblue](0,2)(2,2)(2,3)(0,3)
\psgrid[gridlabels=0pt,subgriddiv=1]
\psline[linewidth=2pt,linecolor=black](0,2)(1,1)
\psline[linewidth=2pt,linecolor=black](1,1)(2,2)
}}
\end{pspicture} = 
\psset{unit=0.35cm}
\begin{pspicture}[shift=-1.2](0,0)(2,3)
\rput(0,3){\psscalebox{1 -1}{
\pspolygon[fillstyle=solid,fillcolor=lightestblue](0,0)(2,0)(2,1)(0,1)
\pspolygon[fillstyle=solid,fillcolor=lightlightblue](0,1)(2,1)(2,3)(0,3)
\psgrid[gridlabels=0pt,subgriddiv=1]
\psline[linewidth=2pt,linecolor=black](0,2)(1,1)
\psline[linewidth=2pt,linecolor=black](1,1)(2,2)
}}
\end{pspicture} = 
\psset{unit=0.35cm}
\begin{pspicture}[shift=-1.2](0,0)(2,3)
\rput(0,0){\psscalebox{1 1}{
\pspolygon[fillstyle=solid,fillcolor=lightestblue](0,0)(2,0)(2,1)(0,1)
\pspolygon[fillstyle=solid,fillcolor=lightlightblue](0,1)(2,1)(2,3)(0,3)
\psgrid[gridlabels=0pt,subgriddiv=1]
\psline[linewidth=2pt,linecolor=black](0,2)(1,1)
\psline[linewidth=2pt,linecolor=black](1,1)(2,2)
}}
\end{pspicture} = 1 
& 
\psset{unit=0.35cm}
\begin{pspicture}[shift=-1.2](0,0)(2,3)
\rput(0,0){\psscalebox{1 1}{
\pspolygon[fillstyle=solid,fillcolor=lightlightblue](0,0)(2,0)(2,1)(0,1)
\pspolygon[fillstyle=solid,fillcolor=lightestblue](0,1)(2,1)(2,2)(0,2)
\pspolygon[fillstyle=solid,fillcolor=lightlightblue](0,2)(2,2)(2,3)(0,3)
\psgrid[gridlabels=0pt,subgriddiv=1]
\psline[linewidth=2pt,linecolor=black](0,2)(1,1)
\psline[linewidth=2pt,linecolor=black](1,1)(2,2)
}}
\end{pspicture} &=
\psset{unit=0.35cm}
\begin{pspicture}[shift=-1.2](0,0)(2,3)
\rput(0,3){\psscalebox{1 -1}{
\pspolygon[fillstyle=solid,fillcolor=lightlightblue](0,0)(2,0)(2,1)(0,1)
\pspolygon[fillstyle=solid,fillcolor=lightestblue](0,1)(2,1)(2,2)(0,2)
\pspolygon[fillstyle=solid,fillcolor=lightlightblue](0,2)(2,2)(2,3)(0,3)
\psgrid[gridlabels=0pt,subgriddiv=1]
\psline[linewidth=2pt,linecolor=black](0,2)(1,1)
\psline[linewidth=2pt,linecolor=black](1,1)(2,2)
}}
\end{pspicture} = \frac{3}{2}
\end{align*}\\[-26pt]
\begin{align*}
\psset{unit=0.35cm}
\begin{pspicture}[shift=-0.75](0,0)(2,2)
\rput(0,0){\psscalebox{1 1}{
\pspolygon[fillstyle=solid,fillcolor=lightlightblue](0,0)(2,0)(2,2)(0,2)
\psgrid[gridlabels=0pt,subgriddiv=1]
\psline[linewidth=2pt,linecolor=black](0,0)(2,2)
}}
\end{pspicture} &=  
\psset{unit=0.35cm}
\begin{pspicture}[shift=-0.75](0,0)(2,2)
\rput(0,2){\psscalebox{1 -1}{
\pspolygon[fillstyle=solid,fillcolor=lightlightblue](0,0)(2,0)(2,2)(0,2)
\psgrid[gridlabels=0pt,subgriddiv=1]
\psline[linewidth=2pt,linecolor=black](0,0)(2,2)
}}
\end{pspicture} = \frac{1}{2} 
&
\psset{unit=0.35cm}
\begin{pspicture}[shift=-0.75](0,0)(2,2)
\rput(0,0){\psscalebox{1 1}{
\pspolygon[fillstyle=solid,fillcolor=lightlightblue](0,0)(2,0)(2,1)(0,1)
\pspolygon[fillstyle=solid,fillcolor=lightestblue](0,1)(2,1)(2,2)(0,2)
\psgrid[gridlabels=0pt,subgriddiv=1]
\psline[linewidth=2pt,linecolor=black](0,0)(2,2)
}}
\end{pspicture} &=  
\psset{unit=0.35cm}
\begin{pspicture}[shift=-0.75](0,0)(2,2)
\rput(2,0){\psscalebox{-1 1}{
\pspolygon[fillstyle=solid,fillcolor=lightlightblue](0,0)(2,0)(2,1)(0,1)
\pspolygon[fillstyle=solid,fillcolor=lightestblue](0,1)(2,1)(2,2)(0,2)
\psgrid[gridlabels=0pt,subgriddiv=1]
\psline[linewidth=2pt,linecolor=black](0,0)(2,2)
}}
\end{pspicture} =  
\psset{unit=0.35cm}
\begin{pspicture}[shift=-0.75](0,0)(2,2)
\rput(2,2){\psscalebox{-1 -1}{
\pspolygon[fillstyle=solid,fillcolor=lightlightblue](0,0)(2,0)(2,1)(0,1)
\pspolygon[fillstyle=solid,fillcolor=lightestblue](0,1)(2,1)(2,2)(0,2)
\psgrid[gridlabels=0pt,subgriddiv=1]
\psline[linewidth=2pt,linecolor=black](0,0)(2,2)
}}
\end{pspicture} =  
\psset{unit=0.35cm}
\begin{pspicture}[shift=-0.75](0,0)(2,2)
\rput(0,2){\psscalebox{1 -1}{
\pspolygon[fillstyle=solid,fillcolor=lightlightblue](0,0)(2,0)(2,1)(0,1)
\pspolygon[fillstyle=solid,fillcolor=lightestblue](0,1)(2,1)(2,2)(0,2)
\psgrid[gridlabels=0pt,subgriddiv=1]
\psline[linewidth=2pt,linecolor=black](0,0)(2,2)
}}
\end{pspicture} =  \frac{5}{4}
\end{align*}
\caption{\label{fig:n=2energies}The gauged local energies of the $n=3$ RSOS models in the interval $0<\lambda<\pi/3$. In this gauge, the local energies take the values $0,\tfrac{1}{4},\tfrac{1}{2},\tfrac{3}{4},1,\tfrac{5}{4},\tfrac{3}{2}$.}
\end{figure}
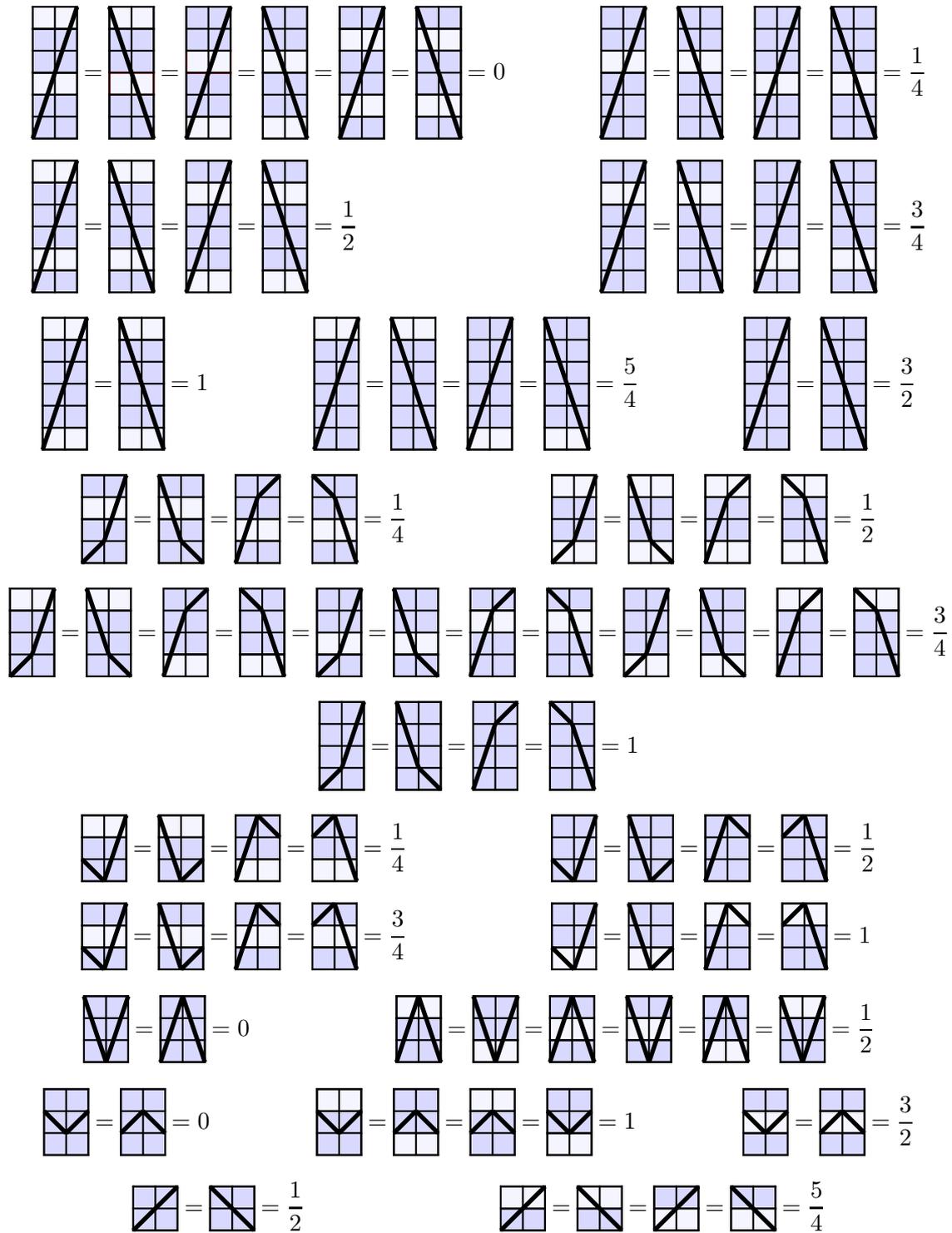

For $n=3$, we follow the same procedure as in the $n=2$ case with the same multiplicative gauge $g_a = w^{a(a \lambda - \pi)/4\pi}$. We perform a conjugate modulus transformation on the weights and further normalize by $\exp(3u(\lambda-u)/\varepsilon)$ to obtain the conjugate modulus face weights as written in (\ref{conjmod3}).

\goodbreak
Taking the low-temperature limit $x\rightarrow0$ gives the  local energy functions
\begin{subequations}
\begin{align}
H(a\pm3,a,a\pm3) &= \pm(h_{a\pm1}+h_{a\pm2}+h_{a\pm3})\\
H(a\pm3,a,a\pm1) &=   H(a\pm1,a,a\pm3) = \half \pm(h_{a\pm1} + h_{a\pm2})\\
H(a\pm3,a,a\mp1) & = H(a\mp1,a,a\pm3) = 1 \pm h_{a\pm1}\\
H(a\pm3,a,a\mp3) &= {\mbox{$\textstyle \frac{3}{2}$}}\\
H(a\pm1,a,a\pm1) &= \begin{cases}
1 \pm h_{a\pm1}, & h_{a+2}=h_{a+1}=h_{a-1}+1=h_{a-2}+1\\
\pm h_{a\pm2}, & \text{otherwise}
\end{cases}\\
H(a\pm1,a,a\mp1) &= \half + h_{a+1}-h_{a-1} 
\end{align}
\end{subequations}
We impose an additive gauge $G_a$ which assures the local energy functions are non-negative
\be
H'(a,b,c) = H(a,b,c) + 2 G_{b} - G_a - G_c \geq0
\ee
One such gauge is
\be
G_a = \tfrac{1}{4}\sum_{k=1}^{a-1}(h_{k+1}+h_{k}),
\qquad
G_{a+1} - G_{a} = \tfrac{1}{4}(\flexpr{a+1}+h_a)
\ee
resulting in the local energies
\begin{subequations}
\begin{align}
H(a\pm3,a,a\pm3)&= \pm\half (h_{a\pm3}-h_a)
 \\
H(a\pm3,a,a\pm1)&= H(a\pm1,a,a\pm3) =\half\pm\mbox{$\frac{1}{4}$}(-2 h_a+h_{a\pm1}+2 h_{a\pm2}-h_{a\pm3})\\
H(a\pm3,a,a\mp1)&=H(a\mp1,a,a\pm3)=1 \pm \mbox{$\frac{1}{4}$}(h_{a\mp1}+2 h_{a\pm1}-2 h_{a\pm2}-h_{a\pm3})\\
H(a\pm3,a,a\mp3)&= \mbox{$\frac{3}{2}$} + \mbox{$\frac{1}{4}$} (h_{a-3}+2 h_{a-2}+2 h_{a-1}-2 h_{a+1}-2 \
h_{a+2}-h_{a+3})\\
H(a\pm1,a,a\pm1)&=\begin{cases}
				\pm(h_{a\pm2}-h_{a\pm1}),  &h_{a+1}=h_a=h_{a-1}\\
				\pm\frac{1}{2} (h_{a\pm1}-h_a)+1, & h_{a+1}=h_{a-1}+1
			\end{cases}\\
H(a\pm1,a,a\mp1)&=\half + \mbox{$\frac{3}{4}$}(h_{a+1}-h_{a-1}) 
\end{align}
\end{subequations}
Evaluating these local energy functions on all possible two step paths and band shadings gives the local energies shown in Figure~\ref{fig:n=2energies}.

\section{Conjectured Finitized Bosonic Branching Functions}

\subsection{Finitized bosonic branching functions}

In this section we present conjectured finitized bosonic branching functions for the nonunitary minimal cosets $\mathcal{M}(M,M'\!,n)$, generalizing the unitary case considered by Schilling \cite{Schilling96}. 
For simplicity, we assume throughout that the system size $N$ is even so that $b-a$ is even.
The finitized bosonic branching functions are written in terms of $q$-multinomials~\cite{Schilling96,Warnaar97}. We use the notation $T_\ell^{(n)}(N,\mu)$ as in (2.12) of Schilling~\cite{Schilling96}
\be
T_\ell^{(n)}(N,\mu) = \sideset{}{'}\sum_{\tilde{v}\in\mathbb{Z}^{n+1}_{\geq0}} q^{vC^{-1}v-e_\ell C^{-1}v}(q)_N \prod_{i=0}^{n}\frac{1}{(q)_{v_i}}
\ee
where $\tilde{v}=(v_0,v_1,\ldots,v_n)$, $v=(v_1,v_2,\ldots,v_{n-1})$ are integer vectors and the primed sum indicates a sum over all \mbox{$v_i\in\mathbb{Z}_{\geq0}$}, $i=0,1,2,\ldots,n$ subject to the constraints
\be
v_0=\tfrac{N}{2}-\tfrac{\mu}{n}-e_1C^{-1}v, \qquad v_n = \tfrac{N}{2}+\tfrac{\mu}{n}-e_{n-1}C^{-1}v, \qquad \mu= -\tfrac{nN}{2}, -\tfrac{nN}{2}+1,\ldots,\tfrac{nN}{2}
\ee
The matrix $C^{-1}$ in the quadratic form is the inverse of the Cartan matrix $C$ with entries $C_{a,b}=2-\delta(a,b\!+\!1)-\delta(a,b\!-\!1)$ and the vectors $\{e_j\}_{j=1}^{n-1}$ are $(n\!-\!1)$-dimensional standard basis vectors
\be
C^{-1}_{i,j}=\begin{cases}
(n-i)j/n, &j\leq i\\
(n-j)i/n, &j>i
\end{cases}\qquad\qquad
(e_i)_j = \begin{cases}
\delta(i,j), & i=1,2,\ldots,n-1\\
0, &\text{otherwise}
\end{cases}
\ee
Alternatively, the $q$-multinomials can be defined recursively
\be
T^{(n)}_\ell(N,\mu) = \sum_{k=0}^{\ell-1} q^{(\ell-k)(\frac{N}{2}+\frac{\mu}{n})}\,T^{(n)}_{n-k}(N\!-\!1,\frac{n}{2}\!+\!\mu\!-\!k)
+ \sum_{k=\ell}^n q^{(k-\ell)(\frac{N}{2}-\frac{\mu}{n})}\,T^{(n)}_{n-k}(N\!-\!1,\frac{n}{2}\!+\!\mu\!-\!k)
\ee
subject to
\be
T^{(n)}_\ell(0,\mu)=\delta(\mu,0)
\ee

Explicitly, generalizing Schilling~\cite{Schilling96}, our conjectured finitized bosonic branching functions are
\begin{align}
b_{r,s,\ell}^{M,M'\!,n;(N)}(q) &\cong \sum_{j=-\infty}^{\infty}\left\{ q^{\frac{j}{n}(jMM'+M'r-Ms)}\,T^{(n)}_{(n+b-c)/2}(N,\half(b-a)+j M')\right.\nn
&\qquad\qquad\quad\left.-q^{\frac{1}{n}(jM+r)(jM'+s)}\,T^{(n)}_{(n+b-c)/2}(N,\half(b+a)+jM')\right\}
\label{finbrsl}
\end{align}
where the symbol $\cong$ indicates that the identification holds up to leading powers of $q$. In this formula
\be
M = nm-(n-1)m',\qquad M'=m',\qquad \ell=\half[n+(-1)^{h_b}(b-c)]
\ee
Observe that $\mbox{gcd}\big(\frac{M'-M}{n},M'\big)=\mbox{gcd}(m'\!-\!m,m')=1$ whenever $\mbox{gcd}(m,m')=1$. 
In the unitary case, all the bands are shaded with $m=m'\!-\!1$, $\lambda = \frac{\pi}{n}$, $M=m'\!-\!n$, $h_a\equiv 0$ so this formula reduces to the formula in Schilling~\cite{Schilling96}. If $h_b$ is odd, we interchange $b$ and $c$. 
After this interchange, if it is required, the quantum numbers and boundary conditions are related by
\be
s=a,\qquad \rho=\half (b+c-n),\qquad {\ell}=\half(n+b-c)=b-\rho
\label{quantum}
\ee
with $r$ uniquely determined by $r=r(\rho)=r^{m,m'\!,n}(\rho)$. 
For many minimal models ${\cal M}(M,M',n)$ and exhaustive boundary conditions, we have checked symbolically in Mathematica that the conjecture correctly reproduces the one-dimensional sums, up to the leading powers of $q$, for $n=1,2,3$ out to system size $N=14$. In every case, as guaranteed by the one-dimensional sums, the resulting $q$-polynomials have nonnegative coefficients. 

Finitizations similar to (\ref{finbrsl}), but involving $q$-supernomials and multiple finitization parameters $\vec L=(L_1,L_2,\ldots,L_n)$, have been proposed by Schilling and Warnaar~\cite{SchillingWarnaar}. 
Setting $L_i=N\delta(i,n)$ for $i=1,2,\ldots,n$ and $\ell=0$, the $q$-supernomials reduce to the $q$-multinomials $T_0^{(n)}(N,\mu)$. Relaxing the Takahashi length restrictions on $a,b$ in these cases, it follows that the finitizations of \cite{SchillingWarnaar} coincide with (\ref{finbrsl}). In these and other cases, these authors have identified the associated fermionic forms and proven bosonic equals fermionic type identities. However, a simple relationship between the $q$-supernomial and $q$-multinomial finitizations in the sectors with $\ell\ne 0,n$ is not known.

Setting $q=1$ in (\ref{finbrsl}) gives the correct counting of states. To take the limit $N\to\infty$ to obtain the full branching functions, we use (2.16) of Schilling~\cite{Schilling96}
\be
T_\ell^{(n)}(\mu)=\lim_{N\to\infty,\text{$N$ even}}T_\ell^{(n)}(N,\mu) = \frac{1}{(q)_\infty}\sum_{v\in\mathbb{Z}^{n-1}_{\geq0},\frac{\mu}{n}+e_1C^{-1}v\in\mathbb{Z}} q^{vC^{-1}v-e_\ell C^{-1}v} \prod_{i=1}^{n-1}\frac{1}{(q)_{v_i}}
\ee
which only depends on $\mu$ mod $n$ with $T_\ell^{(n)}(\mu)=T_\ell^{(n)}(\ell-\mu)$. We therefore find
\begin{align}
b_{r,s,\ell}^{M,M'\!,n}(q) &\cong\sum_{0\le m\le n/2\atop m={\ell}/2\text{\,mod\,1}}T^{(n)}_{{\ell}}(m+\half{\ell}) \nn
&\qquad\times\ \bigg[ \sum_{j\in\mathbb{Z}\atop m_{r-s}(j)=\pm m\,\text{mod\,$n$}}q^{\frac{j}{n}(jMM'+M'r-Ms)}-\sum_{j\in\mathbb{Z}\atop m_{r+s}(j)=\pm m\,\text{mod\,$n$}}q^{\frac{1}{n}(jM+r)(jM'+s)}\bigg]
\label{brsl}
\end{align}
where
\bea
m_a(j):=a/2+jM'\label{maj}
\eea
and we use (\ref{quantum}) and
\bea
\half(b\mp a)+jM'=\half(\rho\mp s)+jM'+\half{\ell}=m+\half{\ell} \text{\ \ mod $n$},\qquad \rho=r \text{\ \ mod $n$}
\label{bminusa}
\eea
We prove the result $\rho=r$ mod $n$ in Appendix~\ref{sec:shadedNbands}.  
Note that with $N$ even, from (\ref{brsl}) to (\ref{bminusa}), $r-s+\ell$ must be even. Also, using (\ref{quantum}), $b-a=\rho-s+\ell$ is also even. 
Note also that, if $n$ is odd, then $b$ and $c$ have opposite parities and it is only possible to get from $a$ to one of $b$ or $c$ in an even number of steps $N$. So $b$ is uniquely determined by the condition 
$b-a=0$ mod 2 and it is the $b$ in (\ref{quantum}). If $n$ is even, it is possible to get to either $b$ or $c$ in $N$ steps. In this case, interchanging $b$ and $c$ is equivalent to $\ell\leftrightarrow n-\ell$.

Lastly, to obtain the branching functions in the form (\ref{branchfns}), we use (3.30) of Schilling~\cite{Schilling96} which asserts that $T^{(n)}_\ell(m+\half\ell)\cong c_{2m}^\ell(q)$. Explicitly, up to leading powers of $q$, $T^{(n)}_\ell(m+\half\ell)$ is the $\mathbb{Z}_n$ parafermionic string function
\bea
c_{2m}^\ell(q)=q^{\frac{\hat{c}}{24}-\frac{1}{4n}+\frac{n}{4MM'}+\frac{\ell+1}{2n}-\frac{(\ell+1)^2}{2n(n+2)}-\frac{1}{8}}\,T^{(n)}_\ell(m+\half\ell),\qquad \ell=0,1,\ldots,n
\eea
with
\bea
\hat{c}=1-\frac{6n}{MM'}+\frac{2(n-1)}{n+2}
\eea
Note that, in this formula, $m$ can take half-integer values and $2m$ is the parafermionic index.

\subsection{Logarithmic limit and finitized Kac characters}
\label{SectLogLimit}

Following \cite{RasLogLimit} and \cite{PR2013}, the Kac characters of the logarithmic minimal models ${\cal LM}(p,p')$~\cite{PRZ} and their $n\times n$ fusion hierarchies~\cite{PRT2014} are given by taking the {\em logarithmic limit}. Symbolically 
\bea
\lim_{m,m'\to\infty, \  {m'\over m}\to {p'\over p}+} {\cal M}(M,M'\!,n)={\cal LM}(P,P'\!,n),\qquad 1\le p<p',\quad \mbox{$p,p'$ coprime}\label{logLimit}
\eea
where
\bea
(M,M')=\big(nm-(n\!-\!1)m',m'\big),\qquad (P,P')=\big(np-(n\!-\!1)p',p'\big)
\eea
The (one-sided) limit is taken through coprime pairs $(m,m')$ with ${m'\over m}>{p'\over p}$. The one-sided limit is needed to ensure the sequences of minimal model ground states converge to the correct logarithmic minimal model ground states. Formally, the logarithmic limit is taken in the continuum scaling limit after the thermodynamic limit. 
The equality indicates the identification of the spectra of the chiral CFTs. In principle, the Jordan cells appearing in the reducible yet indecomposable representations of the logarithmic minimal models should emerge in this limit but there are subtleties~\cite{RasLogLimit}.

Since finitized characters give the spectrum generating functions for finite truncated sets of conformal energies, the logarithmic limit can be applied directly to finitized characters. Assuming $0<|q|<1$ and taking the logarithmic limit of the finitized branching functions ({\ref{finbrsl}), we find that up to leading powers of $q$, the finitized Kac characters are
\begin{align}
\chi_{r,s,\ell}^{P,P'\!,n;(N)}(q) &\cong T^{(n)}_{(n+b-c)/2}\big(N,\half(b-a)\big)-q^{\frac{1}{n}rs}\,T^{(n)}_{(n+b-c)/2}\big(N,\half(b+a)\big)
\label{finlogchi}
\end{align}
where the quantum numbers are related to the boundary conditions by (\ref{elldef}) and (\ref{elltilde}) with $h_a=\sfloor{\frac{a(p'-p)}{p'}}$.
Taking the thermodynamic limit gives
\begin{align}
\chi_{r,s,\ell}^{P,P'\!,n}(q)=\lim_{N\to\infty,\text{$N$ even}}\chi_{r,s,\ell}^{n;(N)}(q) &\cong T^{(n)}_{(n+b-c)/2}\big(\half(b-a)\big)-q^{\frac{1}{n}rs}\,T^{(n)}_{(n+b-c)/2}\big(\half(b+a)\big)
\label{blogchiT}
\end{align}
or equivalently
\begin{align}
\chi_{r,s,\ell}^{P,P'\!,n}(q) &\cong c_{r-s}^\ell-q^{\frac{1}{n}rs}\,c_{r+s}^\ell,\qquad r,s\in\mathbb{N},\quad \ell=0,1,\ldots,n
\label{blogchi}
\end{align}
Since the string functions vanish for $r+s+\ell$ odd, we must have $r+s=\ell$ mod 2. The dependence on $P,P'$ only enters through the leading powers of $q$ as specified in \cite{PR2013}. 
For the logarithmic superconformal minimal models with $n=2$ and $r=1$, the finitized Kac characters agree with those of \cite{PRT2014}. In this case, the counting of states given by trinomials reduces to generalized Motzkin and Riordan numbers in accord with the counting of the fused Temperley-Lieb link states.

\section{Conclusion}

In this paper we conjecture the identification (\ref{conjecture}), in the continuum scaling limit, of the $n\times n$ fused $\mbox{RSOS}(m,m')$ lattice models with the higher-level minimal model cosets ${\cal M}(M,M'\!,n)$ 
at fractional level $k=nM/(M'-M)-2$ with $(M,M')=\big(nm-(n-1)m',m'\big)$. 
This implies that the central charges of the $n\times n$ fused $\mbox{RSOS}(m,m')$ models are
\be
{\cal M}(M,M'\!,n):\qquad c=c^{m,m'\!,n}=\frac{3 n }{n+2}\Big[1-\frac{2 (n+2) (m'-m)^2}{m' \big(m n-m'(n-1)\big)}\Big]\label{ccosets}
\ee
The conjecture agrees with known results in the unitary cases ($m=m'\!-\!1$). It is also supported in nonunitary cases ($2\le m\le m'\!-\!2$) by our explicit calculation of Baxter's one-dimensional sums for $n=1,2,3$. 
Specifically, up to leading powers of $q$, we find that the one-dimensional sums give finitized branching functions. Indeed in many cases, using Mathematica out to system size $N=14$, the resulting 
$q$-series are confirmed to converge towards the full branching functions. 

Separately, generalizing the work of Schilling~\cite{Schilling96}, the bosonic forms of the finitized branching functions (\ref{finbrsl}) are conjectured for all nonunitary cases with $n\ge 1$. 
These finitized bosonic forms give the correct counting of states and reproduce the full branching functions in the thermodynamic limit $N\to\infty$. 
The explicit form of these finitized bosonic branching functions allows us to take the logarithmic limit. In this way, conjectured bosonic forms of the finitized Kac characters (\ref{finlogchi}) are obtained for the higher-level fused 
logarithmic minimal models ${\cal LM}(P,P'\!,n)$ thus extending the recent conjectures~\cite{PRT2014} restricted to the $n=2$ logarithmic superconformal minimal models ${\cal LM}(P,P'\!,2)$. 

All of the cosets (\ref{ccosets}) are realized with $\tfrac{n-1}{n}\,m'<m<m'$ corresponding to the interval $0<\lambda<\pi/n$. Outside of this interval, there are no shaded $n$-bands to support the level-$n$ ground states. 
It would therefore be of interest to extend the considerations of this paper to the full interval $0<\lambda<\pi$. 
The level of rigour could also be improved by calculating, in the low-temperature limit, the local energies $H(a,b,c)$ valid for all $n$ in a common gauge. It should then be possible to extend the proof of Schilling~\cite{Schilling96} 
to rigourously establish the equality of the one-dimensional sums with the finitized bosonic branching functions. We hope to return to these issues in a later paper.

\section*{Acknowledgments} This paper is dedicated to Rodney Baxter on the occasion of his 75th birthday. Elena Tartaglia is supported by an Australian Postgraduate Award. We thank Ole Warnaar for helpful comments and 
encouragement. 

\appendix

\section{Elliptic Functions}
\label{AppA}

We summarize the definitions and properties of the elliptic functions used throughout this paper.
The standard elliptic theta function $\vartheta_1(u,t)$~\cite{GR} is
\begin{align}
\vartheta_1(u,t)&=2t^{1/4}\sin u
\prod_{n=1}^\infty (1-2t^{2n}\cos 2u+t^{4n})(1-t^{2n})
\end{align}
Its conjugate modulus transformation is
\begin{align}
\vartheta_1(u,e^{-\varepsilon})&=\sqrt\frac{\pi}{\varepsilon}\, e^{-(u-\pi/2)^2/\varepsilon}E(e^{-2 \pi u/\varepsilon}, e^{-2\pi^2/\varepsilon} )
\end{align}
where
\bea
E(w,p)=\sum_{k=-\infty}^{\infty} (-1)^kp^{k(k-1)/2}w^k=\prod_{n=1}^\infty(1-p^{n-1}w)(1-p^nw^{-1})(1-p^n)
\eea
The elliptic $\vartheta_1(u)=\vartheta_1(u,t)$ function satisfies the fundamental identity
\begin{align}
&\vartheta_1(u+x)\vartheta_1(u-x)\vartheta_1(v+y)\vartheta_1(v-y) - \vartheta_1(u+y)\vartheta_1(u-y)\vartheta_1(v+x)\vartheta_1(v-x) \nn
&\qquad\qquad\qquad= \vartheta_1(x-y)\vartheta_1(x+y)\vartheta_1(u+v)\vartheta_1(u-v)
\end{align}

\section{Counting of Contiguous Shaded Bands}
\label{sec:shadedNbands}

\subsection{Counting of shaded $n$-bands}

Fix $m$ and $m'$ with $2\le m<m'$, $\mbox{gcd}(m,m')=1$ and consider a walk on the $A_{m'-1}$ Dynkin diagram. 
The bands $(a,a\!+\!1)$ at heights 
$a=\lfloor{{r m'\over m}}\rfloor$ are shaded while the other bands are unshaded. 
An \emph{$n$-band} consists of $n$ contiguous bands, where each band is shaded or unshaded. If all the bands in an $n$-band are shaded, we call it a \emph{shaded $n$-band}. If all the bands in an $n$-band are unshaded, we call it an \emph{unshaded $n$-band}. Otherwise, we call it a \emph{mixed $n$-band}.

Let us assume that $\lambda=(m'-m)\pi/m'<\pi/n$, that is,
\be
nm-(n-1)m'> 0
\label{eq:fusionAssumption}
\ee
and define
\begin{align}
s_n &:= \# \text{ shaded } n\text{-bands}\\
t_n &:= \#\ n\text{-bands with exactly one unshaded 1-band}
\end{align}
In this section, 
we prove that the number $s_n$ of shaded $n$-bands is
\bea
s_n =M-1=n m - (n-1)m' -1
\label{eq:fusedCounting}
\eea
More specifically, we show that
\begin{subequations}
\begin{align}
&\# \text{ 1-bands} = m'-2 = s_n + t_n + n-1 \label{eq:nBandCounting2}\\
&t_n=n\times \# \text{ unshaded 1-bands} = n(m'-m-1) \label{eq:nBandCounting1}
\end{align}
\label{eq:nBandCounting}
\end{subequations}
Solving gives the required result \eqref{eq:fusedCounting}.

We will need the elementary properties of floor functions
\be
\floor{n+x} = n + \floor{x},\qquad
\floor{-x} = -\lceil x \rceil = -1-\floor{x},\qquad n\in \mathbb{Z}, \ x,y\in\mathbb{R}
\label{eq:floorProps}
\ee
Using these, we obtain the two implications
\be
\floor{x} - \floor{y} = n\ \Rightarrow\ n-1 < x-y < n+1, \qquad \floor{x}-\floor{y} \leq n\ \Rightarrow\ x-y<n+1
\label{eq:floorProps2}
\ee
To see the first result, let $x = \floor{x} + r_x, \ y=\floor{y}+r_y$ where $0\le r_x,r_y<1$. Since $\floor{x}-\floor{y}=n$, we find
\be
x - y = n+r_x-r_y, \qquad -1 <r_x-r_y<1
\ee
which gives the required bounds on $x-y$. The second result follows easily from the first. If $\floor{x}-\floor{y} =n$, then this follows from the first result. If not, then there must exist an integer $k\le n-1$ such that $\floor{x}-\floor{y} = k$. We can then apply the first result to obtain $x - y < k+1 \leq n$.

Next we use these properties of floor functions to prove two preliminary results. The first preliminary result is that there are only two possible types of $n$-bands: shaded $n$-bands and $n$-bands with exactly one unshaded 1-band. Setting 
\be
h_{a+1}-h_a=\begin{cases}
0,&\mbox{$(a,a+1)$ shaded}\\
1,&\mbox{$(a,a+1)$ unshaded}
\end{cases}\qquad h_a=\Big\lfloor\frac{a(m'-m)}{m'}\Big\rfloor
\ee
this is equivalent to showing that for all $n$-bands $(a,a+n)$
\be
0\le h_{a+n}-h_a=0,1
\ee
Assuming the converse, that is $h_{a+n}-h_a\geq2$ for some $n$-band starting at height $a$, we use properties \eqref{eq:floorProps} to show
\begin{align}
2\leq h_{a+n}-h_a &= \Bfloor{(a+n)\frac{(m'-m)}{m'}} - \Bfloor{a\frac{(m'-m)}{m'}} =
a+n-1-\floor{(a+n)\frac{m}{m'}} -a +1 + \floor{a\frac{m}{m'}} \nn
&=n-\floor{(a+n)\frac{m}{m'}} +\floor{a\frac{m}{m'}}  
\end{align}
Rearranging and applying the second property of \eqref{eq:floorProps2} gives
\begin{align}
\floor{(a+n)\frac{m}{m'}} -\floor{a\frac{m}{m'}} \leq n-2\ \Rightarrow\ (a+n)\frac{m}{m'} -a\frac{m}{m'} < n-1
\ \Rightarrow\ nm - (n-1)m' < 0
\end{align}
which directly contradicts assumption \eqref{eq:fusionAssumption}.

The second preliminary result needed is that, for $n\ge 2$, there are always $n\!-\!1$ shaded 1-bands at the top and bottom of the $A_{m'-1}$ diagram. 
To show that there are always $n\!-\!1$ shaded 1-bands at the bottom, we use (\ref{eq:floorProps}) and consider
\be
0\le h_n-h_1=h_n=\Bfloor{n\frac{(m'-m)}{m'}}=n-1-\floor{\frac{nm}{m'}}
\ee
But now, from (\ref{eq:fusionAssumption})
\be
\frac{nm}{m'}>n-1\ \Rightarrow\ \floor{\frac{nm}{m'}}\ge n-1\ \Rightarrow\ h_n-h_1\le 0
\ee
We conclude that $h_n-h_1=0$. 
A similar calculation shows that $h_{m'-1}=h_{m'-n}$ and proves that there are always $n\!-\!1$ shaded 1-bands at the top of the diagram.

Finally, we derive \eqref{eq:nBandCounting}. We show that \eqref{eq:nBandCounting1} is true in two steps. Firstly, 
\be
t_n := \#\ n\text{-bands with exactly one unshaded 1-band} = n\times \# \text{ unshaded 1-bands}
\ee
since, by scanning up the $A_{m'-1}$ diagram from bottom to top and looking at each consecutive $n$-band, each time there is an unshaded 1-band, it gets counted exactly $n$ times. This is because there is only at most one shaded 1-band per $n$-band and there are no unshaded 1-bands in the top and bottom $(n\!-\!1)$-bands. Secondly, the total number of 1-bands is $m'-2$ and there are $m-1$ shaded 1-bands, so there must be $m'-m-1$ unshaded 1-bands. Combining these two expressions gives \eqref{eq:nBandCounting1}.
To obtain \eqref{eq:nBandCounting2}, we use the fact that only two types of $n$-bands occur: shaded $n$-bands and $n$-bands with exactly one unshaded 1-band. When counting all the 1-bands by looking at $n$-bands, we count every 1-band $n$ times, except for the $n\!-\!1$ 1-bands at the top and bottom. Starting at the bottom, the first 1-band gets counted once, the second 1-band twice and so on for the first $n\!-\!1$ 1-bands.
Similarly, if we start at the top and work our way down. This means that overall we are undercounting in the first 1-band, by $n\!-\!1$ 1-bands, and in the second 1-band by $n\!-\!2$ 1-bands and so on through to undercounting by 1 in the last 1-band. This happens at both the top and bottom. Hence, to count all 1-bands $n$ times, we need the boundary (last) term in the following expression
\be
n\times \# \text{1-bands} = n s_n + n t_{n} + 2 \sum_{i=1}^{n-1} i = n s_n + n t_{n} + n(n-1)
\ee
Dividing both sides by $n$ and using the fact that there are $m'\!-\!2$ 1-bands in total, gives \eqref{eq:nBandCounting2}.

\subsection{Proof that $\rho=r$ mod $n$}

For the counting of shaded $n$-bands with $\lambda<\pi/n$, we prove the result $\rho=r$ mod $n$. For $n=1$ the result is trivial, so we can assume that $n\ge 2$. 
Each $n$-band must be a shaded $n$-band or it contains precisely one unshaded 1-band. 
We proceed iteratively in steps:\\[4pt]
1. Consider the lowest $n$-band and set $r=\rho=1$ where $r$ counts the shaded $n$-bands from the bottom and $\rho$ labels the current position which is a candidate for the position of a shaded $n$-band. 
The lowest $n$-band must be a shaded $n$-band (labelled by $r=1$) or only contain one unshaded 1-band at the top.\\[4pt]
2. (i) If both the current bottom $n$-band and the current bottom $(n\!+\!1)$-band are shaded, the current $n$-band is a ground state $n$-band labelled by the current value of $r$ at a height given by the current value of $\rho$. 
Increment $r\mapsto r+1$, $\rho\mapsto \rho+1$ and remove the bottom 1-band. Since the top $n\!-\!1$ 1-bands in the removed $n$-band were shaded, the next $n$-band up must be a shaded $n$-band or only contain one unshaded 1-band at the top.\\[2pt]
(ii) If the current bottom $n$-band is shaded and the current bottom $(n\!+\!1)$-band is not shaded, the current $n$-band is a ground state $n$-band labelled by the current value of $r$ at a height given by the current value of $\rho$. 
Increment $r\mapsto r+1$, $\rho\mapsto \rho+n+1$ and remove the bottom $(n\!+\!1)$-band. There are no further shaded $n$-bands involving the 1-bands that are removed. Since the top 1-band removed is unshaded, the next $n$-band up must be a shaded $n$-band or only contain one unshaded 1-band at the top.\\[2pt]
(iii) If the current bottom $n$-band contains an unshaded 1-band, it must occur at the top of this $n$-band. This is not a ground state $n$-band.
Increment $r\mapsto r$, $\rho\mapsto \rho+n$ and remove the bottom $n$-band. There are no further shaded $n$-bands involving the 1-bands that are removed. 
The next $n$-band up must be a shaded $n$-band or only contain one unshaded 1-band at the top.\\[4pt]
3. Iterate step 2 until the top of the $A_{m'-1}$ diagram is reached and all 1-bands have been removed.\\[4pt]
At each step we see that $\rho=r$ mod $n$.

\section{$n=2,3$ Face Weights}
\label{AppFusedWts}

Expressions for the $n\times n$ fused RSOS face weights have be obtained in \cite{DJKMO}. In this appendix, we list explicitly the 19 and 44 face weights for $n=2$ and $n=3$. 
The number of face weights for general $n=1,2,3,\ldots$ is given by the octahedral numbers~\cite{Sloane}
\be
\tfrac{1}{3}\big((n+1)^3+n+1\big)=6,19,44,85,146,\ldots
\ee
Adjacent heights $a,b=1,\ldots,m'\!-\!1$ satisfy
	\be
	|a-b| = \begin{cases}
	0,2,4,\ldots, n, & n \text{ even}\\
	1, 3,5,\ldots, n, & n \text{ odd}
	\end{cases}
	\qquad\qquad
	n+2\le a+b\le 2m'-n-2
	\ee


\allowdisplaybreaks
\subsection{Explicit $2\times2$ fused face weights}
The normalized $2\times2$ fused RSOS face weights are
\be
\Wtwo{a}{b}{c}{d}{u}=\frac{1}{\eta^{2,2}(u)}\;
\psset{unit=1cm}
\begin{pspicture}[shift=-1](-0.2,-0.2)(2.2,2.2)
\rput(0,0){\facegrid{(0,0)}{(2,2)}}
\rput(0.5,0.5){\small $u\!-\!\lambda$}
\rput(1.5,0.5){\small $u$}
\rput(1.5,1.5){\small $u\!+\!\lambda$}
\rput(0.5,1.5){\small $u$}
\rput(0,1){\mydot}
\rput(1,1){\mydot}
\rput(1,0){\mydot}
\rput(-0.15,-0.15){$a$}
\rput(2.15,-0.15){$b$}
\rput(2.15,2.15){$c$}
\rput(-0.15,2.15){$d$}
\rput(2,1){$\times$}
\rput(1,2){$\times$}
\end{pspicture}
\ee
The black dots indicate sums over all allowed heights at the site. The crosses indicate that the weight is independent of the allowed heights on these sites. The fused weights all have a common factor which is removed
\be
\eta^{2,2}(u) = s(2\lambda) s(u) s(u-\lambda)
\ee

The explicit formulas for all 19 types of weights are 
\begin{subequations}
\label{eq:BoltzWeights}
\begin{align}
\Wtwo{a}{a\mp2}{a}{a\pm2}{u} &= \frac{s(u-2 \lambda ) s(u-\lambda )}{s(2 \lambda )} \label{eq:bw01}\\[6pt]
\Wtwo{a}{a\pm2}{a}{a}{u} &=
\Wtwo{a}{a}{a}{a\pm2}{u} = -\frac{s(u-\lambda ) s((a\pm1) \lambda \mp u)}{s((a\pm1) \lambda )} \label{eq:bw03} \\[6pt]
\Wtwo{a\pm2}{a}{a}{a}{u} &= -\frac{s((a\mp1) \lambda )s(u)  s(a \lambda \pm u)}{s(2 \lambda ) s(a \lambda ) s((a\pm1)\lambda )} \label{eq:bw04} \\[6pt]
\Wtwo{a}{a}{a\pm2}{a}{u} &= -\frac{s(2 \lambda ) s((a\pm2) \lambda )s(u) s(a \lambda \pm u)}{s((a-1) \lambda ) s((a+1) \lambda )} \label{eq:bw05} \\[6pt]
\Wtwo{a\pm2}{a}{a\mp2}{a}{u} &= \frac{s((a\mp2) \lambda ) s((a\mp1) \lambda ) s(u)s(\lambda +u)}{s(2 \lambda )   s(a \lambda ) s((a\pm1) \lambda )}\label{eq:bw06}\\[6pt]
\Wtwo{a\pm2}{a}{a\pm2}{a}{u} &= \frac{s(a \lambda \pm u) s((a\pm1) \lambda \pm u)}{s(a \lambda ) s((a \pm 1) \lambda )}\label{eq:bw07}\\[6pt]
\Wtwo{a}{a}{a}{a}{u} &=\frac{s(a\lambda\pm u)s((a\pm 1)\lambda\mp u)}{s(a\lambda)s((a\pm1)\lambda)}+\frac{s((a\pm 1)\lambda)s((a\mp2\lambda))s(u)s(u-\lambda)}{s(2\lambda)s(a\lambda)s((a\mp1)\lambda)}
\label{eq:bw08}\\[6pt]
\Wtwo{a}{a\pm2}{a\pm2}{a}{u} &=\Wtwo{a}{a}{a\pm2}{a\pm2}{u}= \frac{s((a\pm3) \lambda ) s(u) s(u-\lambda )}{s(2 \lambda ) s((a\pm1) \lambda )} \label{eq:bw09}
\end{align}
\end{subequations}

\subsection{Explicit $3\times 3$ fused face weights}
The normalized $3\times3$ fused RSOS face weights are 
\be
\Wthree{a}{b}{c}{d}{u}=\frac{1}{\eta^{3,3}(u)}\;
\psset{unit=0.9cm}
\begin{pspicture}[shift=-1.5](-0.2,-0.2)(3.2,3.2)
\rput(0,0){\facegrid{(0,0)}{(3,3)}}
\rput(0.5,0.5){\small $u\!\!-\!\!2\lambda$}
\rput(1.5,0.5){\small $u\!-\!\lambda$}
\rput(2.5,0.5){\small $u$}
\rput(0.5,1.5){\small $u\!-\!\lambda$}
\rput(1.5,1.5){\small $u$}
\rput(2.5,1.5){\small $u\!+\!\lambda$}
\rput(0.5,2.5){\small $u$}
\rput(1.5,2.5){\small $u\!+\!\lambda$}
\rput(2.5,2.5){\small $u\!\!+\!\!2\lambda$}
\multirput(0,0)(0,1){3}{\multirput(1,0)(1,0){2}{\mydot}}
\rput(0,1){\mydot}\rput(0,2){\mydot}
\rput(-0.15,-0.15){$a$}
\rput(3.15,-0.15){$b$}
\rput(3.15,3.15){$c$}
\rput(-0.15,3.15){$d$}
\rput(3,1){$\times$}
\rput(3,2){$\times$}
\rput(1,3){$\times$}
\rput(32,3){$\times$}
\end{pspicture}
\ee
with normalization
\be
\eta^{3,3}(u) = s(2\lambda)s(3\lambda)s(u-2 \lambda)s^2(u-\lambda)s^2(u)s(u+\lambda)
\ee

\begin{subequations}
\begin{align}
\Wthree{a}{a\pm3}{a}{a\pm3}{u}&=\frac{s((a\pm1)\lambda\mp u) s((a\pm2) \lambda \mp u) s((a\pm3) \lambda \mp u)}{s((a\pm1) \lambda )s((a\pm2) \lambda ) s((a\pm3) \lambda )}\\[6pt]
\Wthree{a}{a\pm3}{a}{a\mp3}{u}&=\frac{s(\lambda -u) s(2 \lambda -u) s(3 \lambda-u)}{s(2 \lambda ) s(3 \lambda )}\\[6pt]
\Wthree{a}{a\pm3}{a}{a\pm1}{u}&=\Wthree{a}{a\pm1}{a}{a\pm3}{u}=\frac{s(\lambda -u)   s((a\pm1) \lambda\mp u) s((a\pm2) \lambda \mp u)}{s((a\pm 1) \lambda ) s((a\pm2) \lambda)}\\[6pt]
\Wthree{a}{a\pm3}{a}{a\mp1}{u}&=\Wthree{a}{a\mp1}{a}{a\pm3}{u}=\frac{s(\lambda -u) s(2 \lambda -u) s((a\pm1) \lambda\mp u)}{s(2 \lambda ) s((a\pm1) \lambda )}\\[6pt]
\Wthree{a}{a\pm1}{a}{a\pm1}{u}&=\frac{s((a\pm1) \lambda\mp u) s((a\pm 1) \lambda\pm u) s((a\pm2) \lambda \mp u)}{s((a\pm1) \lambda )^2 s((a\pm2) \lambda )}\nn
&-\frac{s(2 \lambda) s((a-2) \lambda ) s((a+2) \lambda )s(\lambda - u)  s(u) s((a\pm1) \lambda\mp u)}{s(3 \lambda ) s((a\mp1) \lambda ) s((a\pm1) \lambda   )^2}\\[6pt]
      \Wthree{a}{a\pm1}{a}{a\mp1}{u}&=\frac{s(2 \lambda )^2 s((a\mp2) \lambda ) s(\lambda - u) s(a \lambda \pm u) s((a\pm1) \lambda \mp u)}{s(3 \lambda ) s((a\mp1) \lambda )^2   s((a\pm1) \lambda )}\nn
      &-\frac{s((a\mp3) \lambda ) s((a\pm1) \lambda ) s(2 \lambda -u) s(\lambda - u) s(\lambda +u) }{s(2 \lambda ) s(3 \lambda   ) s((a\mp1) \lambda )^2}\\
\Wthree{a}{a\pm3}{a\pm6}{a\pm3}{u}&=-\frac{s((a\pm4) \lambda ) s((a\pm5) \lambda ) s((a\pm6) \lambda )s(u)  s(\lambda +u) s(2 \lambda +u)}{s(2 \lambda ) s(3 \lambda ) s((a\pm1) \lambda ) s((a\pm2) \lambda ) s((a\pm3) \lambda )}\\
\Wthree{a}{a\pm3}{a\pm4}{a\pm3}{u}&=\frac{s((a\pm4) \lambda ) s((a\pm5) \lambda ) s(u) s(\lambda +u) s((a\pm3) \lambda \mp u)}{s(2 \lambda ) s(3 \lambda ) s((a\pm1) \lambda ) s((a\pm2) \lambda ) s((a\pm3) \lambda )}\\
\Wthree{a}{a\pm3}{a\pm4}{a\pm1}{u}&=\Wthree{a}{a\pm1}{a\pm4}{a\pm3}{u}=-\frac{s((a\pm4) \lambda ) s((a\pm5) \lambda ) s(u) s(u-\lambda ) s(\lambda +u)}{s(2 \lambda ) s(3 \lambda ) s((a\pm1) \lambda ) s((a\pm2) \lambda )}\\
\Wthree{a}{a\pm3}{a\pm2}{a\pm3}{u}&=-\frac{s((a\pm4) \lambda ) s(u) s((a\pm2) \lambda \mp u) s((a\pm3) \lambda \mp u)}{s(3 \lambda ) s((a\pm 1) \lambda ) s((a\pm2) \lambda ) s((a\pm3) \lambda )}\\
\Wthree{a}{a\pm3}{a\pm2}{a\pm1}{u}&=\Wthree{a}{a\pm1}{a\pm2}{a\pm3}{u}=\frac{s((a\pm4) \lambda ) s(u) s(u-\lambda ) s((a\pm2) \lambda \mp u)}{s(3 \lambda ) s((a\pm1) \lambda ) s((a\pm2) \lambda )}\\
\Wthree{a}{a\pm3}{a\pm2}{a\mp1}{u}&=\Wthree{a}{a\mp1}{a\pm2}{a\pm3}{u}=-\frac{s((a\pm4) \lambda ) s(2 \lambda-u ) s(\lambda-u )s(u)}{s(2 \lambda ) s(3 \lambda ) s((a\pm1) \lambda )}\\
\Wthree{a}{a\pm1}{a\pm4}{a\pm1}{u}&=\frac{s(3\lambda)s((a\pm3)\lambda)(s(a\pm4)\lambda)s(u)s(u+\lambda)s((a\pm 1)\lambda\pm u)}{s(2\lambda)s((a-1)\lambda)s((a+1)\lambda)s((a\pm2)\lambda)}\\
\Wthree{a}{a\pm1}{a\pm2}{a\pm1}{u}&=-\frac{s(a \lambda ) s((a\pm3) \lambda ) s((a\pm4) \lambda ) s(u)^2 s(u-\lambda )}{s(2 \lambda ) s(3 \lambda ) s((a\pm1) \lambda )^2 s((a\pm2) \lambda )}\nn
&-\frac{s((a\pm3) \lambda )s(u)  s(a \lambda \pm u) s((a\pm1) \lambda\mp u)}{s((a\mp1) \lambda ) s((a\pm1) \lambda )^2}\\
\Wthree{a}{a\pm1}{a\pm2}{a\mp1}{u}&=\Wthree{a}{a\mp1}{a\pm2}{a\pm1}{u}=\frac{s((a\pm3)\lambda)s(u)s(u-\lambda)s(a\lambda\pm u)}{s((a-1)\lambda)s((a+1)\lambda)}\\
\Wthree{a}{a\pm1}{a\mp2}{a\pm1}{u}&=-\frac{s(3\lambda)s((a\mp 2)\lambda)s(u)s(a\lambda\mp u)s((a\pm 1)\lambda\mp u)}{s((a-1)\lambda)s((a+1)\lambda)s((a\pm 2)\lambda)}
\end{align}
\end{subequations}

\section{$n=2,3$ Diagonal Conjugate Modulus Face Weights}

In this appendix, we list the diagonal conjugate modulus face weights for $n=2$ and $n=3$. 

\subsection{Explicit $2\times2$ conjugate modulus face weights}

The explicit $2\times 2$ diagonal conjugate modulus face weights are
\begin{subequations}
\begin{align}
\Wt{a}{a\mp2}{a}{a\pm2} &= \frac{g_a^2}{g_{a-2}g_{a+2}}\ w\ \frac{E(x^2w^{-1})E(xw^{-1})}{E(x^2)E(x)}\\[4pt]
\Wt{a}{a\pm2}{a}{a} &= \Wt{a}{a}{a}{a\pm2} = \frac{g_a}{g_{a\pm2}}\ \frac{E(x w^{-1})E(x^{a\pm1}w^{\mp1})}{E(x)E(x^{a\pm1})}\\[4pt]
\Wt{a\pm2}{a}{a\pm2}{a} &= \frac{g_{a\pm2}^2}{g_a^2} \ \frac{E(x^aw^{\pm1})E(x^{a\pm1}w^{\pm1})}{E(x^a)E(x^{a\pm1})}\\[4pt]
\Wt{a}{a}{a}{a} &= w x\, \frac{E(x^{a-1})E(x^{a+2})E(w^{-1})E(xw^{-1})}{E(x)E(x^2)E(x^a)E(x^{a+1})} + \frac{E(x^{a-1}w)E(x^aw^{-1})}{E(x^{a-1})E(x^a)}
\end{align}\label{conjmod2}
\end{subequations}

\subsection{Explicit $3\times3$ conjugate modulus face weights}

The explicit $3\times 3$ diagonal conjugate modulus face weights are
\begin{subequations}
\begin{align}
\Wthree{a}{a\pm3}{a}{a\pm3}{u}&= \frac{g_a^2}{g_{a\pm3}^2}
\frac{E(x^{a\pm1}w^{\mp1}) E(x^{a\pm2}w^{\mp1})   E(x^{a\pm3}w^{\mp1})}{E(x^{a\pm1}) E(x^{a\pm2})   E(x^{a\pm3})}\\
\Wthree{a}{a\pm1}{a}{a\pm3}{u}&=\Wthree{a}{a\pm3}{a}{a\pm1}{u}= \frac{g_a^2}{g_{a\pm1}g_{a\pm3}}
\frac{w^{1/2} E(xw^{-1}) E(x^{a\pm1}w^{\mp1}) E(x^{a+2}w^{-1})}{E(x) E(x^{a\pm1}) E(x^{a\pm2})}\\
\Wthree{a}{a\mp1}{a}{a\pm3}{u}&=\Wthree{a}{a\pm3}{a}{a\mp1}{u}= \frac{g_a^2}{g_{a\mp1}g_{a\pm3}}
\frac{w E(xw^{-1}) E(x^2w^{-1}) E(x^{a\pm1}w^{\mp1})}{E(x) E(x^2)   E(x^{a\pm1})}\\
\Wthree{a}{a\mp3}{a}{a\pm3}{u}&=\frac{g_a^2}{g_{a+3}g_{a-3}}
\frac{w^{3/2} E(xw^{-1}) E(x^2w^{-1}) E(x^3w^{-1})}{E(x) E(x^2) E(x^3)}\\
   \Wthree{a}{a\pm1}{a}{a\pm1}{u}&=\frac{g_a^2}{g_{a\pm1}^2} \left(
   \frac{E(x^{a\pm1}w^{\mp1}) E(x^{a\pm1}w^{\pm1} )   E(x^{a\pm2}w^{\mp1})}{E(x^{a\pm1})^2 E(x^{a\pm2})}\right.\nn
   &\left.-\frac{x E(x^2) E(x^{a+2}) E(x^{a-2}) E(w)  E(xw^{-1})   E(x^{a\pm1}w^{\mp1})}{E(x)^2 E(x^3) E(x^{a\mp1})   E(x^{a\pm1})^2} \right)\\
   \Wthree{a}{a\mp1}{a}{a\pm1}{u}&=\frac{g_a^2}{g_{a+1}g_{a-1}}
   \left(\frac{w^{1/2} E(x^2)^2 E(x^{a\pm2}) E(xw^{-1}) E(x^{a\mp1}w^{\pm1}) E(x^aw^{\mp1})}{E(x)^2 E(x^3) E(x^{a\mp1}) E(x^{a\pm1})^2}\right. \nn
   &\left.-\frac{w^{1/2} x E(x^{a\mp1}) E(x^{a\pm3}) E(xw^{-1}) E(xw) E(x^2w^{-1})}{E(x) E(x^2) E(x^3) E(x^{a\pm1})^2}\right)
   \end{align}\label{conjmod3}
   \end{subequations}

\bigskip\bigskip
\goodbreak

\end{document}